\documentclass[a4paper,12pt]{report}

\usepackage{hbni}
\usepackage{srcltx}
\usepackage{titlepage_hbni}

\usepackage{amsmath, amssymb}
\usepackage[bf]{caption}
\usepackage{remreset}
\makeatletter\@removefromreset{footnote}{chapter}\makeatother

\usepackage[all]{xy}
\usepackage{float}
\usepackage{graphicx}
\usepackage{subfig}
\usepackage[linktoc=page]{hyperref}

\title{Study of Early Universe in an M Theoretic Model}
\degree{DOCTOR OF PHILOSOPHY}
\subject{Physical Sciences}
\author{Samrat Bhowmick}

\declaration{
I, hereby declare that the investigation presented
in the thesis has been carried out by me. The work is
original and the work has not been submitted earlier as a
whole or in part for a degree/diploma at this or any other
Institution or University. 
}
\dedication{
{\ldots\ldots to my parents and to the nature \ldots\ldots}
}
\acknowledgements{

First of all I would like to thank my advisor, Prof. S. Kalyana Rama for 
guidance as well as encouragement during this work, and for introducing me to
some interesting aspect of string theory and cosmology. 

I would also like to thank Prof. Sanatan Digal for helping me in my work,
specifically his advice enabled me to develop an understanding of various
numerical technique. Besides my thank extends to all my teachers for helping
me to develop an understanding of physics. I also thank IMSc string theory
group including students, post docs and faculties, for various discussions.

I offer my special regards to all my friends of IMSc for their constant
moral support during my stay in IMSc, which made my life comfortable. Lastly
I would like to thank IMSc for providing me financial support for whole period 
of stay here.
}

\advisor{S Kalyana Rama}

\abstract{
In this thesis we study early universe in the frame work of M theory.
We assume that the early universe is homogeneous, anisotropic, and is dominated
by the mutually BPS intersecting branes of M theory. Certain class of black
holes can be described by string/M theory have similar structure of intersecting
BPS branes configurations. 
We are motivated by such black holes to make a similar model for 
early universe. 

But due to the lack of knowledge of the exact brane dynamics, we use
U duality symmetry of the M theory to get an equation of states. We also 
verify the equations of states obtained by duality also hold for known
case like black holes. Then we solve Einstein equations to get evolution
of early universe.

In particular  We assume that the early universe is homogeneous, anisotropic, 
and is dominated by the mutually BPS $22'55'$ intersecting branes of M theory. 
The spatial directions are all taken to be toroidal. Using analytical and 
numerical methods, we study the evolution of such an universe. We find that, 
asymptotically, three spatial directions expand to infinity and the remaining 
spatial directions reach stabilised values. From string theory perspective, 
the dilaton is hence stabilised also. We give a physical 
description of the stabilisation mechanism. 

Any stabilised values can be 
obtained by a fine tuning of initial brane densities.
The constant sizes depend on certain imbalance among initial values. One 
naturally obtains $M_{11} \simeq M_s \simeq M_4 \text{ and } 
g_s \simeq 1$ within a few 
orders of magnitude. Smaller numbers, for example $M_s \simeq 10^{- 16} M_4$, 
are also possible but require fine tuning. 

In some sense our $22'55'$ configuration is special. We give some example of 
other configurations for which stabilisation can not be achieved. We give their
asymptotic time evolution. We find only $22'55'$ and its U dual configuration
can achieved stabilisation of 7 spacelike dimensions.

Also, from the perspective of four 
dimensional spacetime, the effective four dimensional Newton's constant $G_4$ is 
now time varying. Its time dependence will follow from explicit solutions. 
We find in the present case that, asymptotically, $G_4$ exhibits characteristic 
log periodic oscillations. 

\section*{List of Papers}
\begin{itemize}
  \item 
  { S. ~Bhowmick, S.~Digal, and S.~{Kalyana Rama}},\\
  {\em ``{Stabilisation of Seven
  (Toroidal) Directions and Expansion of the remaining Three in an M theoretic
  Early Universe Model}''},
  \href{http://dx.doi.org/10.1103/PhysRevD.79.101901}{{\em Phys. Rev.}
  {\bfseries D79}, 101901 (2009) }, \\
  \href{http://arxiv.org/abs/0810.4049}{{\ttfamily arXiv:0810.4049 [hep-th]}}.
  \item 
  { S. ~Bhowmick and S.~{Kalyana Rama}}, \\
  {\em ``{From 10+1 to 3+1 dimensions in an
  early universe with mutually BPS intersecting branes}''}, \\
  \href{http://dx.doi.org/10.1103/PhysRevD.82.083526}{{\em Phys. Rev.}
  {\bfseries D82}, 083526 (2010) }, \\
  \href{http://arxiv.org/abs/1007.0205}{{\ttfamily arXiv:1007.0205 [hep-th]}}.
\end{itemize}
}
\begin{document}
\maketitle
\pagenumbering{roman}
\tableofcontents

\listoffigures
\listoftables

\newpage
\thispagestyle{empty}

\pagenumbering{arabic}


\chapter{\bf Introduction}
\section{Background}
In physics one of the problems, mankind has been wondering for all time is 
``how universe evolves?'' In last century advancement of physics gave us enormous
amount of knowledge to address such question in a scientific framework. So
the study of evolution of universe began, which is known as {\em cosmology}.

It is understood that gravity plays main role in evolution of universe,
because it is long ranged and unlike electromagnetic force, it is only
attractive. After the discovery of general relativity people understand that,
spacetime has a rich geometric structure and is dynamic in nature. This
dynamics is governed by Einstein equation. 

It is assumed in cosmology that spacetime is homogeneous and isotropic. 
On the basis of this fact a well known model has been made after the work of
Friedmann, Robertson and Walker. This is known as FRW model. In this model 
matter is taken as homogeneous and isotropic fluids distributed over the 
whole space like surface. FRW model is very successful to describe 
universe for large interval of {\em time}, except very early universe, and 
possibly present day universe (late time acceleration). 

Another striking discovery in physics of last century is quantum mechanics.
It describes motion of atom and subatomic particle with extraordinary
precision. Nevertheless GR, the theory of gravity is classical. So a
challenge posed to physicist is to make a quantum theory of gravity. 
Toward the end of last century string/M theory \cite{gsw,pol,bbs} emerges as
the most promising candidate for quantum gravity. 

String/M -theory unifies all the forces of the nature. In string theory
fundamental building blocks of nature are one or higher dimensional objects,
(open and closed string -- 1 dimensional, $Dp$-branes -- $p$ dimensional, 
$NS5$-branes -- 5 dimensional) instead of point like particle. In M-theory
there are two types of objects $M2$ brane and $M5$ branes which are 2 and 5 
dimensional object respectively. But a new problem appears in this new theory
-- spacetime now become higher dimensional. In string theory spacetime is $9+1$
dimensional and in M theory spacetime is $10+1$ dimensional. 
On the other hand the large spacetime we observe is $3+1$. So other 7 
dimensions of M theory must be curled up to a very tiny scale, beyond the reach 
of present day experiment. So now cosmology has to explain evolution of 
all 10 space like dimension and how dynamically 7 of them stabilised 
whereas other 3 continues to expand. We studied this problem in this thesis.

\section{Preliminaries}
M theory is a 11 dimensional theory which consists of $M2$ branes and $M5$ 
branes.  Various string theories are various limits of M theory. Until very
recently almost nothing was known about full quantum mechanics of M-branes.
Even now we do not have any clear idea. But 11-dimensional supergravity which is 
supposed to be low energy limit of M theory is well understood.

11 dim supergravity is basically a theory of gravity with super symmetry.
It is governed by the action 
\begin{equation}
 \label{sugra11}
 S = \frac{1}{16 \pi G_{11}} \int d^{11}x \, \sqrt{-g} \left( R
   - \frac{1}{2\times 4!} F_4^2 \right) \;+\; S_{CS} \;+\; S_{f}\; .
\end{equation}
Where $F_4$ is a 4 form field strength such that, 
$F_4 = d C_3$. $S_f$ is fermionic part of 
the action and $S_{CS}$ is Chern-Simon term. $M2$ brane and $M5$ brane are 
respectively, electrically and magnetically
charged under $F_4$. This theory contains black hole solutions. 
Black holes are made of stacks of $M2$ or $M5$ branes or intersecting
combination of them. These black hole solutions of single $M2$ branes or $M5$ 
branes preserve all dynamical supersymmetries, that is half of the 
supersymmetries of the action. For intersection, if they
follow certain rules then they also preserve same amount of supersymmetries. 
These configurations are called BPS configuration.
This black hole solutions are well known and will be discussed in some details
in later chapter. We are motivated by this black hole set up to make a model 
for early universe.  

Another novel feature of M theory is duality symmetries. For details of this 
duality see for example \cite{duality,pol,bOrtin}. 
M theory,
compactified on one of the ten space like dimension gives type 
IIA string theory. In string theory there are S and T duality which connect
various string theories. For example T duality on type IIA theory
transforms it to a type IIB theory and vice versa.

Unlike point partials closed string can wind aground compact direction. So
beside momentum it has another quantum number called winding number. Let
$p$ and $w$ be the momentum and winding of a closed string. It wraps
a compact direction of radius $R$. Then T duality is a symmetry
where 
\begin{eqnarray}
 p &\longleftrightarrow& w \nonumber \\
 R &\longleftrightarrow& \frac{\alpha'}{R} \;.
\end{eqnarray}
Where $\alpha'$ is string length. In case of open string T duality transforms
a $Dp$ brane $D(p-1)$ or $D(p+1)$ depending on whether T duality is
applied along the brane direction or normal to it respectively. In this case 
also  $R \longleftrightarrow \frac{\alpha'}{R}$. 

In type IIB string theory there is another symmetry called S duality. This is 
a symmetry between strong coupling and weak coupling. That is under S duality
$\text{coupling constant} \longleftrightarrow (\text{coupling constant})^{-1}$.

Those symmetries also hold in supergravity level, in some cases with a slightly 
larger symmetry group. When 11 dimensional supergravity compactified on
an $S^1$, it gives type IIA supergravity. Which is governed by
\begin{eqnarray}
 \label{sugraIIA}
 S_{IIA} = \frac{1}{16 \pi G_{10}} \int &d^{10}x& \, \sqrt{-g} 
 \left[ e^{-2\phi }\left( R + 4 \partial_\mu \phi \partial^\mu \phi
 - \frac{1}{2} H_3^2 \right) 
 -\frac{1}{2}(F_2^2+\tilde{F}_4^2) \right] \nonumber \\
 &+& S_{CS}\;+\; S_{f}\; .
\end{eqnarray}
Where $\phi$ is dilaton, a scalar field
 which comes from compactified dimension of M theory.
$H_3 (= dB_2)$ is NS-NS 3-form field strength and $F_p (= dA_{p-1})$'s are 
R-R field strength. $\tilde{F}_4$ is 
\begin{equation}
 \tilde{F}_4 =dA_3 - A_1 \wedge F_3 \;.
\end{equation}
$S_{CS}$ is Chern-Simons term, given by
\begin{equation}
 S_{CS} = -\frac{1}{2 \times 16 \pi G_{10}} \int B_2 \wedge F_4 \wedge F_4 \;.
\end{equation}
$S_{f}$ is the action for fermionic fields.

This theory contains $Dp$ branes with $p$ even. They are charged under
RR fields, $F_{p+2}$. Under T duality IIA SuGra becomes IIB SuGra
theory. IIB theory also contains $Dp$ branes with $p$ odd. 
Again S duality is a symmetry of IIB theory with symmetry group 
$SL(2,\mathbb{R})$. Under these duality operations metric, NS-NS and R-R 
fields changes. In Appendix {\bf \ref{ap:TS}} we listed the rules
for this changes.

\section{Present Work}
In early universe, temperatures and densities reach Planckian
scales. Its description then requires a quantum theory of
gravity. A promising candidate for such a theory is string/M
theory. When the temperatures and densities reach string/M
theory scales, the appropriate description is expected to be
given in terms of highly energetic and highly interacting
string/M theory excitations 
\cite{sbh1,sbh2,sbh3,sbh4,sbh4a,sbh5,sbh6,k0603,Bowick,k0610}.
\footnote{ Only string theory is considered in these
references. But their arguments can be extended for M theory
also leading to similar conclusions.} In this context, it has been
proposed in an earlier work an entropic principle according
to which the final spacetime configuration that emerges from
such high temperature string/M theory phase is the one that has
maximum entropy for a given energy. This principle implies,
under certain assumptions, that the number of large spacetime
dimensions is $3 + 1 \;$ \cite{k0610}.

High densities and high temperatures also arise near black hole
singularities. In string theory black holes are made of branes. These 
branes have their excitations. So in terms of this excitations black holes have
temperature. Since branes stays near singularity, high temperature of black 
holes near singularity has meaning.
Therefore, it is reasonable to expect that the
string/M theory configurations which describe such regions of
black holes will describe the early universe also.

Consider the case of black holes. Various properties of a class
of black holes have been successfully described using mutually
BPS intersecting configurations of string/M theory branes.
\footnote{\label{foot:bps} Mutually BPS intersecting configurations means, for
example, that in M theory two stacks of 2 branes intersect at a
point; two stacks of 5 branes intersect along three common
spatial directions; a stack of 2 branes intersect a stack of 5
branes along one common spatial direction; waves, if present,
will be along a common intersection direction; and each stack of
branes is smeared uniformly along the other brane directions.
See \cite{bps} for more details and for other such string/M
theory configurations.} Black hole entropies are calculated from
counting excitations of such configurations, and Hawking
radiation is calculated from interactions between them.

In the extremal limit, such brane configurations consist of only
branes and no antibranes. In the near extremal limit, they
consist of a small number of antibranes also. It is the
interaction between branes and antibranes which give rise to
Hawking radiation. String theory calculations are tractable and
match those of Bekenstein and Hawking in the extremal and near
extremal limits. But they are not tractable in the far extremal
limit where the numbers of branes and antibranes are comparable.
However, even in the far extremal limit, black hole dynamics is
expected to be described by mutually BPS intersecting brane
configurations where they now consist of branes, antibranes, and
other excitations living on them, all at non zero temperature
and in dynamical equilibrium with each other 
\cite{hm1,hm2,hm3,hm4,hm5,hm6,hm7,hm8,hm9,hm10}. 
For the sake of brevity, we will refer to such far
extremal configurations also as brane configurations even though
they may now consist of branes and antibranes, left moving and
right moving waves, and other excitations.

The entropy $S$ of N stacks of mutually BPS intersecting brane
configurations, in the limit where $S \gg 1 \;$, is expected to
be given by
\begin{equation}\label{sn}
S \sim \prod_I \sqrt{n_I + \bar{n}_I} 
\sim {\cal E}^{\frac{N}{2}} 
\end{equation}
where $n_I$ and $\bar{n}_I \;$, $I = 1, \cdots, N \;$, denote
the numbers of branes and antibranes of $I^{th}$ type, ${\cal
E}$ is the total energy, and the second expression applies for
the charge neutral case where $n_I = \bar{n}_I$ for all $I \;$.
The proof for this expression is given by comparing it in
various limits with the entropy of the corresponding black holes
\cite{hm1, hm2}, see also \cite{hm1,hm2,hm3,hm4,hm5,hm6,hm7,hm8,hm9,hm10,cm,
Mathur}.
For $N \le 4
\;$ and when other calculable factors omitted here are restored,
this expression matches that for the corresponding black holes
in the extremal and near extremal limit and, in the models based
on that of Danielsson et al \cite{hm3}, matches upto a numerical
factor in the far extremal limit \cite{hm1} -- \cite{Mathur}
also. However, no such proof exists for $N > 4 \;$ since no
analogous object, black hole or otherwise, is known whose
entropy is $\propto {\cal E}^* \;$ with $* > 2 \;$.

Note that, in the limit of large ${\cal E} \;$, the entropy
$S({\cal E})$ is $\ll {\cal E}$ for radiation in a finite volume
and is $ \sim {\cal E}$ for strings in the Hagedorn regime. In
comparison, the entropy given in (\ref{sn}) is much larger when
$N > 2 \;$. This is because the branes in the mutually BPS
intersecting brane configurations form bound states, become
fractional, 
\footnote{Fractionation of branes states is familiar phenomena in string
theory \cite{fract}, see also \cite{fuzz}. Consider for example a string of 
length $L_T$ with an
wave on it. it wraps a compact direction $n$ times. Each cycle has a length
say $L$, so $L_T=nL$. Total momentum of the wave say $\frac{n_p}{L}$. This 
momentum can be distributed all of these cycles. So strings along each of these
cycles behaves like a fractional string and total entropy, 
$S \propto \sqrt{n \, n_p}$. This is called fractionation. 
Similar thing happens in branes also.}
and support very low energy excitations which lead
to a large entropy. Thus, for a given energy, such brane
configurations are highly entropic.

Another novel consequence of fractional branes is the following.
According to the `fuzz ball' picture for black holes
\cite{fuzz}, the fractional branes arising from the bound states
formed by intersecting brane configurations have non trivial
transverse spatial extensions due to quantum dynamics. The size
of their transverse extent is of the order of Schwarzschild
radius of the black holes. Therefore, essentially, the region
inside the `horizon' of the black hole is not empty but is
filled with fuzz ball whose fuzz arise from the quantum dynamics
of fractional strings/branes.

Chowdhury and Mathur have recently extended the fuzz ball
picture to the early universe \cite{cm, Mathur}. They have argued
that the early universe is filled with fractional branes arising
from the bound states of the intersecting brane configurations,
and that the brane configurations with highest $N$ are
entropically favourable, see equation (\ref{sn}).

However, as mentioned below equation (\ref{sn}) and noted also
in \cite{cm, Mathur}, the entropy expression in (\ref{sn}) is proved
in various limits for $N \le 4 \;$ only and no proof exists for
$N > 4$. Also, we are not certain of the existence of any system
whose entropy $S({\cal E})$ is parametrically larger than ${\cal
E}^2$ for large ${\cal E} \;$. See related discussions in
\cite{k0702, k0707}. Therefore, in the following we will assume
that $N \le 4 \;$. Then, a homogeneous early universe in
string/M theory may be taken to be dominated by the maximum
entropic $N = 4 \;$ brane configurations distributed uniformly
in the common transverse space.

Such $N = 4$ mutually BPS intersecting brane configurations in
the early universe may then provide a concrete realisation of
the entropic principle proposed earlier by one of us to
determine the number $(3 + 1)$ of large spacetime dimensions
\cite{k0610}. Indeed, in further works \cite{k0702, k0707, bdr,br10},
using M theory symmetries and certain natural assumptions, it has
been shown that these configurations lead to three spatial
directions expanding and the remaining seven spatial directions
stabilising to constant sizes.

In this thesis, we assume that the early universe in M theory is
homogeneous and anisotropic and that it is dominated by $N = 4$
mutually BPS intersecting brane configurations. \footnote{ There
is an enormous amount of work on the study of early universe in
string/M theory. For a small, non exhaustive, sample of such
works, see 
\cite{m,k97,bv,gas,3+1,gm,im1,im2,burt1,gz,sudipta1,sudipta2,sudipta3}.}
In this context, it is
natural to assume that all spatial directions are on equal
footing to begin with. Therefore we assume that the ten
dimensional space is toroidal. We then present a thorough
analysis of the evolution of such an universe.

The corresponding energy momentum tensor $T_{A B} \;$ has been
calculated in \cite{cm} under certain assumptions. However,
general relations among the components of $T_{A B}$ may be
obtained \cite{k0707} using U duality symmetries of M theory
which are, therefore, valid more generally. 
We show in this
thesis that these U duality relations alone imply, under a
technical assumption, that the $N = 4$ mutually BPS intersecting
brane configurations with identical numbers of branes and
antibranes will asymptotically lead to an effective $(3 + 1)$
dimensional expanding universe.

In order to proceed further, and to obtain the details of the
evolution, we make further assumptions about $T_{A B} \;$. We
then analyse the evolution equations in D dimensions in general,
and then specialise to the eleven dimensional case of interest
here.

We are unable to solve explicitly the relevant equations.
However, applying the general analysis mentioned above, we
describe the qualitative features of the evolution of the $N =
4$ brane configuration. In the asymptotic limit, three spatial
directions expand as in the standard FRW universe and the
remaining seven spatial directions reach constant, stabilised
values. These values depend on the initial conditions and can be
obtained numerically. Also, we find that any stabilised values
may be obtained, but requires a fine tuning of the initial brane
densities.

Using the analysis given here, we present a physical description
of the mechanism of stabilisation of the seven brane directions.
The stabilisation is due, in essence, to the relations among the
components of $T_{A B} \;$ which follow from U duality
symmetries, and to each of the brane directions in the $N = 4
\;$ configuration being wrapped by, and being transverse to,
just the right number and kind of branes. This mechanism is very
different from the ones proposed in string theory or in brane
gas models \cite{bv} -- \cite{gm} to obtain large $3 + 1$
dimensional spacetime. (See section {\bf I A} below also.)

In the asymptotic limit, the eleven dimensional universe being
studied here can also be considered from the perspective of four
dimensional spacetime. One then obtains an effective four
dimensional Newton's constant $G_4 \;$ which is now time
varying. Its precise time dependence will follow from explicit
solutions of the eleven dimensional evolution equations.

We find that, in the case of $N = 4$ brane configuration, $G_4
\;$ has a characteristic asymptotic time dependence : the
fractional deviation $\delta G_4 \;$ of $G_4 \;$ from its
asymptotic value exhibits log periodic oscillations given by
\begin{equation}\label{fG4}
\delta G_4 \; \propto \; 
\frac{1}{t^\alpha} \; 
\; Sin (\omega \; ln \; t + \phi ) \; \; . 
\end{equation}
The proportionality constant and the phase angle $\phi \;$
depend on initial conditions and matching details of the
asymptotics, but the exponents $\alpha \;$ and $\omega \;$
depend only on the configuration parameters.  Such log periodic
oscillations seem to be ubiquitous and occur in a variety of
physical systems \cite{braaten, kol, sornette}.  But, to our
knowledge, this is the first time it appears in a cosmological
context. One expects such a behaviour to leave some novel
imprint in the late time universe, but its implications are not
clear to us.

Since we are unable to solve the evolution equations explicitly,
we analyse them using numerical methods. We present the results
of the numerical analysis of the evolution. We illustrate the
typical evolution of the scale factors showing stabilisation and
the log periodic oscillations mentioned above. By way of
illustration, we choose a few sets of initial values and present
the resulting values for the sizes of the stabilised directions
and ratios of the string/M theory scales to the effective four
dimensional scale.

We also discuss critically the implications of our assumptions.
As we will explain, many important dynamical questions must be
answered before one understands how our known $3 + 1$
dimensional universe may emerge from M theory. Until these
questions are answered and our assumptions justified, our
assumptions are to be regarded conservatively as amounting to a
choice of initial conditions which are fine tuned and may not
arise naturally.

\section{Organisation of the Thesis}
The organisation of this thesis is as follows: 

In chapter {\bf \ref{chap:2}} we discuss various black hole solution and 
their energy momentum tensor. Then we discuss U duality relations in 
M theory and string theory. 

In first part of chapter {\bf \ref{chap:2}} (sections {\bf \ref{sec:BH}} to
{\bf \ref{sec:2255}}) we discuss some known result of black holes in
M theory. In the second part (sections {\bf \ref{sec:Ud}} to
{\bf \ref{sec:stringd}}) we discuss U duality in M and string theory, and 
their applications in black hole solution.

In chapter {\bf \ref{chap:3}}, evolution of early universe has been discussed.
In section {\bf \ref{sec:EoMC}}, we present equations of motion fro our 
cosmological model. In section {\bf \ref{sec:Ucos}}, we discuss consequences
of U duality in our cosmological model, and consequently in section 
{\bf \ref{sec:genr}} and {\bf \ref{sec:TAB}}, we give a general result and make
our ansatz for $T_{AB}$. 

In section {\bf \ref{sec:GEvol}} to section {\bf \ref{sec:asevol}}, we
present first a general analysis of $D$ dimensional evolution equations and
then we specialise to eleven dimensional case of $N=4$ intersecting 
brane configurations and described various result mentioned above. 

In section {\bf \ref{sec:Mec}}, we present mechanism of stabilisation
in details and then in section {\bf \ref{sec:size}}, we discuss stabilised
values of the brane directions, their ranges, and the necessity of fine tuning.

In section {\bf \ref{sec:no2255}}, we discuss consequences of some example of
intersecting configurations which are not $22'55'$, and we showed in these cases
stabilisation can not be achieved.

In section {\bf \ref{sec:G(t)}}, we present the four dimensional
perspective and the time variations of $G_4 \;$. In section {\bf
\ref{sec:Num}}, we present the results of numerical analysis. In section
{\bf \ref{sec:concl}}, we conclude by presenting a brief summary, a few
comments on the assumptions made, and by mentioning a few issues
which may be studied further.


\chapter{\bf  Black Holes and Duality in M-Theory and String Theory} 
\label{chap:2}
U-duality is a symmetry of M-theory which consists of T-duality, S-duality of 
string theory and 
dimensional reduction and dimension upliftment. In certain cases of supergravity
solutions this symmetries can be used to get relations among various metric 
component. Theses relations can be used to get relations among various 
components of energy-momentum tensors. We will describe the 
procedure in detail in this chapter. In our model for early
universe we use these relations to find equations of states.

We here discuss intersecting M branes system. In this chapter we study black 
holes made by mutually BPS intersecting M-branes. In chapter 
{\bf \ref{chap:3}} we discuss mutually BPS intersecting M-branes dominated
universe. To be specific,
we consider $22'55'$ configurations.\footnote{ In our notation,
$22'55'$ denotes two stacks each of 2 branes and 5 branes, all
intersecting each other in a mutually BPS configuration.
Similarly for other configurations, {\em e.g.} $55'5''W$ denotes
three stacks of 5 branes intersecting in a mutually BPS
configuration with a wave along the common intersection
direction.} 

Such system is dictated by a $(10 + 1)$ dimensional effective action
given, in the standard notation, by
\begin{equation}
\label{s11}
S_{11} = \frac{1}{16 \pi G_{11}} \; \int d^{11} x \; 
\sqrt{-g} \; R + S_{br} 
\end{equation}
where $S_{br}$ is the action for the fields corresponding to the
branes. The corresponding equations of motion are given, in the
standard notation and in units where $8 \pi G_{11} = 1 \;$, by
\footnote{ In the following, the convention of summing over
repeated indices is not always applicable. Hence, we will always
write the summation indices explicitly. Unless indicated
otherwise, the indices $A, B, \cdots$ run from $0$ to $10$, the
indices $i, j, \cdots$ from $1$ to $10$, and the indices $I, J,
\cdots$ from $1$ to $N$.}
\begin{equation}\label{r11}
R_{A B} - \frac{1}{2} \; g_{A B} R = T_{A B} 
\; \; \; , \; \; \; \; 
\sum_A \nabla_A T^A_{\; \; \; B} = 0 
\end{equation}
where $A = (0, i) \;$ with $i = 1, 2, \cdots, 10 \;$ and $T_{A
B}$ is the energy momentum tensor corresponding to the action
$S_{br} \;$, defined by
\begin{equation}
 \delta_g S_{br} = \frac{1}{2}
 \;  \int d^{11} x \; \sqrt{-g} \; T_{AB} \; \delta g^{AB} \;,
\end{equation}
where $\delta_g$ is the variation with respect to gravitational fields
$g_{AB}$. Equations of motion for the brane fields can be found varying the
action with respect to brane fields. 
\begin{equation}
 \frac{\partial S}{\partial \,(\text{Brane fields})} = 0.
\end{equation}
If we know brane fields explicitly we can find equation of motion. For example
in black hole cases we show them in next section.

\section{Black Brane Energy Momentum Tensor and Solutions} \label{sec:BH}
For black hole case, $T_{A B}$ is obtained from the action for
higher form gauge fields. That is $S_{br}$ is known. 
With a suitable ansatz for the metric,
equations of motion (\ref{r11}) can be solved to obtain black hole
solutions. 

To explain the process consider 11-dimensional supergravity
action. The bosonic part of the action is
 \begin{equation}
 \label{Mact}
 S = \frac{1}{16 \pi G_{11}} \int d^{11}x \, \sqrt{-g} \left( R
   - \frac{1}{2\times 4!} F_4^2 \right)\; ,
\end{equation}
where $F_4$ is a 4 form field, $F_4 = d C_3$. So $S_{br}$ is given by
\begin{equation}
 \label{bhsbr}
 S_{br} = \frac{1}{16 \pi G_{11}} \; \int d^{11} x \; \sqrt{-g}
        \left( - \frac{1}{2\times 4!} F_4^2 \right)\; .
\end{equation}\
The existence of 3-form gauge 
potential suggests that the theory contains 2-branes which couples
to 3-form via
\begin{equation}
 Q_2 \int d^3 \xi \; C_{MNL} \frac{\partial x^M}{\partial \xi^a}\, 
\frac{\partial x^N}{\partial \xi^b}\, \frac{\partial x^L}{\partial \xi^c}
\end{equation}
where $\xi^a$ are world volume coordinates of 2-brane. Equations of motion for
$F_4$ are
\begin{equation}
 \label{maxwell}
 \nabla_M F^{MNPQ} = 0 \;,
\end{equation}
where right hand side is set to zero because fermionic currents corresponding to
branes are set to zero. 
This 4-form field also satisfy Bianchi identity,
\begin{equation}
 \label{bianchi}
 \nabla_M F_{NPQL} + \nabla_L F_{MNPQ} + \nabla_Q F_{LMNP} + \nabla_P F_{QLMN}
 + \nabla_N F_{PQLM} = 0 \;. 
\end{equation}

The 
theory given by (\ref{Mact}) contains a 2 dimensional and a 5 dimensional 
objects $M2$ and $M5$ brane.
$M2$ branes are electrically charged and $M5$ branes are magnetically charged 
under $F_4$.
For this matter field, $F_4$ energy-momentum tensor, $T_{AB}$ is given by
\begin{equation}
 \label{bhTAB}
 T_{AB} = \frac{1}{48} \; \left[4 F_{AMNP} \, F_B{}^{MNP}
          -\frac{1}{2} g_{AB} \, F_4^2 \right] \; .
\end{equation}
Second equation of equation (\ref{r11}),
\begin{equation}
 \label{tconv}
 \sum_A \nabla_A T^A_{\; \; \; B} = 0 
\end{equation}
now follows from equations 
(\ref{maxwell}), (\ref{bianchi}) and (\ref{bhTAB}). to show this consider
\begin{eqnarray*}
 \sum_A \nabla_A T^A_{\; \; \; B} 
&=& 4 (\nabla_A F^{AMNP})F_{BMNP} \\
&+& 4 F^{AMNP} (\nabla_A F_{BMNP}) - g^A{}_B \; (\nabla_A F_{MNPL}) F^{MNPL}\;.
\end{eqnarray*}
First term is 0 by equation (\ref{maxwell}). Last two terms can be written as
\begin{equation*}
 - F^{MNPL} \; 
 \left( \nabla_L F_{BMNP} +  \nabla_P F_{LBMN} + \nabla_N F_{PLBM}
 + \nabla_M F_{NPLB} + \nabla_B F_{MNPL} \right) \;,
\end{equation*}
which is zero by equation (\ref{bianchi}).

This theory contains black brane solution, which are solutions with
solitonic objects.  
These black holes are actually made of stack of $M2$ or $M5$ branes
or BPS intersecting combinations of them. With a suitable ansatz for metric 
and fields Einstein equation can be solved to obtain black hole solutions. 
First we calculate energy momentum tensor for various intersecting brane 
configurations. To do that we make suitable ansatz for metric and justify it.

\section{General Black Brane solutions}\label{sec:gbbs}
Now we are in a position to get black hole solution for BPS configuration.
*[reference]
To get the solutions, let the spacetime coordinates be $x^A = (r,
x^\alpha) \;$ where $x^\alpha = (x^0, x^i, \theta^a) \;$ with
$x^0 = t \;$, $i = 1, \cdots, q \;$, $a = 1, \cdots, m \;$, and
$q + m = 9 \;$. The $x^i$ directions may be taken to be
toroidal, some or all of which are wrapped by branes, and
$\theta^a \;$ are coordinates for an $m$ dimensional space of
constant curvature given by $\epsilon = \pm 1$ or $0 \;$.
The
metric and brane fields depend only on $r$ coordinate, 
and defined by $r^2=\sum_{\alpha=q+1}^{q+m}(x^\alpha)^2$. We write
the line element $d s$, in an obvious notation, as
\begin{equation}\label{dsbh}
d s^2 = - e^{2 \lambda_0} d t^2 
+ \sum_i e^{2 \lambda^i} (d x^i)^2 + e^{2 \lambda} d r^2 
+ e^{2 \sigma} d \Omega^2_{m, \epsilon} \; \; .
\end{equation} 
Black hole solutions are given by $\epsilon = +1$. 
But the
analysis is true for any maximally symmetric non-compact space.

The independent non vanishing components of $T^A_{\; \; B} \;$
are given by $T^r_{\; \; r} = f \;$ and $T^\alpha_{\; \; \alpha}
= p_\alpha \;$ where $\alpha = (0, i, a) \;$. These components
can be calculated explicitly using the action $S_{br} \;$. For
example, for an electric $p$-brane along $(x^1, \cdots, x^p) \;$
directions, they are given by (see equation (\ref{bhTAB}))
\begin{equation}\label{electric}
p_0 = p_\parallel = - p_\perp = - p_a = f 
= \frac{1}{4} \; F_{0 1 \cdots p r} \; 
F^{0 1 \cdots p r} 
\end{equation} 
where $p_\parallel \; = p_i$ for $i = 1, \cdots, p \;$, $p_\perp
\; = p_i$ for $i = p + 1, \cdots, q \;$, and note that $f \;$ is
negative. For mutually BPS $N$ intersecting brane
configurations, it turns out 
\cite{addn1,addn2,addn3,addn4,addn5,addn6,addn7,addn8,addn9,addn10} that
the respective energy momentum tensors $T^A_{\; \; B} \;$ and
$T^A_{\; \; B (I)} \;$ obey conservation equations separately. 
We explained this point in detail in previous section.
\begin{equation}\label{tabI}
T^A_{\; \; B} = \sum_I T^A_{\; \; B (I)} 
\; \; , \; \; \; 
\sum_A \nabla_A  T^A_{\; \; B (I)} = 0 \; \; . 
\end{equation}

Equations of motion may now be obtained from equations
(\ref{r11}) and (\ref{tabI}). After some manipulations, they may
be written as follows:
\begin{eqnarray}
\Lambda_r^2 - \sum_\alpha (\lambda^\alpha_r)^2 & = & 
2 \; f + \epsilon \; m (m - 1) e^{- 2 \sigma} \label{t21bh} \\
\lambda^\alpha_{r r} + \Lambda_r \lambda^\alpha_r & = & 
- \; p_\alpha + \frac{1}{9} \; (f + \sum_\beta p_\beta)
\nonumber \\
& & + \; \epsilon \; (m - 1) e^{- 2 \sigma} \; \delta^{\alpha a}
\label{t22bh} \\
f_r + f \Lambda_r - \sum_\alpha p_\alpha \lambda^\alpha_r & = & 0
\label{t23bh}
\end{eqnarray}
where $\Lambda = \sum_\alpha \lambda^\alpha = \lambda^0 + \sum_i
\lambda^i + m \sigma \;$ and the subscripts $r$ denote
$r$-derivatives. 
One can see equation (\ref{t23bh}), which is same as (\ref{tconv}) in this case
is consequence of equations of motion for 4-form gauge field and definition
of energy momentum tensor.
In case of intersecting branes $p_\alpha = \sum_I p_{\alpha\,(I)}$, 
$f = \sum_I f_I$ and equation (\ref{t23bh}) may be written as
\begin{equation}
 \label{t23bhI}
 f_{I\,r} + f_{I} \Lambda_r - \sum_\alpha p_{\alpha\,(I)} 
\lambda^\alpha_r = 0\;.
\end{equation}
This is because of, as we already claimed, $T^A{}_{B\,(I)}$ obey conservation
equation separately.
See \cite{addn9} particularly, whose set up and
the equations of motions are closest to the present ones.

The other equations of motion are the equations for 4-form field., which is 
same as equations (\ref{maxwell}) and (\ref{bianchi}). Equation (\ref{t23bhI})
can be derived from equations (\ref{maxwell}) and (\ref{bianchi}).

If we assume $p_{\alpha \,(I)} = - (1 - u^I_\alpha) \;f_I \;$, then equation
(\ref{t23bhI}) can be solved. The solution is found to be
\begin{equation}
\label{f_I}
f_I = - \; e^{l^I - 2 \Lambda} \; \; , \; \; \; 
l^I = \sum_\alpha u^I_\alpha \lambda^\alpha +l^I_0 \;.
\end{equation}
Now we define the matrices $G_{\alpha \beta}$ and ${\cal G}^{I J}$ as
\begin{equation}
\label{defnG}
 G_{\alpha \beta} = 1 - \delta_{\alpha \beta} 
\; \; \; , \; \; \; \;
{\cal G}^{I J} = \sum_{\alpha, \beta} 
G^{\alpha \beta} \; u^I_\alpha \; u^J_\beta \;  ,
\end{equation}
where $G^{\alpha\beta}$ is the inverse of $G_{\alpha\beta}$ and is given by
\begin{equation}
 G^{\alpha\beta} = \frac{1}{9} -\delta^{\alpha\beta} \;.
\end{equation}
If we define a 
new coordinate $\tau$ by $d\tau = e^{- \Lambda} \; dr$, equation (\ref{t22bh})
becomes 
\begin{eqnarray}
\label{lambdatautau}
e^{-2\Lambda} \; \lambda^\alpha_{\tau \tau} =  
- \; p_\alpha + \frac{1}{9} \; (f + \sum_\beta p_\beta)
+ \; \epsilon \; (m - 1) e^{- 2 \sigma} \; \delta^{\alpha a} \;.
\label{t22bhtau} 
\end{eqnarray}
If one multiplies both side by $u_\alpha$ and take a sum over $\alpha$ and then 
use equation (\ref{f_I}) in (\ref{t22bhtau}) one finds
\begin{equation}
\label{genlI}
l^I_{\tau \tau} = - \; \sum_J {\cal G}^{I J} \; e^{l^J} 
+ \sum_{a \in \Omega} u_a \; \epsilon \; (m - 1) \; e^{2 (\Lambda - \sigma)}
\; \; .
\end{equation}
Specific values of ${\cal G}^{I J}$ depend on 
intersecting configuration. In case 
of black holes we can calculate them from explicit calculation of
energy momentum tensor or by using duality. In section {\bf \ref{sec:Ubh}} we
discuss this point in detail.

If the components of energy momentum tensor follow a relation like
\begin{equation}
 \label{rlnbh}
 \sum_\alpha c_\alpha 
\left( -p_\alpha + \frac{1}{9} \; (f + \sum_\beta p_\beta) \right) = 0
\end{equation}
then this immediately implies a relation among metric components $\lambda^\alpha$
and $\sigma$. We will see, for various brane solutions
from explicit calculation and also using duality same type of relations come.
For example, when $\alpha \ne a$, then equations (\ref{lambdatautau}) and
(\ref{rlnbh}) implies 
\begin{equation}
 \label{weakbh}
 \sum_\alpha c'_\alpha \lambda^\alpha =0 \;.
\end{equation}
Explicit intersecting configurations give values of $c_\alpha$ and $c'_\alpha$.
We will see some examples in the next section.

\section{Energy-Momentum Tensor of Black Brane Solution} \label{sec:bbsl}
In this section we calculate Energy-Momentum Tensor for various black brane
configurations. First we calculate for $M2$ and $M5$ brane case. Then 
we also give energy-momentum tensor of various intersecting combinations of
them, both BPS and non-BPS. 

We show here that, for BPS intersecting brane configurations, total energy 
momentum tensor is just sum of energy momentum
tensors created by individual set of branes. On the other hand we give an 
example of 2 sets of intersecting $M2$ branes configuration, which
does not follow BPS rule, total energy momentum tensor is not just
sum over that of individual brane solutions.

In BPS cases we also notice, certain relations hold among 
various components of
$T^A{}_B$. In a later section 
we will show that, these relations can be obtained by using U duality 
without computing explicit form of $T^A{}_B$.

\subsection{$M2$ Branes} \label{sec:2}
Consider a stack of $M2$ branes along $(x^1,x^2)$ directions. $x^1$ and $x^2$ 
are taken to be compact. $x^3$ and $x^4$ are also taken to be compact.
In this case the solution is known, it is given by in extremal case
\begin{equation}
 \label{M2soln1}
 ds^2 = H_2^{-\frac{2}{3}} \left( -dt^2 + \sum_{i=1}^2(dx^i)^2 \right)
      + H_2^{\frac{1}{3}}\left(\sum_{i=3}^4(dx^i)^2+dr^2+r^2 d\Omega^2_7 
      \right) \;,
\end{equation}
where $H_2(r) = 1-\frac{1}{r^4}$ and $r^2 = \sum_{i=5}^{10} (x^i)^2$. 
In general
line element of this solution can be taken to be 
\begin{equation}
\label{bbraneanz}
 ds^2 = -e^{2\lambda^0(r)}\,dt^2+\sum_{i=1}^4 e^{2\lambda^i(r)}\,(dx^i)^2
      + e^{2\lambda(r)} \left(dr^2+r^2 d\Omega^2_5 \right) \;.
\end{equation}
Ansatz for field, $C_{MNP}$ is 
\begin{equation}
 C_{012} = f(r) \;,
\end{equation}
which gives 
\begin{equation}
 F_{012r} = f'(r) = E(r)\;,
\end{equation}
where ${}'$ indicate
derivative with respect to $r$.
This actually means $M2$ branes are electrically charged under the field $F_4$.
Energy momentum tensor is given by
\begin{equation}
 \label{emt2}
 T^0{}_0 = T^\parallel{}_\parallel = -T^\perp{}_\perp = T^r{}_r = 
 -T^a{}_a = -\frac{1}{4}\; e^{-2(\lambda^0+\lambda^1+\lambda^2+\lambda)} 
 (E(r))^2\;,
\end{equation}
where indices $\parallel$ and $\perp$ indicate parallel and
perpendicular to the brane directions
and $a$ indicates directions in $\Omega_5$ respectively. 

As we have shown in last section that, conservation equation follows
from equations (\ref{maxwell}), (\ref{bianchi}) and (\ref{bhTAB})
here also 
conservation equation $\sum_A \nabla_A T^A{}_B = 0$ is satisfied.

One can see from equations (\ref{emt2}) and (\ref{t22bh}) that the 
constraining relation
among scale factor, mentioned before turns out to be
\begin{equation}
 \label{sf2}
 \lambda^0 = \lambda^\parallel = - 2\lambda^\perp \;.
\end{equation}

\subsection{BPS Intersection of 2 Sets of $M2$ Branes} \label{sec:22'}

Now consider 2 sets of intersecting $M2$ branes along $(x^1,x^2)$ and 
$(x^3,x^4)$. we denote first set by 2 and second set by $2'$. Black brane 
solution of intersecting branes was first identified in \cite{intM}, then
many solutions were quickly constructed and governing rules of their existence 
were studied. For a review of intersecting branes solution see book 
\cite{bOrtin} and reference there in.
This intersecting configuration follows BPS rules. BPS intersecting
solution the intersection rules are discussed in Appendix {\bf \ref{ap:Ege0}}.
For this 
configuration our metric ansatz remains same as above. 
\begin{equation*}
 ds^2 = -e^{2\lambda^0(r)}\,dt^2+\sum_{i=1}^4 e^{2\lambda^i(r)}\,(dx^i)^2
      + e^{2\lambda(r)} \left(dr^2+r^2 d\Omega^2_5 \right) \;.
\end{equation*}
Nevertheless, we have now two sets of electrically charged branes, so our 
ansatz for $C_{MNP}$ gets changed. Now we have
\begin{eqnarray}
 C_{012} &=& f_2(r) \nonumber \\
 C_{034} &=& f_{2'}(r) \;.
\end{eqnarray}
So now non vanishing components are $C_{012}$ and $C_{034}$ and their cyclic
permutations.
Above potential gives field strength of the form
\begin{eqnarray}
 F_{012r} &=& f_2'(r) = E_2(r)\nonumber \\
 F_{034r} &=& f_{2'}'(r) =E_{2'}(r) \;.
\end{eqnarray}
Energy momentum tensors in this case found to be 
\begin{eqnarray}
  \label{emt22}
  T^0{}_0 &=& -\frac{1}{4}\; e^{-2(\lambda^0+\lambda)}
 \left\{e^{-2(\lambda^1+\lambda^2)}(E_2(r))^2 + 
 e^{-2(\lambda^3+\lambda^4)}(E_{2'}(r))^2\right\} \nonumber \\
  T^1{}_1 = T^2{}_2 &=&-\frac{1}{4}\; 
  e^{-2(\lambda^0+\lambda)}
 \left\{e^{-2(\lambda^1+\lambda^2)}(E_2(r))^2 - 
  e^{-2(\lambda^3+\lambda^4)}(E_{2'}(r))^2\right\} \nonumber \\
  T^3{}_3 = T^4{}_4 &=& \frac{1}{4}\; 
  e^{-2(\lambda^0+\lambda)}
 \left\{e^{-2(\lambda^1+\lambda^2)}(E_2(r))^2 - 
 e^{-2(\lambda^3+\lambda^4)}(E_{2'}(r))^2\right\} \nonumber \\
  T^r{}_r &=& -\frac{1}{4}\; e^{-2(\lambda^0+\lambda)}
 \left\{e^{-2(\lambda^1+\lambda^2)}(E_2(r))^2 + 
 e^{-2(\lambda^3+\lambda^4)}(E_{2'}(r))^2\right\} \nonumber \\
  T^a{}_a &=& \frac{1}{4}\; e^{-2(\lambda^0+\lambda)} 
 \left\{e^{-2(\lambda^1+\lambda^2)}(E_2(r))^2 + 
 e^{-2(\lambda^3+\lambda^4)}(E_{2'}(r))^2\right\} \;.
\end{eqnarray}

One can see easily from above equations, that total energy momentum tensors are
just the sum of individual brane configurations,
\begin{equation*}
 T^A{}_B = \sum_I \; T^A{}_{B\,(I)} \; .
\end{equation*}
From general analysis of section {\bf \ref{sec:BH}} 
we can say that conservation equation $\sum_A \nabla_A T^A{}_B = 0$ is 
satisfied for total energy momentum tensor as well as individual energy
momentum tensors. Last conclusion is confirmed by the fact that, 
conservation of $T^A{}_{B\,(I)}$ is verified in section {\bf \ref{sec:2}}.

Like in previous subsection here also a relation among scale factors 
comes out using equations (\ref{emt22}) and equation of motion.
\begin{eqnarray}
 \label{sf22}
 \lambda^1 &=& \lambda^2 \nonumber \\
 \lambda^3 &=& \lambda^4 \nonumber \\ 
 2\lambda^1+2\lambda^3+\lambda^0 &=& 0\;.
\end{eqnarray}

\subsection{$M5$ Branes} \label{sec:5}
In case of $M5$ brane, just like $M2$ brane case metric ansatz is taken of the 
same form, except now 5 of the 10 spacelike dimensions 
$(x^1,x^2,x^3,x^4,x^5)$ are compact and $M5$ branes wrap them. In general
we may take some of the other directions are also compact. In that case they 
will be treated as directions perpendicular to branes and will be in 
same footing as directions of $\Omega_4$.
Metric ansatz is taken to be
\begin{equation}
 ds^2 = -e^{2\lambda^0(r)} dt^2+\sum_{i=1}^5 e^{\lambda^i} (dx^i)^2
      + e^{2\lambda(r)} \left(dr^2+r^2 d\Omega^2_4 \right) \;,
\end{equation}
where now $r^2 = \sum_{i=6}^{10} (x^i)^2$.
$M5$ branes are magnetically charged under the gauge field $F_4$. Its dual
field is a 7-form field $F_7$. It is electrically charged under this 7-form
$F_7$. $F_4$ and $F_7$ are related by
\begin{equation}
 (F_4)_{MNPQ} = 
   \sqrt{-g} \; \epsilon_{012345rMNPQ} (F_7)^{01\cdots r}\;.
\end{equation}
So one may take ansatz for for 3-form potential is of the form
\begin{equation}
 C_{NPQ} = \frac{1}{4} \epsilon_{012345rMNPQ}
 f(r) x^M \;.
\end{equation}
So $F_4$ takes non zero value only when $(MNPQ) \in 
\{7,8,9,10\}$. 
For computational simplicity and to match with standard notations, we
take $f(r)$ as
\begin{equation}
 f(r) = \sqrt{-g\;} \,g^{00} g^{11} g^{22} g^{33} g^{44} g^{55} g^{rr} E(r) \;,
\end{equation}
where $g_{AB}$ is the metric, $g^{AB}$ is its inverse and $g$ is its 
determinant.
With this ansatz energy momentum tensor turns out to be
\begin{equation}
 \label{emt5}
 T^0{}_0 = T^\parallel{}_\parallel =  T^r{}_r = 
 -T^a{}_a = -\frac{1}{4}\; 
 e^{-2(\lambda^0+\lambda^1+\lambda^2+\lambda^3+\lambda^4+\lambda^5+\lambda)} 
 (E(r))^2\;,
\end{equation}
where index $\parallel$ indicates parallel to brane directions
and $a$ indicates directions in $\Omega_4$ respectively.

Here we have only one type of branes.
Again general argument of section {\bf \ref{sec:BH}} is applicable. 
So
\[\sum_A \nabla_A T^A{}_B = 0\] is true.

Equations (\ref{emt5}) and equations of motion imply relations among
scale factors. These equations are same as (\ref{sf2}).
\begin{equation}
 \label{sf5}
 2 \lambda^0 = 2 \lambda^\parallel = - \lambda^\perp \;.
\end{equation}

\subsection{BPS Intersection of $M2$ Branes and $M5$ Branes} \label{sec:25}
In this subsection we give an example of BPS intersecting configuration of
a stack of  $M2$ branes, stretched along 
$(x^1,x^2)$ and that of $M5$ branes are stretched along 
$(x^1,x^3,x^4,x^5,x^6)$. As before all these $x^1 \cdots x^6$ are compact, 
and the system is localised in common transverse space $x^7 \cdots x^{10}$.
Again ansatz for black brane metric is similar to previous cases. It
is taken of the form
\begin{equation}
 \label{bh25anz}
 ds^2 = -e^{2\lambda^0(r)} dt^2+\sum_{i=1}^6 e^{\lambda^i} (dx^i)^2
      + e^{2\lambda(r)} \left(dr^2+r^2 d\Omega^2_3 \right) \;.
\end{equation}
Here $r^2 = \sum_{i=7}^{10}(x^i)^2$.
Under 4-form gauge field $M2$ branes are electrically charged and $M5$
branes are charged magnetically. So we take an ansatz for our gauge potential
as
\begin{eqnarray}
 C_{012} &=& f_2(r) \nonumber \\
 C_{NPQ} &=& \frac{1}{4} \epsilon_{013456rMNPQ} 
 f_5(r)\,x^M \;,
\end{eqnarray}
where again $f_5(r)$ is taken of the form
\begin{equation}
 f_5(r) = \sqrt{-g\;}\,g^{00} g^{11} g^{33} g^{44} g^{55} g^{66} g^{rr} E_5(r) \;.
\end{equation}
Note that, 
here $\{M,N,P,Q\}$ can take value in $x^3$ and $\Omega_3$, 
otherwise $C_{MPQ}$ will be zero. Now the non-zero components of
energy momentum tensor for this
set of fields are
\begin{eqnarray}
 \label{emt25}
  T^0{}_0 &=& -\frac{1}{4}\; e^{-2(\lambda^0+\lambda^1+\lambda)}
 \left\{e^{-2\lambda^2}(E_2(r))^2 + 
 e^{-2(\lambda^3+\lambda^4+\lambda^5+\lambda^6)}(E_5(r))^2\right\}
  \nonumber \\
  T^1{}_1 &=& -\frac{1}{4}\; e^{-2(\lambda^0+\lambda^1+\lambda)}
 \left\{e^{-2\lambda^2}(E_2(r))^2 + 
 e^{-2(\lambda^3+\lambda^4+\lambda^5+\lambda^6)}(E_5(r))^2\right\}
  \nonumber \\
  T^2{}_2 &=& -\frac{1}{4}\; e^{-2(\lambda^0+\lambda^1+\lambda)}
 \left\{e^{-2\lambda^2}(E_2(r))^2 - 
 e^{-2(\lambda^3+\lambda^4+\lambda^5+\lambda^6)}(E_5(r))^2\right\}
  \nonumber \\
  T^3{}_3 &=& T^4{}_4 \;=\; T^5{}_5 \;=\; T^6{}_6 \nonumber \\
  &=& \frac{1}{4}\; e^{-2(\lambda^0+\lambda^1+\lambda)}
 \left\{e^{-2\lambda^2}(E_2(r))^2 - 
 e^{-2(\lambda^3+\lambda^4+\lambda^5+\lambda^6)}(E_5(r))^2\right\}
  \nonumber \\
  T^r{}_r &=& -\frac{1}{4}\; e^{-2(\lambda^0+\lambda^1+\lambda)}
 \left\{e^{-2\lambda^2}(E_2(r))^2 + 
 e^{-2(\lambda^3+\lambda^4+\lambda^5+\lambda^6)}(E_5(r))^2\right\}
  \nonumber \\
  T^a{}_a &=& \frac{1}{4}\; e^{-2(\lambda^0+\lambda^1+\lambda)}
 \left\{e^{-2\lambda^2}(E_2(r))^2 + 
 e^{-2(\lambda^3+\lambda^4+\lambda^5+\lambda^6)}(E_5(r))^2\right\}
  \;.
\end{eqnarray}
So it is again just the sum of individual $T^A{}_B$ created by $M2$ 
branes and $M5$ branes separately. Same argument of section {\bf \ref{sec:BH}}
goes through about
conservation equation. Also
and $T^A{}_{B\,(I)}$'s are conserved separately as they are, for single $M2$
brane and single $M5$ brane case.

Equations (\ref{emt25}) and equation of motion imply constraining relations, like
before, among scale factors.
\begin{eqnarray}
 \label{sf25}
 \lambda^0 &=& \lambda^1 \nonumber \\
 \lambda^3 &=& \lambda^4 \;=\; \lambda^5 \;=\; \lambda^6 \nonumber \\
 \lambda^2+2\lambda^3 &=& 0\;.
\end{eqnarray}

\subsection{Non-BPS Intersection of Branes} \label{nbps}
Now we will consider an almost similar configuration of subsection 
{\bf \ref{sec:22'}},
except now our configuration is non-BPS. In this case lets take two sets of 2 
branes along $(x^1,x^2)$ and $(x^2,x^3)$. It is a non-BPS intersection of 
branes. We take $x^1,x^2$ and $x^3$ as compact. Both set are 
electrically charged. So the 3-form potential components in this case are
\begin{eqnarray}
 C_{012} &=& f_2(r) \nonumber \\
 C_{023} &=& f_{2'}(r) \;.
\end{eqnarray}
So corresponding fields are 
\begin{eqnarray}
 F_{012r} &=& f_2'(r) = E_2(r)\nonumber \\
 F_{023r} &=& f_{2'}'(r) =E_{2'}(r) \;.
\end{eqnarray}
For metric we may start with an ansatz like (\ref{bh25anz}), but it turns 
out that this ansatz is inconsistent. The reason is explained below. Because 
of the first term in the expression of energy momentum tensor 
(equations (\ref{bhTAB})), $T_{13}$ is non-zero, 
\begin{equation}
 T_{13} = \frac{1}{12} g^{00} g^{22} g^{rr} F_{102r} F_{302r} \times 3! \;.
\end{equation}
But since the metric is diagonal, $(R_{13} - \frac{1}{2} g_{13} \, R)$ is zero.
So obviously Einstein equations are not satisfied. Therefore one has to take a 
different ansatz, simplest one is diagonal metric with only $g_{13}$ non-zero.
That is,
\begin{equation}
\label{nbpsanz}
 ds^2 = -e^{2\lambda^0(r)}\,dt^2+\sum_{i=1}^3 e^{2\lambda^i(r)}\,(dx^i)^2
      +2 \, e^{2\lambda^c} dx^1 dx^3 +
      + e^{2\lambda(r)} \left(dr^2+r^2 d\Omega^2_6 \right) .
\end{equation}
With this ansatz it turns out non vanishing components of Einstein tensor
are all diagonal components and $G_{13}$. So now we can equate $G_{MN}$ and
$T_{MN}$. So we will take  the above line element as our ansatz.

We calculate here $T^{M}{}_N$. The non-zero components of
them turn out to be
\begin{eqnarray}
\label{nbpsTab}
 T^0{}_0 &=& \frac{1}{4}\left(g^{00}g^{11}g^{22}g^{rr}F_{012r}F_{012r} 
             + g^{00}g^{22}g^{33}g^{rr}F_{023r}F_{023r}
             + g^{00}g^{13}g^{22}g^{rr}F_{012r}F_{032r} \right) \nonumber \\
 T^1{}_1 &=& \frac{1}{4}\left(g^{00}g^{11}g^{22}g^{rr}F_{012r}F^{012r} 
             - g^{00}g^{22}g^{33}g^{rr}F_{023r}F_{023r}
             + g^{00}g^{13}g^{22}g^{rr}F_{012r}F_{032r} \right) \nonumber \\
 T^2{}_2 &=& \frac{1}{4}\left(g^{00}g^{11}g^{22}g^{rr}F_{012r}F^{012r} 
             + g^{00}g^{22}g^{33}g^{rr}F_{023r}F_{023r}
             + g^{00}g^{13}g^{22}g^{rr}F_{012r}F_{032r} \right) \nonumber \\
 T^3{}_3 &=& \frac{1}{4}\left(-g^{00}g^{11}g^{22}g^{rr}F_{012r}F_{012r} 
             + g^{00}g^{22}g^{33}g^{rr}F_{023r}F_{023r}
             + g^{00}g^{13}g^{22}g^{rr}F_{012r}F_{032r} \right) \nonumber \\
 T^r{}_r &=& \frac{1}{4}\left(g^{00}g^{11}g^{22}g^{rr}F_{012r}F_{012r} 
             + g^{00}g^{22}g^{33}g^{rr}F_{023r}F_{023r}
             + g^{00}g^{13}g^{22}g^{rr}F_{012r}F_{032r} \right) \nonumber \\
 T^a{}_a &=& -\frac{1}{4}\left(g^{00}g^{11}g^{22}g^{rr}F_{012r}F_{012r} 
             + g^{00}g^{22}g^{33}g^{rr}F_{023r}F_{023r}
             + g^{00}g^{13}g^{22}g^{rr}F_{012r}F_{032r} \right) \;, \nonumber \\
\end{eqnarray}
where index $a$ denotes coordinates in $\Omega_6$. There is another component
$T_{13}$. Now because of $T^1{}_3 = g^{11} T_{13} + g^{13} T_{33}$ and 
$T^3{}_1 = g^{33} T_{13} + g^{13} T_{11}$, $T^1{}_3$ and $T^3{}_1$ are not 
symmetric. They turned out to be various combination of $F_{012r}$ and 
$F_{023r}$, and are non-zero.
From above equations one can see clearly that, total energy momentum 
tensor is not just the sum of individual energy momentum tensors
created by each sets of branes separately. For example in equations 
(\ref{nbpsTab}) first two terms in each equation give energy momentum
tensor for individual brane configuration, but the third term is extra.
Also $T_{13}$ is a new component, which was not in case of single
$M2$ brane system. So we may conclude that, for non-BPS intersection
equation (\ref{tabI}) is not satisfied.

\section{$22'55'$ Black Holes} \label{sec:2255}
BPS intersection of multiple branes describes multicharge extremal black holes.
We will now discuss of 2 $M2$ branes and 2 $M5$ branes intersecting 
configuration. This $22'55'$ configuration is the matter
of our cosmological model. Here we take $x^1$ to $x^7$ are compact.
First set of $M2$ branes wrap $(x^1,x^2)$ directions, second set of $M2$ branes
wrap $(x^3,x^4)$, first $M5$ branes wrap $(x^1,x^3,x^5,x^6,x^7)$ and second 
$M5$ branes wrap $(x^2,x^4,x^5,x^6,x^7)$.
$M2$ branes are electrically charged and $M5$ branes are magnetically charged
under $F_4$. Let $Q_{2}$, $Q_{2'}$, $Q_{5}$, $Q_{5'}$ are the 4 
parameter
which determine mass and charges of the extremal black hole solution we get.
So the components of the gauge field potential are
\begin{eqnarray}
 C_{012} &=& H_2^{-1}(r)-1 \\
 C_{034} &=& H_{2'}^{-1}(r)-1 \\
 C_{LMN} &=& \frac{1}{4} 
 \epsilon_{013567rPLMN} f_5(r) \; x^P\\
 C_{ABC} &=& \frac{1}{4} 
 \epsilon_{024567rPABC}f_{5'}(r)\;x^P \;;
\end{eqnarray}
where
\begin{eqnarray}
 f_5(r) &=& \sqrt{-g\;}\,g^{00} g^{11} g^{33} 
 g^{55} g^{66} g^{77} g^{rr} H_5^{-2}(r) \\
 f_{5'}(r) &=& \sqrt{-g\;}\,g^{00} g^{22} g^{44} 
 g^{55} g^{66} g^{77} g^{rr} H_{5'}^{-2}(r) \;\;.
\end{eqnarray}
$H_I$ are the harmonic function, and given by $H_I = 1+\frac{Q_I}{r}$. 
For this configuration geometry turns out to be
\begin{eqnarray}
 \label{2255soln}
 ds^2 &=&  (H_2 H_{2'})^{-2/3} (H_5 H_{5'})^{-1/3} 
[ 
-dt^2 + H_{2'} H_{5'} (dx^1)^2 + H_{2'} H_{5} (dx^2)^2 + \nonumber \\
& & H_{2} H_{5'} (dx^3)^2 + H_{2} H_{5} (dx^4)^2 +
H_{2} H_{2'} \left\{ (dx^5)^2+ (dx^6)^2 +(dx^7)^2  \right\}  \nonumber \\
& & + H_{2} H_{2'} H_{5} H_{5'} \left( dr^2 + r^2 d \Omega_2^2 \right) 
]
\end{eqnarray}

Here one can calculate energy momentum
tensor for each type of branes separately using equation (\ref{bhTAB}). 
They are 
\begin{eqnarray}
 &T^0{}_{0\,(2)} = T^1{}_{1\,(2)} = T^2{}_{2\,(2)} = -T^3{}_{3\,(2)}= \cdots  
  = -T^8{}_{8\,(2)} \nonumber &\\ 
 &= T^r{}_{r\,(2)} = 
 -T^a{}_{a\,(2)} = 
-\frac{1}{4} \,\frac{(H'_2 / H_2)^2 }{(H_2 H_{2'}) ^{1/3}(H_5 H_{5'})^{2/3}} &
\end{eqnarray}
\begin{eqnarray}
 &T^0{}_{0\,(2')} = T^3{}_{3\,(2')} = T^4{}_{4\,(2')} = -T^1{}_{1\,(2')}= \cdots  
  = -T^8{}_{8\,(2')} \nonumber &\\ 
 &= T^r{}_{r\,(2')} = 
 -T^a{}_{a\,(2')} =
-\frac{1}{4} \,\frac{(H'_{2'} / H_{2'})^2 }
{(H_2 H_{2'})^{1/3}(H_5 H_{5'})^{2/3}} &
\end{eqnarray}
\begin{eqnarray}
 &T^0{}_{0\,(5)} = T^1{}_{1\,(5)} = T^3{}_{3\,(5)} = T^5{}_{5\,(5)}= \cdots  
  = T^7{}_{7\,(5)} = \nonumber &\\ 
 &-T^2{}_{2\,(5)} = -T^4{}_{4\,(5)} = T^r{}_{r\,(5)} = 
 -T^a{}_{a\,(5)} = 
-\frac{1}{4} \,\frac{(H'_5 / H_5)^2 }{(H_2 H_{2'}) ^{1/3}(H_5 H_{5'})^{2/3}} &
\end{eqnarray}
\begin{eqnarray}
 &T^0{}_{0\,(5')} = T^2{}_{2\,(5')} = T^4{}_{4\,(5')} = T^5{}_{5\,(5')}= \cdots  
  = T^7{}_{7\,(5')} = \nonumber &\\ 
 &-T^1{}_{1\,(5')} = -T^3{}_{3\,(5')} = T^r{}_{r\,(5')} = 
 -T^a{}_{a\,(5')} = 
-\frac{1}{4} \,\frac{(H'_{5'} / H_{5'})^2 }
{(H_2 H_{2'})^{1/3}(H_5 H_{5'})^{2/3}} &
\end{eqnarray}
Then
one can see they satisfy conservation equation separately. Total $T^{A}{}_B$ is 
just sum of individual contributions as we have seen in previous examples.
\begin{equation}
 T^{A}{}_B = \sum_{I} T^A{}_{B\,(I)} \;\;.
\end{equation}

\section{U Duality Relations In M Theory} \label{sec:Ud}
We now describe the relations which follow from U duality
symmetries, involving chains of dimensional reduction and
uplifting and T and S dualities of string theory. See
\cite{k0707}. To explain the concept let us 
consider a solution of the form
\begin{equation}\label{dsgen11}
 ds_{11}^2 = - e^{2\lambda^0}dt^2 +
 \sum_{\mu=1}^{10} e^{2 \lambda^\mu} (d x^\mu)^2 \;,
\end{equation}
where we assume for $\mu = i, j, k$ $x^i$ are compact and metric does
not depends on them. That is $\lambda^\mu = \lambda^\mu(t,X)$, where $X$ 
includes space like coordinates except $x^i, x^j$ and $x^k$.
Let $\downarrow_k$ and
$\uparrow_k$ denote dimensional reduction and uplifting along
$k^{th}$ direction between M theory and type IIA string theory.
We are applying operation $\downarrow_{k}$. To do that
we write $ds_{11}$ as
\begin{equation}
  \label{downk}
  ds_{11}^2 = e^{-\frac{2}{3} \phi} \; ds_{10}^2 
  + e^{\frac{4}{3} \phi} \, (dx^{k})^2 \;,
\end{equation}
where $ds_{10}$ is 10 dimensional line element of type IIA theory. 
Type IIA string theory metric is given by
\begin{equation}
 \label{dsIIA}
 ds_{10}^2 = - e^{2\lambda'^{0}}dt^2 
 + \sum_{\mu \ne k} e^{2 \lambda'^\mu} (d x^{\mu})^2 \;.
\end{equation}
$\phi$ is dilaton and is independent of $x^i, x^j$ and $x^k$. It is
function of $(t,X)$ only.
If we integrate over $x^{k}$ with above metric we will get type 
IIA super gravity action.
Here comparing equations (\ref{dsgen11}), (\ref{downk}) and 
(\ref{dsIIA}) one can see
$\phi$ and $\lambda'^\mu$ are given by
\begin{eqnarray}
 \label{down_k}
 \phi &=& \frac{3}{2} \lambda^{k}  \nonumber \\
 \lambda'^\mu &=& \lambda^\mu + \frac{1}{3} \phi
 = \lambda^\mu +\frac{1}{2} \lambda^k
\end{eqnarray}

Now in string theory one can perform T duality along a compact direction. It
converts type IIA theory to type IIB theory and back. 
It also converts a $Dp$ brane to $D(p-1)$ or $D(p+1)$ branes depending on
whether T duality is applied along brane or perpendicular to brane 
respectively.
This is a symmetry of 
the IIA/IIB string theory. A solution, for example, (\ref{dsIIA}) breaks it.
So applying this transformation generates new solution. 
We denote T duality operation along $i^{th}$ direction by $T_i$.
S duality is a 
symmetry of type IIB theory. This is asymmetry between
$\mbox{coupling constant} \leftrightarrow (\mbox{coupling constant})^{-1}$.
Rules for the these duality transformations are listed in Appendix
{\bf \ref{ap:TS}}.

Applying a T duality along say, $x^j$ which we denote by $T_j$, generates
a new solution, given by (see equation (\ref{TD1}) and (\ref{TD6}))
\begin{eqnarray}
 {ds'}_{10}^{2} &=& - e^{2\lambda'^{0}}dt^2 
 + \sum_{\mu \ne \{j,k\} } 
 e^{2 \lambda'^\mu} (dx^\mu)^2 + e^{-2 \lambda'^j} (dx^j)^2 \;,
\nonumber \\
 \phi' &=& \phi - \lambda'^j = \lambda^k - \lambda^j \;,
\end{eqnarray}
where equation (\ref{down_k}) has been used. Note that metric along $x^j$, 
$g_{jj}$ goes to $(g_{jj})^{-1}$ according to equation (\ref{TD1}).
This solution is of type IIB theory.
Again application of $T_i$ generate a new solution of IIA theory. 
\begin{eqnarray}
 \label{T_j}
 {ds''}_{10}^2 &=& - e^{2\lambda'^{0}}dt^2 + \sum_{\mu \ne \{i,j,k\} } 
e^{2 \lambda'^\mu} (dx^\mu)^2 + e^{-2 \lambda'^j} (dx^j)^2
+e^{-2 \lambda'^i} (dx^i)^2 \nonumber\\
 \phi'' &=& \phi - \lambda'^i - \lambda'^j = 
\frac{1}{2} \lambda^k -\lambda^i - \lambda^j \;.
\end{eqnarray}
Again $g_{ii}$ goes to $(g_{ii})^{-1}$ according to equation (\ref{TD1}).

Now dimensional upliftment to 11 dimensional theory can be done via
\begin{equation}
  \label{upk}
  ds'^2_{11} = e^{-\frac{2}{3} \phi''} \; {ds''}_{10}^2 
  + e^{\frac{4}{3} \phi''} \, (dx^{k})^2 \;.
\end{equation}
Using equation (\ref{T_j}) in (\ref{upk}) one gets
\begin{equation}
 d s'^2_{11} = - e^{2{\lambda''}^0}dt^2 + \sum_{\mu=1}^{10} 
e^{2 {\lambda''}{^\mu}} (dx^\mu)^2 \;,
\end{equation}
where these $\lambda{''^i}$'s are given in terms of $\lambda^i$'s by
(using equation (\ref{down_k})),
\begin{eqnarray}
 \lambda{''^i} &=& \lambda^j - \frac{2}{3} (\lambda^i+\lambda^j+\lambda^k)
  \nonumber \\
 \lambda{''^j} &=& \lambda^i - \frac{2}{3} (\lambda^i+\lambda^j+\lambda^k)
  \nonumber \\
 \lambda{''^k} &=& \lambda^k - \frac{2}{3} (\lambda^i+\lambda^j+\lambda^k)
  \nonumber \\
 \lambda{''^l} &=& \lambda^l +  \frac{1}{3} (\lambda^i+\lambda^j+\lambda^k) 
\;\;\;\;\;\;\;  \forall \; l \ne \{i,j,k\}
\end{eqnarray}
In general, simplifying notation, we can write,  
application of U duality $\uparrow_k T_i T_j \downarrow_k$ in 
(\ref{dsgen11}),transforms the $\lambda^i$s in the scale factors to
$\lambda'^i$s given by
\begin{eqnarray}
\label{lambdareln}
& & \lambda'^i = \lambda^j - 2 \lambda \; \; , \; \; \;
\lambda'^j = \lambda^i - 2 \lambda \; \; , \; \; \; 
\lambda'^k = \lambda^k - 2 \lambda \nonumber \\
& & \lambda'^l = \lambda^l + \lambda \; \; , \; \; \;
l \ne \{i, j, k\}  \; \; , \; \; \;
\lambda = \frac{\lambda^i + \lambda^j + \lambda^k}{3} \; \; .
\label{uduality}
\end{eqnarray}

\section{Application of U duality relations in Black Holes} \label{sec:Ubh}
Note that, as can be seen from the above steps, the U duality
relations follow as long as the directions involved in the U
duality operations are isometry directions. Since none of the
common transverse directions are involved in obtaining the
relations above, it follows that they are valid even if the
common transverse directions are not compact. Thus the U
duality relations are applicable in such cases also. 

Similarly, the time dependence of $\lambda^i$s played no role in
obtaining the U duality relations here. Hence, these relations
may be expected to arise for the black hole case also. We will describe 
this case in detail in this section.

Consider black holes in $m + 2$ dimensional spacetime described
by mutually BPS intersecting brane configurations in M theory.
The brane action $S_{br} \;$ in equation (\ref{s11}) is the
standard one for higher form gauge fields and given by equation
(\ref{bhsbr}). Corresponding black hole solutions are given in section
{\bf \ref{sec:bbsl}}. So
here we only highlight the points related to U duality
symmetries. Also, for illustration, we consider only 2 branes
and 5 branes.

As mentioned in section {\bf \ref{sec:Ud}}, the method of U duality
symmetries applies here also and leads to the same relations
between $\lambda^i \;$. They are best seen in the extremal
case. (The non extremal case requires further analysis and is
more involved.)

The eleven dimensional line element $d s$ for 2 brane is
\begin{eqnarray}
 \label{M2soln}
 ds^2 &=& H_2^{-\frac{2}{3}} \left( -dt^2 + \sum_{i=1}^2(dx^i)^2 \right)
      + H_2^{\frac{1}{3}}\left(dr^2+r^2 d\Omega^2_7 \right) \;.
\end{eqnarray}
For $M5$ brane $d s$ is
 \begin{eqnarray}
  \label{M5soln}
  ds^2 &=& H_5^{-1/3}\left(-dt^2+\sum_{i=1}^5(dx^i)^2\right)
       + H_5^{2/3} \left(dr^2+r^2 d\Omega^2_4 \right) \; ,
 \end{eqnarray}
where $H_2$ and $H_5$ are harmonic function and function of $r$.

In general line element for 2 or 5 brane can be 
written in the form
\begin{equation}\label{ds2}
d s^2 = - e^{2 \lambda^0} d t^2 
+ \sum_i e^{2 \lambda^i} (d x^i)^2 
\end{equation}
where $(\lambda^0, \lambda^i) $ depend on $r$, the radial
coordinate of the $m + 1$ dimensional transverse space.
For 2
branes and 5 branes, as can be seen from, equation (\ref{M2soln}) 
and (\ref{M5soln}), the $\lambda^i$s may be written as
\begin{eqnarray}
\lambda^1 = \lambda^2 = - \frac{2 \tilde{h}}{6}
& \; \; \; , \; \; \; \; & 
\lambda^3 = \cdots = \lambda^{10} = \frac{\tilde{h}}{6} 
\label{2bh} \\
\lambda^1 = \cdots = \lambda^5 = - \frac{\tilde{h}}{6}
& \; \; \; , \; \; \; \; &
\lambda^6 = \cdots = \lambda^{10} = \frac{2 \tilde{h}}{6} 
\label{5bh}
\end{eqnarray}
where $e^{\tilde{h}} = H = 1 + \frac{Q}{r^{m - 1}} \;$ is the
corresponding harmonic function and $Q$ is the charge. See, for
example, \cite{cvetic} for more detail.

We see now relations among $\lambda$'s [equations (\ref{2bh}) or (\ref{5bh})]
follow from U duality relations. Consider a solution 
of $M2$ brane along $(x^1,x^2)$. Also take $(x^3,x^4,x^5)$ as compact
and are isometry direction. An obvious symmetry implies 
\begin{equation}
 \label{obv2bh1}
 \lambda^1 = \lambda^2 \;.
\end{equation}
It also implies 
\begin{equation}
 \label{obv2bh2}
 \lambda^3 = \lambda^4 = \lambda^5 \;.
\end{equation}
Directions $x^a$ $(\in \{\Omega_4 \; \text{and} \; r\})$ are also transverse to 
brane directions. So we may assume 
\begin{equation}
 \label{obv2bh3}
 \lambda^3 = \lambda^4 = \lambda^5 =\lambda^6 = \lambda^7 = \lambda^8
= \lambda^9 = \lambda^{10} \;.
\end{equation}
Now apply U duality operations $\downarrow_5 T_3 T_4 \uparrow_5$. It transforms
$M2$ brane to $M5$ brane.
\begin{displaymath}
 \xymatrix{
           M2 (12) \ar[r]^{\downarrow_5} & D2 (12) \ar[r]^{T_4} & D3 (124)
           \ar[r]^{T_3} & D4 (1234) \ar[r]^{\uparrow_5} & M5 (12345)
          } \; .
\end{displaymath}
This new solution is $M5$ branes may be given by
\begin{equation}
 d s'^2_{11} = - d \hat{t}^2 + \sum_{i=1}^{10} 
e^{2 \lambda{'^i}(\hat{t})} (d x^i)^2 \;.
\end{equation}
Where using equations (\ref{lambdareln}) we find
these $\lambda{'^i}$'s are given in terms of $\lambda^i$'s by
\begin{eqnarray}
 \lambda{'^3} &=& \lambda^4 - \frac{2}{3} (\lambda^3+\lambda^4+\lambda^5)
  \nonumber \\
 \lambda{'^4} &=& \lambda^3 - \frac{2}{3} (\lambda^3+\lambda^4+\lambda^5)
  \nonumber \\
 \lambda{'^5} &=& \lambda^5 - \frac{2}{3} (\lambda^3+\lambda^4+\lambda^5)
  \nonumber \\
 \lambda{'^i} &=& \lambda^i +  \frac{1}{3} (\lambda^3+\lambda^4+\lambda^5) 
\;\;\;\;\;\;\;\;\;  \forall \; i \ne \{3,4,5\} \;.
\end{eqnarray}
There are also obvious symmetry relation for $M5$ brane. 
\begin{equation}
 \label{obv5bh}
 \lambda'^1 = \lambda'^2 = \lambda'^3 = \lambda'^4 = \lambda'^5 \;,
\;\;\;
 \lambda'^6 = \lambda'^7 = \lambda'^8 = \lambda'^9 = \lambda'^{10} \;.
\end{equation}
So now one can see equations (\ref{2bh}) and (\ref{5bh}) are satisfied by the 
metric components
of black hole solutions.
in short we can write the relations among $\lambda$'s as 
\begin{equation}
 \label{25relnbh}
 \lambda^\parallel + 2 \lambda^\perp = 0 , \;\;\;
 2 \lambda'^\parallel + \lambda'^\perp =0 \;.
\end{equation}
where the superscripts $\parallel \;$ and $\perp \;$ denote
spatial directions along and transverse to the branes
respectively. 
Note that, to find these relation we have use 
duality relations only. Explicit form of $\tilde{h}$ can only be known
by solving equations of motion and putting proper boundary conditions, like
asymptotic flatness.

For the extremal $22'55' \;$ configuration $(12, 34, 13567,
24567) \;$, the transverse space is three dimensional and 
the $\lambda^i$s may be written as \cite{cvetic} (see equation \ref{2255soln}),
\begin{eqnarray}
\lambda^1 & = & \frac{1}{6} \; ( -2 \tilde{h}_1 
+ \tilde{h}_2 - \tilde{h}_3 + 2 \tilde{h}_4 )
\nonumber \\
\lambda^2 & = & \frac{1}{6} \; ( -2 \tilde{h}_1 
+ \tilde{h}_2 + 2 \tilde{h}_3 - \tilde{h}_4 )
\nonumber \\
\lambda^3 & = & \frac{1}{6} \; ( \tilde{h}_1 
- 2 \tilde{h}_2 - \tilde{h}_3 + 2 \tilde{h}_4 )
\nonumber \\
\lambda^4 & = & \frac{1}{6} \; ( \tilde{h}_1 
- 2 \tilde{h}_2 + 2 \tilde{h}_3 - \tilde{h}_4 )
\nonumber \\
\lambda^5 = \lambda^6 = \lambda^7 & = & \frac{1}{6} \; 
( \tilde{h}_1 + \tilde{h}_2 - \tilde{h}_3 
- \tilde{h}_4 ) \nonumber \\
\lambda^8 = \lambda^9 = \lambda^{10} & = & \frac{1}{6} \; 
( \tilde{h}_1 + \tilde{h}_2 + 2 \tilde{h}_3 
+ 2 \tilde{h}_4 ) \label{2255bh}
\end{eqnarray}
where $e^{\tilde{h}_I} = H^I = 1 + \frac{Q_I}{r} \;$ are the
corresponding harmonic functions and $Q_I$s are the charges. 
Furthermore, if 2 and
$2'$ branes are identical then $\tilde{h}_1 = \tilde{h}_2 \;$
and we get $\lambda^1 = \lambda^3 \;$, and similarly other
relations when different sets of branes are identical.

Note that obvious symmetry relations for $22'55'$ black hole are
\begin{equation}\label{obv2255bh}
\lambda^5 = \lambda^6 = \lambda^7
\; \; , \; \; \; \lambda^8 = \lambda^9 = \lambda^{10} \; \; ,
\end{equation}
and the U duality relation comes out following above steps are
\begin{equation}\label{2255relnbh}
\lambda^1 + \lambda^4 + \lambda^5 = 
\lambda^2 + \lambda^3 + \lambda^5 = 0 \; \; . 
\end{equation}
These equations (\ref{obv2255bh}) and (\ref{2255relnbh}) lead to
equation (\ref{2255bh}).

We further illustrate the U duality methods by interpreting a U
duality relation $\sum_i c_i \lambda^i = 0 \;$ as implying a
relation among the components of the energy momentum tensor
$T_{A B} \;$. The relations thus obtained are indeed obeyed by
the components of $T_{A B}$ calculated explicitly using the
corresponding higher form gauge field action $S_{br} \;$ 
in section {\bf \ref{sec:bbsl}}.

Consider now the case of $2$ branes or 5 branes. We assume that
$p_a = p_\perp \;$ which is natural since $\theta^a \;$
directions are transverse to the branes. Applying the U duality
relations in equation (\ref{25relnbh}) then implies, for both 2
branes and 5 branes, the relation
\begin{equation}\label{zbh}
p_\parallel = p_0 + p_\perp + f 
\end{equation}
among the components of their energy momentum tensors.
See equations (\ref{emt2}) and (\ref{emt5}).
Note that
it is also natural to take $p_0 = p_\parallel \;$ since $x^0 = t
\;$ is one of the worldvolume coordinates and may naturally be
taken to be on the same footing as the other ones $(x^1, \cdots,
x^p) \;$. Equation (\ref{zbh}) then implies that $p_\perp = - f
\;$. The relation between $p_\parallel \;$ and $f$ is to be
specified by an equation of state which, in the black hole case,
is that given in equations (\ref{emt2}) and (\ref{emt5}).

For now, however, we take $p_0$ and $p_\parallel \;$ to be
different. Keeping in mind that $f$ is negative, we assume the
equations of state to be of the form $p_{\alpha I} = - (1 -
u^I_\alpha) \;f_I \;$ where $\alpha = (0, i, a) \;$, $I = 1,
\cdots, N \;$, and $u^I_\alpha \;$ are constants, mentioned in section
{\bf \ref{sec:gbbs}}. Here we give explicitly $u_\alpha$'s for 2 and 5
black brane solution. 
\begin{eqnarray} 
2 & : & u_\alpha = 
(\; \; u_0, \; \; u_\parallel, \; \; u_\parallel, 
\; \; u_\perp, \; \; u_\perp, \; \; u_\perp, 
\; \; u_\perp, \; \; u_\perp, \; \; u_\perp, \; \; u_\perp) 
\nonumber \\
5 & : & u_\alpha = 
(\; \; u_0, \; \; u_\parallel, \; \; u_\parallel, \; 
\; \; u_\parallel, \; \; u_\parallel, \; \; u_\parallel, \;
\; \; u_\perp, \; \; u_\perp, \; \; u_\perp, \; \; u_\perp) 
\label{25Wbh}
\end{eqnarray}
where the $I$ superscripts have been omitted here and
$u_\parallel = u_0 + u_\perp \;$ which follows from equation
(\ref{zbh}). Note that $u_\perp = 0 \;$ and $u_0 = u_\parallel =
2 \;$ in the black hole case given in equation (\ref{electric}) or 
specifically for 2 brane and 5 brane in equations (\ref{emt2}) and 
(\ref{emt5}).

Using definition of ${\cal G}^{IJ}$, $l^I$ and $\tau$ equation (\ref{genlI}) 
now becomes 
\begin{equation}\label{c3bh}
l^I_{\tau \tau} = - \; \sum_J {\cal G}^{I J} \; e^{l^J} 
+ u_\perp \; \epsilon \; m (m - 1) \; e^{2 (\Lambda - \sigma)}
\; \; .
\end{equation}
We will see that same type of equation will appear in cosmological case.
Since now we know $u_\alpha$'s, 
using equations (\ref{25Wbh}) and (\ref{defnG}), it is now
straightforward to calculate ${\cal G}^{I J} \;$ for $N$
intersecting brane configurations.  
les first calculate ${\cal G}^{IJ}$ for 2 brane using (\ref{defnG}).
\begin{equation}
{\cal G}^{I J} = 2 u_0 \; (u_\perp - u_0 \delta^{I J}) \; \; . 
\end{equation}
Same expression comes for 5 brane. 
It turns out, for any 
BPS intersecting configuration that, ${\cal G}^{I J} \;$ is of the same form,
namely
\begin{equation}\label{G^IJbh}
{\cal G}^{I J} = 2 u_0 \; (u_\perp - u_0 \delta^{I J}) \; \; . 
\end{equation}
The corresponding ${\cal G}_{I J} \;$ is given by
\begin{equation}\label{G_IJbh}
{\cal G}_{I J} = \frac{1}{2 u^2_0} \; 
( \frac{u_\perp}{N u_\perp - u_0} - \delta_{I J} ) \; \; .
\end{equation}
Now take $p_0 = p_\parallel \;$. Then equation (\ref{zbh}) gives
$p_\perp + f = 0 \;$. In terms of $u_\alpha \;$, we now have
$u_0 = u_\parallel \;$ and $u_\perp = 0 \;$. Clearly, then
${\cal G}^{I J} \propto \delta^{I J} \;$ and equations
(\ref{c3bh}) can be solved for $l^I (\tau)$. See \cite{addn9}
for such solutions, with $u_0 = 2 \;$ as follows from equation
(\ref{electric}), and their analysis.


\section{S and T Dualities in String Theory} \label{sec:stringd}
String theory has S and T duality symmetries. 
T duality transforms type IIA theory to type IIB and back. So it is a symmetry
of IIA and IIB theory combined together. S duality is a symmetry of type IIB 
theory which transform coupling constant to $(\mbox{coupling constant})^{-1}$.
Just like M-theory case, for 
certain supergravity solutions these symmetries can be used to get relations
among various metric components and dilaton, and hence relations among
energy-momentum tensors. To illustrate this, we consider 
a general solution of string theory (type IIA or IIB supergravity).
Line element $ds_{Dp}$ in Einstein frame is 
\begin{equation}\label{dsStrp}
ds_{Dp}^2 = - e^{2\lambda^0_p} Z d t^2 + 
\sum_{i=1}^{p} e^{2 \lambda^\parallel_p} (d x^i)^2+
\sum_{i=p+1}^{q} e^{2 \lambda^\perp_p} (d x^i)^2
+e^{2\sigma} \left( \frac{dr^2}{Z} + r^2 d\Omega_{n+1,\epsilon}^2 \right)
\end{equation}
and dilaton, $\phi_p\,=\,\phi_p(t,r)$. here we assume $x^1, \cdots, x^q$ are
compact and are isometry direction,
and $i=1,\cdots,p$ are directions parallel to Dp-branes. Obvious symmetries
ensure us to take all $\lambda^i$ parallel to branes are equal and we denote 
them as before by $\lambda^\parallel_p$, similarly for $\lambda^\perp_p$. 
Here all $\lambda$s, $Z$ and $\sigma$ are function of $r$.
$d\Omega_{n+1,\epsilon}$ is metric of constant curvature 
$n+1$-dimensional space. 
In fact we don't even need non-compact directions in above form. It can be any
curved geometry.

Here to illustrate duality relation we consider black $p$-brane
solution and this is of the form given in equation (\ref{dsStrp}).
The subscript $Dp$ indicate metric is for $Dp$-branes.
This metric is a of general black $p$-brane solution. $q$ is the total
number of compact directions, and $q+n=7$. This system physically
describes geometry created by $D$-brane localised in space. 
Technically relations we will get in this section can be used when $\lambda$'s 
are function of $t$ only. So if we consider no noncompact direction and $Z=1$
we can as well describe cosmological case.   

This solution is in Einstein frame. To apply duality rules we first convert it
in string frame. String frame metric is
\begin{eqnarray}
\label{dsStrpS}
ds_{s,Dp}^2 &=& - e^{2\lambda^0_p + \frac{\phi_p}{2}} Z d t^2 + 
\sum_{i=1}^{p} e^{2 \lambda^\parallel_p + \frac{\phi_p}{2}} (d x^i)^2+
\sum_{i=p+1}^{q} e^{2 \lambda^\perp_p + \frac{\phi_p}{2}} (d x^i)^2 \nonumber \\
&+& e^{2\sigma + \frac{\phi_p}{2}} \left( \frac{dr^2}{Z} + 
r^2 d\Omega_{n+1,\epsilon}^2 \right)
\end{eqnarray}
Now if we perform a T-duality along $p^{th}$ direction we get $D\,(p-1)$
branes solution. 
\begin{eqnarray}
\label{dsStrpST}
ds_{s,Dp-1}^2 &=& - e^{2\lambda^0_p + \frac{\phi_p}{2}} Z d t^2 + 
\sum_{i=1}^{p-1} e^{2 \lambda^\parallel_p + \frac{\phi_p}{2}} (d x^i)^2+
e^{-(2 \lambda^\parallel_p + \frac{\phi_p}{2})} (d x^p)^2 \nonumber \\
&+& \sum_{i=p+1}^{q} e^{2 \lambda^\perp_p + \frac{\phi_p}{2}} (d x^i)^2 
+ e^{2\sigma + \frac{\phi_p}{2}} \left( \frac{dr^2}{Z} + 
r^2 d\Omega_{n+1,\epsilon}^2 \right) \\
\label{phiT}
\phi_{p-1} &=& \phi_p -\frac{1}{2} 
ln (e^{2 \lambda^\parallel_p + \frac{\phi_p}{2}}) = \frac{3 \phi_p}{4} - 
\lambda^\parallel_p \;.
\end{eqnarray}
Note that in $p^{th}$ direction metric component changes sign according to
equation (\ref{TD1}). Equation (\ref{phiT}) can be found using equation
(\ref{TD6}).

The Einstein frame metric for $D(p-1)$ solution can be found by multiplying 
$e^{-\phi_{p-1}/2}$ to $ds_{s,Dp-1}^2$.
\begin{eqnarray}\label{dsStrp-1E}
 ds_{Dp-1}^2 &=& e^{-\phi_{p-1}/2} ds_{s,Dp-1}^2 \nonumber \\
	     &= & - e^{2\lambda^0_{p-1}}Z d t^2 +
\sum_{i=1}^{p-1} e^{\frac{5 \lambda^\parallel_p}{2} + 
\frac{\phi_p}{8}} (d x^i)^2 + 
e^{-\frac{3 \lambda^\parallel_p}{2} - \frac{7\phi_p}{8})} (d x^p)^2
\nonumber \\
&+& \sum_{i=p+1}^{q} e^{2 \lambda^\perp_p + 
\frac{\lambda^\parallel_p}{2}\frac{\phi_p}{8}} (d x^i)^2   +ds_\perp^2 \;,
\end{eqnarray}
where $ds_\perp^2$ is metric for transverse space. 
Equations (\ref{dsStrpST}) and (\ref{phiT} have been used to get above 
expression.
This line element can be written as 
\begin{equation}\label{dsStrp-1}
ds_{Dp-1}^2 = -e^{2\lambda_0} Z dt^2 + 
\sum_{i=1}^{p-1} e^{2 \lambda^\parallel_{p-1}}
(d x^i)^2+ \sum_{i=p}^{q} e^{2 \lambda^\perp_{p-1}} (d x^i)^2
+ds_\perp^2 \;,
\end{equation}
where 
$\lambda_p$ can be given in terms of $\lambda_{p-1}$ by
 \begin{eqnarray}
 \label{lambdareln1}
  2\lambda^\parallel_{p-1} &=& \frac{5 \lambda^\parallel_p}{2}+\; 
\frac{\phi_p}{8}  \\
 \label{lambdareln2}
  2\lambda^\perp_{p-1} &=&
  -\frac{3}{2} \lambda^\parallel_p -\frac{7}{8} \phi_p =
  2\lambda^\perp_{p}+
 \frac{\lambda^\parallel_{p}}{2}+\frac{\phi_p}{8} 
 \end{eqnarray}
 and 
 \begin{eqnarray}\label{phireln}
  \phi_{p-1}=\frac{3 \phi_p}{2}-\lambda^\parallel_p \;.
 \end{eqnarray}
Simplification of equation (\ref{lambdareln2}) shows that,
\begin{equation}
 \label{strreln}
 \lambda_p^\perp \;+\; \lambda_p^\parallel \;+\; \frac{\phi_p}{2} \;=\;0\;.
\end{equation}
Now consider a $D3$ brane, S-duality of $D3$ brane gives $D3$ brane. So S
duality rule (\ref{SD4}) gives $\phi_3=-\phi_3$, which implies $\phi_3=0$.
Consider $D2$ brane solution now, $\phi_2 = -\lambda^\parallel_3$, (using
equation (\ref{phireln})). Equation (\ref{strreln}) gives 
$\lambda^\parallel_3 = - \lambda^\perp_3$. Denoting $\lambda^\perp_3$ by
$\lambda$, one find
\begin{equation}
 \lambda_p^\perp=\lambda\;,\;\;\, 
\lambda_p^\parallel=-\lambda\;,\;\;\;
\phi_p = 0 \times \lambda.
\end{equation}

In general using induction it is easy to show that,
\begin{equation}
 \lambda_p^\perp=\frac{p+1}{4}\lambda\;,\;\;\, 
\lambda_p^\parallel=-\frac{7-p}{4}\lambda\;,\;\;\;
\phi_p = (3-p) \lambda \;,
\end{equation}
where $\lambda$ is now the only parameter determining full line element.
If one uses different set of duality operations one can also find similar
relations for $F1$-string or $NS5$-brane. In general
\begin{equation}
\label{strrln}
\lambda_p^\perp=\frac{p+1}{4}\lambda\;,\;\;\, 
\lambda_p^\parallel=-\frac{7-p}{4}\lambda\;,\;\;\;
\phi_p = z(3-p) \lambda,
\end{equation}
where $z=1$ for $Dp$-brane and $z=-1$ for $F1$-string or $NS 5$-brane.

Just like before we take energy-momentum tensors has only diagonal component.
\begin{equation}
 T^\mu{}_\nu \;=\; diag(P_0,\, P_\parallel,\, P_\perp,\, P_r,\, P_a), 
\end{equation}
where $P_\parallel$ for $i=1,2,\cdots,p$; $P_\perp$ for $i=p+1,\cdots,q$; 
$T^r{}_r=P_r$ and $T^{\theta_a}{}_{\theta_a}=
P_a$. It is natural to assume $P_\perp=P_a$ because all of them
are transverse to brane direction but for the time being we keep this notation.
We also have another component, $T_\phi$ coming from $\phi$-variation of action
\begin{equation}
 \delta_\phi S = \int d^{10}x \; \sqrt{-g} \; T_\phi \;\delta\phi\; ,
\end{equation}
where $\delta_\phi$ denote variation with respect to $\phi$.
Equation of motions are now Einstein equations (\ref{r11}) together with
\begin{equation}\label{Tfhi}
 \nabla^2 \phi = -T_\phi \;.
\end{equation}
The equations of motion for the black brane case (metric given by 
(\ref{dsStrp})) turns out to be
\begin{eqnarray}
 \label{strlambda}
 \lambda''^i + L' \lambda'^i 
&=& \frac{1}{Z} e^{2\sigma} \, (P - P_i) \\
 \label{strphi}
 \phi'' + L' \phi' &=& -\frac{1}{Z} e^{2\sigma} \, T_\phi
\end{eqnarray}
where $L=\frac{Z'}{Z} + \frac{n+1}{r} + n \sigma + \lambda^0 + \Lambda$ and
$P= \frac{1}{8} (P_0+\sum P_i + P_r + (n+1)P_a)$.

Using equation (\ref{strrln}) in equation (\ref{strlambda}) one finds
\begin{equation}
 - \frac{p+1}{7-p} = \frac{P-P_\perp}{P-P_\parallel} \;,
\end{equation}
which on simplification gives 
\begin{equation}
 P_\parallel+(7-q)P_\perp = P_0 + P_r + (n+1) P_a \;.
\end{equation}
Similarly use of equation (\ref{strrln}) in equation (\ref{strphi})
gives 
\begin{equation}
 T_\phi =- \frac{3-p}{2} z \left(P_0+P_r+(n+1)P_a)-(8-q)P_\perp\right)
\end{equation}

Just like black brane in M-theory case taking $P_\perp=P_a$, one can do
the same calculation to solve for black $D$-brane or black string solution.
Also note that in cosmological case there is no transverse metric and $q=9$.
Then above relation will translate to
\begin{eqnarray}
&& \rho+ P_\parallel-2P_\perp = 0 \\
&& T_\phi =- \frac{3-p}{2} z \left( -\rho+P_\perp \right) ,
\end{eqnarray}
where $P_0$ is taken as $(-\rho)$.
Then again same analysis can be done as in both 
cosmological case and black hole case of M-theory to 
get same conclusions.


\chapter{\bf Evolution of Early Universe} \label{chap:3}

In this chapter we discuss evolution of early universe made of 
mutually BPS intersecting
$M2$ and $M5$ branes. To be specific,
we consider $22'55'$ configurations. 
To do this we use duality relations discuss in
previous chapter. Before going into details analysis we give a general 
consequences U duality in our cosmological context. 

\section{Equations of Motion for Our Cosmological Model}\label{sec:EoMC}
For cosmological case, $T_{A B}$ is often determined
using equations of state of the dominant constituent of the
universe. Such equations of state may be obtained if the
underlying physics is known; or, one may assume a general ansatz
for them and proceed. \footnote{ This is similar to the FRW
case. Equation of state $p = \frac{\rho}{3} \;$ for radiation,
or $p = 0$ for pressureless dust, may be obtained from the
physics of radiation or of massive particles; or, one may assume
a general ansatz $p = w \rho \;$ and proceed.}

$T_{A B}$ for intersecting branes in the early universe has been
calculated in \cite{cm} assuming that the branes and antibranes
in the intersecting brane configurations are non interacting and
that their numbers are all equal, {\em i.e.} $n_I = \bar{n}_I$
for $I = 1, 2, \cdots, N$ and $n_1 = \cdots = n_N \;$. However,
general relations among the components of $T_{A B}$ may be
obtained \cite{k0707} using U duality symmetries of M theory,
involving chains of dimensional reduction and uplifting and T
and S dualities of string theory, using which 2 branes and 5
branes or $22'55'$ and $55'5''W$ configurations can be
interchanged. Such relations are valid more generally, for
example even when $n_I \;$ and $\bar{n}_I \;$ are all different.

These general relations on the equations of state are
sufficient to show, under a technical assumption, that the $N =
4$ mutually BPS intersecting brane configurations with identical
numbers of branes and antibranes, {\em i.e.} with $n_1 = \cdots
= n_4 \;$ and $\bar{n}_1 = \cdots = \bar{n}_4 \;$, will
asymptotically lead to an effective $(3 + 1)$ dimensional
expanding universe. To obtain the details of the evolution,
however, we need further assumptions and an ansatz of the type
$p = w \rho \;$ \cite{k0707, bdr, br10}. 

We now present the details. Let the line element $d s$ be given
by 
\begin{equation}\label{ds}
d s^2 = - d t^2 + \sum_i e^{2 \lambda^i} (d x^i)^2 
\end{equation}
where $e^{\lambda^i} \;$ are scale factors and, due to
homogeneity, $\lambda^i$ are functions of the physical time $t$
only. (Parametrising the scale factors as $e^{\lambda^i} \;$
turns out to be convenient for our purposes.)  It then follows
that $T_{A B}$ depends on $t$ only and that it is of the form
\begin{equation}\label{tabdiag}
T^A_{\; \; \; B} = diag( - \rho, \; p_i ) \; \; . 
\end{equation}
We assume that $\rho > 0 \;$. From equations (\ref{r11}) one now
obtains
\begin{eqnarray}
\Lambda_t^2 - \sum_i (\lambda^i_t)^2 & = & 2 \; \rho 
\label{t21} \\
\lambda^i_{t t} + \Lambda_t \lambda^i_t & = & 
p_i + \frac{1}{9} \; (\rho - \sum_j p_j) 
\label{t22} \\
\rho_t + \rho \Lambda_t + \sum_i p_i \lambda^i_t & = & 0 
\label{t23}
\end{eqnarray}
where $\Lambda = \sum_i \lambda^i \;$ and the subscripts $t$
denote time derivatives. Note, from equation (\ref{t21}), that
$\Lambda_t$ cannot vanish. Hence, if $\Lambda_t > 0$ at an
initial time $t_0$ then it follows that $e^\Lambda$ increases
monotonically for $t > t_0 \;$. We assume the evolution to be
such that $e^\Lambda \to \infty$ eventually.

In the context of early universe in M theory, it is natural to
assume that all spatial directions are on equal footing to begin
with. Therefore we assume that the ten dimensional space is
toroidal. Further, we assume that the early universe is
homogeneous and is dominated by the $22'55'$ configuration
where, with no loss of generality, we take two stacks each of $2
\;$ branes and $5 \;$ branes to be along $(x^1, x^2) \;$, $(x^3,
x^4) \;$, $(x^1, x^3, x^5, x^6, x^7) \;$, and $(x^2, x^4, x^5,
x^6, x^7) \;$ directions respectively, and take these
intersecting branes to be distributed uniformly along the common
transverse space directions $(x^8, x^9, x^{10}) \;$. Note that
the total brane charges must vanish, {\em i.e.} $n_I =
\bar{n}_I$ for all $I \;$, since the common transverse space is
compact. We denote this $22'55'$ configuration as $(12, 34,
13567, 24567) \;$. The meaning of this notation is clear and,
below, such a notation will be used to denote other
configurations also. 

\section{Equations of States as Consequence of U Duality}\label{sec:Ucos}
Application of such U duality operations generate new solution from old one.
We discuss this technique in section {\bf \ref{sec:Ubh}}, but for we repeat the 
steps since here they are our main tools to find equations of states.
Lets take an example of $M2$ brane along $(x^1,x^2)$.
We consider cosmological solution of the form,
\begin{equation}\label{ds11}
 ds_{11}^2 = -dt^2 +
 \sum_{i=1}^{10} e^{2 \lambda^i(t)} (d x^i)^2 \;,
\end{equation}
where we assume all the special directions are compact and toroidal.
Here obvious symmetry
implies 
\begin{equation}
 \label{obv2}
 \lambda^1 = \lambda^2 ,
\;\;\;
 \lambda^3 = \lambda^4 = \lambda^5 = 
  \lambda^6 = \lambda^7 = \lambda^8 = \lambda^9 \;.
\end{equation}
If we apply a chain of operations $\downarrow_5 T_3 T_4 \uparrow_5$ in the 
way mentioned in previous section, {\bf \ref{sec:Ud}},
we can generate a $M5$ brane solution.
\begin{displaymath}
 \xymatrix{
           M2 (12) \ar[r]^{\downarrow_5} & D2 (12) \ar[r]^{T_4} & D3 (124)
           \ar[r]^{T_3} & D4 (1234) \ar[r]^{\uparrow_5} & M5 (12345)
          } \; .
\end{displaymath}
This new solution is $M5$ branes may be given by
\begin{equation}
 d s'^2_{11} = - d \hat{t}^2 + \sum_{i=1}^{10} 
e^{2 \lambda{'^i}(\hat{t})} (d x^i)^2 \;.
\end{equation}
Where using equations (\ref{lambdareln}) we find
these $\lambda{'^i}$'s are given in terms of $\lambda^i$'s by
\begin{eqnarray}
 \lambda{'^3} &=& \lambda^4 - \frac{2}{3} (\lambda^3+\lambda^4+\lambda^5)
  \nonumber \\
 \lambda{'^4} &=& \lambda^3 - \frac{2}{3} (\lambda^3+\lambda^4+\lambda^5)
  \nonumber \\
 \lambda{'^5} &=& \lambda^5 - \frac{2}{3} (\lambda^3+\lambda^4+\lambda^5)
  \nonumber \\
 \lambda{'^i} &=& \lambda^i +  \frac{1}{3} (\lambda^3+\lambda^4+\lambda^5) 
\;\;\;\;\;\;\;\;\;  \forall \; i \ne \{3,4,5\} \;.
\end{eqnarray}
Note that in case of our cosmological solution we can always redefine our 
time coordinate to go to comoving frame, as we have redefine time as $\hat{t}$
here.

Obvious symmetry of $M5$ brane solution demands
\begin{equation}
 \label{obv5}
 \lambda'^1 = \lambda'^2 = \lambda'^3 = \lambda'^4 = \lambda'^5 \;,
\;\;\;
 \lambda'^6 = \lambda'^7 = \lambda'^8 = \lambda'^9 = \lambda'^{10} \;.
\end{equation}
So using equations (\ref{obv2}), (\ref{obv5}) and the relations among 
$\lambda'^i$ and $\lambda^i$ one can show that
\begin{equation}
 \label{25reln}
 \lambda^\parallel + 2 \lambda^\perp = 0 , \;\;\;
 2 \lambda'^\parallel + \lambda'^\perp =0 \;.
\end{equation}
where the superscripts $\parallel \;$ and $\perp \;$ denote
spatial directions along and transverse to the branes
respectively. 

Putting back these relations in the equation of motion (\ref{t22}) one can show
\begin{equation} \label{prho25}
 p_\parallel = 2 p_\perp -\rho \;.
\end{equation}
In general $M2$ brane or $M5$ brane or wave above relation turns out to be
\begin{equation}
 \label{z}
 p_\parallel = z(\rho - p_\perp)+ p_\perp \; ,
\end{equation}
where $z=-1$ for 2 and 5 branes and $=+1$ for waves. 

Similarly one can go through all such operations in intersecting brane
configuration to generate new solutions. In case of our $22'55'$ configuration
$(12, 34, 13567, 24567) \;$ the obvious symmetry relations are
\begin{equation}\label{obv2255}
22'55' \; \; : \; \; \; \; \lambda^5 = \lambda^6 = \lambda^7
\; \; , \; \; \; \lambda^8 = \lambda^9 = \lambda^{10} \; \; .
\end{equation}
Proceeding as in the case of 2 and 5 branes above, and using the
U duality $\uparrow_5 T_1 T_2 \downarrow_5 \;$ which relates the
$22'55'$ and $W55'5''$ configurations, one obtains two more
relations given by \cite{k0707}
\begin{equation}\label{2255reln}
\lambda^1 + \lambda^4 + \lambda^5 = 
\lambda^2 + \lambda^3 + \lambda^5 = 0 \; \; . 
\end{equation} 
Clearly, the U duality relations in equations (\ref{obv2255bh})
and (\ref{2255relnbh}) are valid here also.
Then again like before, use of equation (\ref{t22}) yields relations among
various components of energy momentum tensor. 
\begin{eqnarray}
& p_5=p_6=p_7 & \\
& p_8=p_9=p_{10} & \\
& p_1+p_4+p_5 = p_2+p_3+p_5 = 0 &
\end{eqnarray}

In general, for an $N$ intersecting brane configuration, the U
duality symmetries will lead to $10 - N$ relations among the
$\lambda^i$s, leaving only $N$ of them independent. These
relations are of the form $\sum_i c_i \lambda^i = 0 \;$ where
$c_i$ are constants. Clearly, such a relation can be violated by
constant scaling of $x^i$ coordinates. Hence, we interpret it as
implying a relation among the components of $T_{A B} \;$. In
view of equation (\ref{t22}), we interpret a U duality relation
$\sum_i c_i \lambda^i = 0 \;$ as implying that
\begin{equation}\label{weak}
\sum_i c_i f^i = 0  
\; \; \; , \; \; \; \; 
f^i = p_i + \frac{1}{9} \; (\rho - \sum_j p_j) \; \; . 
\end{equation} 
Substituting equation (\ref{weak}) in equation (\ref{t22}), it
follows upon an integration that 
\begin{equation}\label{K}
\sum_i c_i \lambda^i_t = c \; e^{- \Lambda} 
\end{equation}
where $c$ is an integration constant. If $\sum_i c_i \lambda^i_t
= 0 \;$ initially at $t = t_0 \;$ then $c = 0 \;$. In such cases
then $\sum_i c_i \lambda^i_t = 0 \;$ for all $t \;$ and, hence,
$\sum_i c_i \lambda^i = v \;$ where $v$ is another integration
constant.

In general $\sum_i c_i \lambda^i_t \ne 0 \;$ initially at $t =
t_0 \;$ and, hence, $c \ne 0 \;$. Let the evolution be such that
$e^\Lambda \sim t^\beta \to \infty \;$ in the limit $t \to
\infty \;$. Then it follows from equation (\ref{K}) that $\sum_i
c_i \lambda^i_t \to 0 \;$ in this limit. If, furthermore, $\beta
> 1 \;$ then equation (\ref{K}) also gives
\begin{equation}\label{L} 
\sum_i c_i \lambda^i = v + {\cal O}(t^{1 - \beta}) \to v
\end{equation} 
where $v$ is an integration constant. If $\beta \le 1 \;$ then
$\sum_i c_i \lambda^i \;$ is a function of $t \;$. We will see
later that, in the solutions we obtain with further assumptions,
$\beta$ turns out to be $ > 1$ for $N > 1 \;$.

Note that, in previous chapter, for black hole case we showed that
U duality relations hold by calculating $T_{AB}$ explicitly.
On the other hand here we use U duality to get equations of states.


\section{A General Result}\label{sec:genr}
We now consider a general result for the $22'55'$ configuration
that follows from the U duality relations alone \cite{k0707}.
The $\lambda^i$s for this configuration obey the relations given
in equations (\ref{obv2255}) and (\ref{2255reln}). Note that a
suitable U duality, for example $\uparrow_6 T_4 T_5 \downarrow_6
\;$, can transform 2 branes and 5 branes into each other. Hence,
we will refer to two types of branes as being identical if they
have identical numbers of branes and antibranes, {\em i.e.}
$I^{th}$ type is identical to $J^{th}$ type if $n_I = n_J$ and
$\bar{n}_I = \bar{n}_J \;$.

Consider the case when 2 and $2'$ branes in the $22'55'$
configurations are identical. This will enhance the obvious
symmetry relations. It is easy to see that we now have one more
independent relation $\lambda^1 = \lambda^3 \;$. If, instead, 5
and $5'$ branes are identical, then the extra independent
relation is $\lambda^1 = \lambda^2 \;$. Similarly, if 2 and $5'$
branes are identical then, after a few steps involving U duality
$\uparrow_6 T_4 T_5 \downarrow_6 \;$ which interchanges 2 and
$5'$ branes, it follows that the extra independent relation is
$\lambda^2 = \lambda^5 \;$.

Now if all the four types of branes in the $22'55'$
configuration are identical, {\em i.e.} if $n_1 = \cdots = n_4
\;$ and $\bar{n}_1 = \cdots = \bar{n}_4 \;$, then, we have three
extra independent relations
\begin{equation}\label{3extra}
\lambda^1 = \lambda^2 = \lambda^3 = \lambda^5 \; \; .
\end{equation}

Combined with equations (\ref{obv2255}) and (\ref{2255reln}), we
get $\lambda^1 = \cdots = \lambda^7 = 0 \;$ which is to be
interpreted as $f^1 = \cdots = f^7 = 0 \;$, see equation
(\ref{weak}). Hence, as described earlier, it follows for $i =
1, \cdots, 7 \;$ that if $\lambda^i_t = 0$ initially at $t = t_0
\;$ then $\lambda^i_t = 0$ and $\lambda^i = v^i$ for all $t \;$
where $v^i \;$ are constants. Or, if $e^\Lambda \sim t^\beta \to
\infty \;$ in the limit $t \to \infty \;$ with $\beta > 1 \;$,
it then follows for $i = 1, \cdots, 7\;$ that $\lambda^i_t \to 0
\;$ and $\lambda^i \to v^i \;$ in this limit. Obtaining the
values of the asymptotic constants $v^i \;$, however, requires
knowing the details of evolution. It also follows similarly that
$e^{\lambda^i} \sim e^{ \frac{\Lambda} {3}} \to \infty \;$ for
$i = 8, 9, 10 \;$. It is straightforward to show that same
results are obtained for the equivalent $55'5''W$ configuration
also.

Thus, assuming either that $\lambda^1_t = \cdots = \lambda^7_t =
0 \;$ initially at $t = t_0 \;$ or that $e^\Lambda \sim t^\beta
\to \infty \;$ in the limit $t \to \infty \;$ with $\beta > 1
\;$, we obtain that the $N = 4$ mutually BPS intersecting brane
configurations with identical numbers of branes and antibranes,
{\em i.e.} with $n_1 = \cdots = n_4 \;$ and $\bar{n}_1 = \cdots
= \bar{n}_4 \;$, will asymptotically lead to an effective $(3 +
1)$ dimensional expanding universe with the remaining seven
spatial directions reaching constant sizes. This result follows
as a consequence of U duality symmetries alone, which imply
relations of the type given in equation (\ref{weak}) among the
components of the energy momentum tensor $T_{A B} \;$. This
result is otherwise independent of the details of the equations
of state, and also of the ansatzes for $T_{A B} \;$ we make in
the following in order to proceed further.

\section{Ansatz for $T_{A B} \;$} \label{sec:TAB}
The dynamics underlying the general result given above may be
understood in more detail, and the asymptotic constants $v^i \;$
can be obtained, if an explicit solution for the evolution is
available. In the following, we will make a few assumptions
which enable us to obtain such details.

Consider now the case of $2$ branes or 5 branes only. From the U
duality relations, it follows from equation (\ref{prho25}), that
$p_\parallel = - \rho + 2 p_\perp \;$.
For the case of waves, the relation is given by equation (\ref{z}).
A similar relation for black hole is derived in equation (\ref{zbh}).
In general, $\rho \;$, $p_\parallel \;$,
and $p_\perp \;$ are functions of the numbers $n$ and $\bar{n}$
of branes and antibranes, satisfying the U duality relations
(\ref{z}). If $n = \bar{n} \;$ then $p_\parallel$ and $p_\perp$
may be thought of as functions of $\rho \;$ satisfying equation
(\ref{z}) \cite{k0707}.

Consider now mutually BPS $N$ intersecting brane
configuration. In the black hole case, 
following the discussion of section {\bf \ref{sec:BH}} and {\bf \ref{sec:Ubh}},
it turns out that the
energy momentum tensors $T^A_{\; \; B (I)} \;$ of the $I^{th}$
type of branes are mutually non interacting and separately
conserved \cite{addn1} -- \cite{addn10}. That is,
\begin{equation}\label{tabI1}
T^A_{\; \; B} = \sum_I T^A_{\; \; B (I)} 
\; \; , \; \; \; 
\sum_A \nabla_A  T^A_{\; \; B (I)} = 0 \; \; . 
\end{equation}
We assume that this is the case in the context of early universe
also where $T^A_{\; \; B} = diag (-\rho, \; p_i) \;$, $T^A_{\;
\; B (I)} = diag (-\rho_I, \; p_{i I}) \;$, $\rho_I > 0 \;$, and
$(\rho_I, \; p_{i I}) \;$ satisfy the U duality relations in
(\ref{z}) for all $I \;$. Equations (\ref{tabI}) now give
\begin{eqnarray}
& & \rho = \sum_I \rho_I \; \; , \; \; \; \;
p_i = \sum_I p_{i I} \label{add} \\
& & (\rho_I)_t + \rho_I \Lambda_t 
+ \sum_i p_{i I} \lambda^i_t = 0 \; \; . \label{t23I}
\end{eqnarray}
We have verified explicitly for a variety of mutually BPS $N$
intersecting brane configurations that equations (\ref{z}) and
(\ref{add}) are sufficient to satisfy all the relations of the
type $\sum_i c_i f^i = 0 \;$ implied by U duality
symmetries. See \cite{k0707} for more details.

To solve the evolution equations (\ref{t21}), (\ref{t22}),
(\ref{add}), and (\ref{t23I}), we need the functions $\rho_I
\;$, $p_{\parallel I} \;$, and $p_{\perp I} \;$. To proceed
further, we assume that $n_I = \bar{n}_I \;$ for all $I
\;$. This is necessary if, as is the case here, the common
transverse directions are compact and hence the net charges
must vanish. Then $p_{\parallel I}$ and $p_{\perp I}$ may be
thought of as functions of $\rho_I \;$ satisfying equation
(\ref{z}).

It is natural to expect that $p_{\perp I}(\rho_I)$ is the same
function for waves, 2 branes, and 5 branes since they can all be
transformed into each other by U duality operations which do not
involve the transverse directions. We assume that this is the
case. We further assume that this function $p_{\perp}(\rho)$ is
given by
\begin{equation}\label{prho}
p_\perp = (1 - u) \; \rho 
\end{equation}
where $u$ is a constant. Such a parametrisation of the equation
of state, instead of the usual one $p = w \rho \;$, leads to
elegant expressions as will become clear in the following, see
\cite{im1, im2} also. The results of \cite{cm} correspond to the
case where $u = 1 \;$. Here, we assume only that $0 < u < 2
\;$. The constant $u$ is arbitrary otherwise.

It now follows that $p_{i I} \;$ in equation (\ref{add}) are of
the form $p_{i I} = (1 - u^I_i) \; \rho_I \;$ and that the
constants $u^I_i \;$ can be obtained in terms of $u$ using
equations (\ref{z}) and (\ref{prho}). Thus, for 2 branes, 5
branes, and waves, we have $u_\perp = u \;$, $u_\parallel = (1 -
z) \; u \;$, and hence
\begin{eqnarray}
2 & : & u_i = 
(\; \; 2, \; \; 2, \; \; 1, \; \; 1, \; \; 1, 
\; \; 1, \; \; 1, \; \; 1, \; \; 1, \; \; 1) \; u 
\nonumber \\
5 & : & u_i = 
(\; \; 2, \; \; 2, \; \; 2, \; \; 2, \; \; 2, 
\; \; 1, \; \; 1, \; \; 1, \; \; 1, \; \; 1) \; u 
\nonumber \\
W & : & u_i = 
(\; \; 0, \; \; 1, \; \; 1, \; \; 1, \; \; 1, 
\; \; 1, \; \; 1, \; \; 1, \; \; 1, \; \; 1) \; u 
\label{25W}
\end{eqnarray}
where the $I$ superscripts have been omitted since $N = 1 \;$.
Similarly, $u^I_i \;$ for the $22'55' \;$ configuration are
given by
\begin{eqnarray} 
2 & : & u_i^1 = 
(\; \; 2, \; \; 2, \; \; 1, \; \; 1, \; \; 1, 
\; \; 1, \; \; 1, \; \; 1, \; \; 1, \; \; 1) \; u 
\nonumber \\
2' & : & u_i^2 = 
(\; \; 1, \; \; 1, \; \; 2, \; \; 2, \; \; 1, 
\; \; 1, \; \; 1, \; \; 1, \; \; 1, \; \; 1) \; u 
\nonumber \\
5 & : & u_i^3 = 
(\; \; 2, \; \; 1, \; \; 2, \; \; 1, \; \; 2, 
\; \; 2, \; \; 2, \; \; 1, \; \; 1, \; \; 1) \; u 
\nonumber \\
5' & : & u_i^4 = 
(\; \; 1, \; \; 2, \; \; 1, \; \; 2, \; \; 2, 
\; \; 2, \; \; 2, \; \; 1, \; \; 1, \; \; 1) \; u \; \; .
\label{I} 
\end{eqnarray}
This completes our ansatz for the energy momentum tensor $T_{A
B}$ for the intersecting brane configurations in the early
universe.

\section{General Analysis of Evolution Equations} \label{sec:GEvol}
The evolution of the universe can now be analysed. In this
section, we first present the analysis in a general form which
is applicable to a $D$ dimensional homogeneous, anisotropic
universe. We specialise to the intersecting brane configurations
in the next section.

The $D$ dimensional line element $d s$ is given by equation
(\ref{ds}), now with $i = 1, 2, \cdots, D - 1 \;$. The total
energy momentum tensor $T_{A B}$ of the dominant constituents of
the universe is given by equation (\ref{tabdiag}).  The
equations of motion for the evolution of the universe is given,
in units where $8 \pi G_D = 1 \;$, by equations (\ref{t21}) --
(\ref{t23}) with $9$ in equation (\ref{t22}) now replaced by $D
- 2 \;$. Defining
\begin{equation}\label{Gij}
G_{i j} = 1 - \delta_{i j} \; \; , \; \; \; 
G^{i j} = \frac{1}{D - 2} - \delta_{i j} \; \; , 
\end{equation} 
the equations (\ref{t21}) and (\ref{t22}), with $9$ replaced by
$D - 2 \;$, may be written compactly as
\begin{eqnarray}
\sum_{i, j} G_{i j} \lambda^i_t \; \lambda^j_t & = & 
2 \; \rho \label{a1} \\
\lambda^i_{t t} + \Lambda_t \lambda^i_t & = & 
\sum_j G^{i j} \; (\rho - p_j) \label{a2}
\end{eqnarray}
where $i, j, \cdots \;$ run from $1$ to $D - 1 \;$.

Let the universe be dominated by $N$ types of mutually non
interacting and separately conserved matter labelled by $I = 1,
\cdots, N \;$. Then the corresponding energy momentum tensors
$T_{A B (I)} \;$ and their components $\rho_I \;$ and $p_{i I}
\;$ satisfy equations (\ref{tabI1}) -- (\ref{t23I}). 

Further, let the equations of state be given by $p_{i I} = (1 -
u^I_i) \; \rho_I \;$ where $u^I_i \;$ are constants. Equations
(\ref{t23I}), (\ref{a1}), and (\ref{a2}) may now be simplified
and cast in various useful forms as follow. 

Using $p_{i I} = (1 - u^I_i) \; \rho_I \;$, equation
(\ref{t23I}) can be integrated to give
\begin{equation}\label{lI}
\rho_I = e^{l^I - 2 \Lambda} \; \; , \; \; \;
l^I = \sum_i u^I_i \lambda^i +l^I_0 
\end{equation} 
where $l^I_0 \;$ are integration constants. Further using
equations (\ref{add}) and (\ref{lI}), equations (\ref{a1}) and
(\ref{a2}) become
\begin{eqnarray}
\sum_{i, j} G_{i j} \lambda^i_t \; \lambda^j_t & = & 
2 \; \sum_J e^{l^J - 2 \Lambda} \label{b1} \\
\lambda^i_{t t} + \Lambda_t \lambda^i_t & = & 
\sum_J u^{i J} \; e^{l^J - 2 \Lambda} \label{b2}
\end{eqnarray}
where $u^{i J} = \sum_j G^{i j} \; u^J_j \;$. Let the initial
conditions at an initial time $t_0 \;$ be given, with no loss of
generality, by
\begin{equation}\label{ic}
\left( \lambda^i, \; \lambda^i_t, \; l^I, \; l^I_t, \; \rho_I
\right)_{t = t_0} = 
\left( 0, \; k^i, \; l^I_0, \; K^I, \; \rho_{I 0} \right) 
\end{equation}
where 
\begin{equation}\label{rhoki}
\rho_{I 0} = e^{l^I_0} \; \; , \; \; \; 
K^I = \sum_i u^I_i k^i \; \; , \; \; \; 
\sum_{i, j} G_{i j} k^i k^j = 2 \; \sum_J e^{l^J_0} \; \; .
\end{equation}
Equations (\ref{b1}) and (\ref{b2}) may now be solved for the $D
- 1$ variables $\lambda^i \;$ with the above initial conditions.
Or, instead, these equations may be manipulated so that one
needs to solve for $N$ variables $l^I$ only, see equations
(\ref{tau}), (\ref{c3}), (\ref{c5}), and (\ref{c6}) below. We
now perform these manipulations.

First define a variable $\tau(t) \;$ as follows:
\begin{equation}\label{tau}
d \tau = e^{- \Lambda} \; d t \; \; , \; \; \; 
\tau(t_0) = 0 \; \; .
\end{equation} 
Then, for $\lambda^i(t) \;$ or equivalently $\lambda^i(\tau(t))
\;$, we have
\begin{equation}\label{fttau}
\lambda^i_t = e^{- \Lambda} \; \lambda^i_\tau \; \; , \; \; \;
\lambda^i_{t t} + \Lambda_t \lambda^i_t = 
e^{- 2 \Lambda} \; \lambda^i_{\tau \tau}
\end{equation}
where the subscripts $\tau$ denote $\tau$--derivatives. Note
that the initial values at $\tau(t_0) = 0 \;$ remain unchanged
since $\Lambda = 0 \;$, and hence $\lambda^i_t = \lambda^i_\tau
\;$ at $t = t_0 \;$. Equations (\ref{b1}) and (\ref{b2}) now
become
\begin{eqnarray}
\sum_{i, j} G_{i j} \lambda^i_\tau \; \lambda^j_\tau & = & 
2 \; \sum_J e^{l^J} \label{c1} \\
\lambda^i_{\tau \tau} & = & \sum_J u^{i J} \; e^{l^J} 
\; \; . \label{c2}
\end{eqnarray}

Also, from $l^I = \sum_i u^I_i \lambda^i +l^I_0 \;$, it follows
that
\begin{equation}\label{c3}
l^I_{\tau \tau} = \sum_J {\cal G}^{I J} \; e^{l^J} 
\end{equation}
where 
\begin{equation}\label{cG^IJ}
{\cal G}^{I J} = \sum_i  u^I_i \; u^{i J} 
= \sum_{i, j} G^{i j} \; u^I_i \; u^J_j \; \; .
\end{equation}
We assume that ${\cal G}_{I J} \; $ exists such that $\sum_J
{\cal G}_{I J} \; {\cal G}^{J K} = \delta_I^{\; \; K} \; $, {\em
i.e.} that the matrix ${\cal G}$ formed by ${\cal G}^{I J} \; $
is invertible. \footnote{This is not always the case. For
example, $u^I_i = u^I \;$ for all $i \;$ in the isotropic
case. Then ${\cal G}^{I J} \propto u^I u^J \;$ and ${\cal G}$ is
not invertible. This is not a problem, it just means that the
set of variables $l^I$ can be reduced to a smaller independent
set; one then proceeds with the smaller set.} Then, from
equation (\ref{c3}), we have
\begin{equation}\label{c4}
\sum_J {\cal G}_{I J} \; l^J_{\tau \tau} = e^{l^I} \; \; .
\end{equation} 
Note that equation (\ref{c3}) is of the form equation (\ref{c3bh}). But here
we can not calculate ${\cal G}^{I J}$ explicitly because we do not know
explicit form of $T^A{}_{B\,(I)}$ now. But using equation of states and
duality relations we will be able to calculate it, it turns out 
${\cal G}^{I J}$ is not diagonal in this case.
Substituting this expression for $e^{l^I}$ into equation
(\ref{c2}), then integrating it twice and incorporating the
initial conditions given in equation (\ref{ic}), we get
\begin{equation}\label{c5}
\lambda^i = \sum_I u^i_I \; (l^I - l^I_0) + \; L^i \; \tau
\; \; \; , \; \;\; \; 
u^i_I = \sum_{j, J} {\cal G}_{I J} \; G^{i j} \; u^J_j
\end{equation}
where $L^i \;$ are integration constants. It follows from
$\lambda^i_\tau (0)$ that $L^i$ are related to initial values
$k^i$ and $K^I$ by $k^i = \sum_I u^i_I \; K^I + \; L^i \;$.
Using this expression for $k^i$ in the relation $K^I = \sum_i
u^I_i k^i \;$, or substituting the expression for $\lambda^i \;$
given in equation (\ref{c5}) into the equation (\ref{lI}) for
$l^I \;$, leads to the following $N$ constraints on $L^i \;$:
\begin{equation}\label{uiLi}
\sum_i u^I_i L^i = 0 \; \; , \; \; \; 
I = 1, 2, \cdots, N \; \; .
\end{equation}
Now, using equations (\ref{c5}) and (\ref{uiLi}), equation
(\ref{c1}) may be written in terms of $l^I \;$ as follows:
\begin{equation}\label{c6}
\sum_{I, J} {\cal G}_{I J} \; l^I_\tau \; l^J_\tau = 
2 \; ( E + \sum_I e^{l^I} )
\; \; , \; \; \; 
2 E = - \; \sum_{i, j} G_{i j} L^i L^j \; \; .
\end{equation}
One may now solve equations (\ref{c3}) and (\ref{c6}) for $N$
variables $l^I(\tau) \;$. Then $\lambda^i (\tau) \;$ are
obtained from equation (\ref{c5}) and $t(\tau) \;$ from equation
(\ref{tau}). Inverting $t(\tau) \;$ then gives $\tau(t) \;$, and
thereby $\lambda^i (t) \;$.

\subsection{$N=1$ Case} \label{sec:Neq1}
Consider the $N = 1 \;$ case. Note that we are still considering
the general $D$ dimensional universe, not the eleven dimensional
one. We assume here that ${\cal G}^{1 1} = {\cal G} > 0 \;$.
Now, as shown in Appendix {\bf \ref{ap:Ege0}}, it follows in general that if
$\sum_i u_i L^i = 0 \;$ and $\sum_{i, j} G^{i j} u_i u_j > 0 \;$
then $E \ge 0 \;$ and $E$ vanishes if and only if $L^i \;$ all
vanish. Since $\sum_i u^1_i L^i = 0 \;$, see equation
(\ref{uiLi}), and we assume that ${\cal G}^{1 1} = \sum_{i, j}
G^{i j} u^1_i u^1_j > 0 \;$, we have $E \ge 0 \;$. We further
assume that $E > 0 \;$, equivalently that $L^i$s do not all
vanish.

Omitting the $I$ labels, equations (\ref{c3}) and (\ref{c6}) for
$l(\tau)$ become
\begin{equation}\label{1c1}
l_{\tau \tau} =  {\cal G} \; e^l 
\; \; \; , \; \; \; 
(l_\tau)^2 = 2 \; {\cal G} \; (E + e^l) \; \; . 
\end{equation}
The initial values are $l_0 = l(0) \;$ and $K = l_\tau(0) \;$
obeying $K^2 = 2 \; {\cal G} \; (E + e^{l_0}) \;$. We take $K >
0 \;$ with no loss of generality. Then the solution for $l(\tau)
\;$ is given by
\begin{equation}\label{1c2}
e^l = \frac{E}{Sinh^2 \; \alpha (\tau_\infty - \tau)} 
\end{equation}
where 
\begin{equation}\label{1alpha}
2 \alpha^2 = {\cal G} E \; \; , \; \; \; 
Sinh^2 \; \alpha \tau_\infty = E \; e^{- l_0} \; \; , \; \; \;
K = 2 \alpha \; Coth \; \alpha \tau_\infty \; \; . 
\end{equation}
The sign of $\alpha$ is immaterial but, just to be definite, we
take it to be positive. The sign of $\tau_\infty \;$ is same as
that of $K \;$, hence $\tau_\infty > 0 \;$. Also, $\lambda^i
(\tau) \;$ and $t(\tau) \;$ may now be obtained but are not
needed here for our purposes.

Note that $e^l \to 4 E \; e^{2 \alpha (\tau - \tau_\infty)} \;$
and vanishes in the limit $\tau \to - \infty \;$, whereas $e^l
\to \frac{2} {{\cal G}} \; (\tau_\infty - \tau)^{- 2} \;$ and
diverges in the limit $\tau \to \tau_\infty \;$. The value of
$\tau_\infty \;$ depends on the initial values $l_0 \;$ and $K$,
or equivalently $E \;$, as given in equations (\ref{1alpha}). It
is finite and can be evaluated exactly. However, if $e^{l_0} \ll
E \;$ then $\tau_\infty \;$ may be approximated in a way that
will be useful later on.

From the exact solution given above, we have $Sinh^2 \; \alpha
\tau_\infty = E \; e^{- l_0} \;$ and $K = 2 \alpha \; Coth \;
\alpha \tau_\infty \;$. In the limit $e^{l_0} \ll E \;$, we then
have $e^{2 \alpha \tau_\infty} \simeq 4 \; E \; e^{- l_0} \;$
and $K \simeq 2 \alpha \;$. It, therefore, follows that
\begin{equation}\label{tauexact}
\tau_\infty \simeq \; \frac{1}{K} \;
(ln \; E - l_0 + ln \; 4) \; \; .  
\end{equation}

In the limit $e^{l_0} \ll E \;$, the evolution of $l(\tau) \;$
may also be thought of as follows. Consider $E$ to be fixed and
$e^{l_0} \;$ to be very small so that $e^{l_0} \ll E \;$. It
then follows from equations (\ref{1c1}) that, at initial times,
$l_{\tau \tau} \;$ is very small and that $l_\tau \simeq \sqrt{2
{\cal G} E} = 2 \alpha \;$ is independent of $e^l \;$. Hence, $l
(\tau)$ evolves as if there is no `force', {\em i.e.} $l (\tau)
\simeq l_0 + K \tau $ where $K = l_\tau (0) \; > 0 \;$ is the
initial `velocity'. Once $e^l$ becomes of ${\cal O} (E) \;$ then
it affects $l_\tau \;$. But, from then on, $e^l$ evolves quickly
and diverges soon after.

This suggests that one may well approximate $\tau_\infty$ by the
time $\tau_a \;$ required for $l \;$, which starts from $l_0 \;$
with a velocity $K \;$ and evolves freely with no force, to
reach $ln \; E \;$ -- namely, to reach a value where $e^l =
e^{l_0 + K \tau_a} = E \;$. In other words, if $e^{l_0} \ll E
\;$ then
\begin{equation}\label{tauapprox}
\tau_\infty \simeq \tau_a = \; \frac{1}{K} \;
(ln \; E - l_0) \; \; . 
\end{equation}
A comparison with equation (\ref{tauexact}) shows that the exact
$\tau_\infty$ which follows from solving the evolution equations
is indeed well approximated by $\tau_a \;$ in equation
(\ref{tauapprox}) in the limit $e^{l_0} \ll E \;$. Note that
$\tau_a$ is calculated using only the initial values, requiring
no knowledge of the exact solution.

\subsection{$N > 1 \;$ Case} \label{sec:Ngt1}
When $N > 1 \;$, the equations of motion can be solved if ${\cal
G}^{I J} \;$ are of certain form \cite{im1} -- \cite{sudipta3}.
For example, if ${\cal G}^{I J} \propto \delta^{I J} \;$ then
the solutions are similar to those in the $N = 1 \;$ case
described above. For general forms of ${\cal G}^{I J} \;$, we
are unable to obtain explicit solutions. Nevertheless, the
general evolution can still be analysed if one assumes suitable
asymptotic forms for the scale factors $e^{\lambda^i} \;$.

It follows from equations (\ref{Gij}) and (\ref{a1}) that
$\Lambda_t \;$ cannot vanish. With no loss of generality, let
$\Lambda_t > 0$ initially at $t = t_0 \;$. Then $e^\Lambda$
decreases monotonically for $t < t_0 \;$, equivalently $\tau < 0
\;$, and increases monotonically for $t > t_0 \;$, equivalently
$\tau > 0 \;$. Further features of the evolution depend on the
structure of ${\cal G}^{I J} \;$ and $u^I_i \;$. In the cases of
interest here, it turns out that $e^\Lambda \;$ and also all
$e^{l^I} \;$ vanish in the limit $\tau \to - \infty \;$, and
diverge in the limit $\tau \to \tau_\infty \;$ where
$\tau_\infty \;$ is finite. We assume such a behaviour and
analyse the asymptotic solutions.


\subsubsection{Asymptotic evolution: $e^\Lambda \to 0 \;$}\label{sec:tto0}
We assume that $(e^\Lambda, e^{l^I}) \to 0 \;$ in the limit
$\tau \to - \infty \;$. Then, equations (\ref{c2}) and
(\ref{c3}) can be solved since their right hand sides depend
only on $e^{l^I}$s which now vanish. Hence, in the limit $\tau
\to - \infty \;$, we write
\begin{equation}\label{0l}
e^{l^I} \; = \; e^{\tilde{c}^I \tau} \; = \; 
t^{\tilde{b}^I} \; \; \; , \; \; \; \; 
e^{\lambda^i} \; = \; e^{\tilde{c}^i \tau} \; = \; 
t^{\tilde{b}^i} 
\end{equation}
which are valid up to multiplicative constants and where
$(\tilde{c}^I, \; \tilde{c}^i, \; \tilde{b}^I, \; \tilde{b}^i)
\;$ are constants. Also, the equalities in the asymptotic
expressions here and in the following are valid only up to the
leading order. Equation (\ref{c5}) now implies that $\tilde{c}^i
= \sum_I u^i_I \tilde{c}^I + L^i \;$. Also, $e^\Lambda = \;
e^{\tilde{c} \tau} \;$ where $\tilde{c} = \sum_i \tilde{c}^i
\;$. Then it follows from equation (\ref{tau}) that $t \sim
e^{\tilde{c} \tau} \;$. Hence,
\begin{equation}\label{bc}
\tilde{b}^I = \frac {\tilde{c}^I} {\tilde{c}} 
\; \; \; , \; \; \; \;
\tilde{b}^i = \frac {\tilde{c}^i} {\tilde{c}}
\; \; \; , \; \; \; \; 
\sum_i \tilde{b}^i = 1 \; \; . 
\end{equation}
Furthermore, equation (\ref{c1}) implies that $(\sum_i
\tilde{b}^i)^2 - \sum_i (\tilde{b}^i)^2 = 0 \;$. Thus the
evolution is of Kasner type in the limit $\tau \to - \infty \;$.
The constants $\tilde{c}^I$s in equations (\ref{0l}) must be
such that the resulting $\sum_i \tilde{b}^i = \sum_i
(\tilde{b}^i)^2 = 1 \;$, but are otherwise arbitrary. In an
actual evolution, however, $\tilde{c}^I$s can be determined in
terms of the initial values $l^I_0 \;$ and $K^I \;$ with no
arbitrariness, but this requires complete solution for $l^I
(\tau) \;$.

\subsubsection{Asymptotic evolution: $e^\Lambda \to \infty \;$}
\label{sec:ttoinfty}
We assume that $e^\Lambda \to \infty \;$ in the limit $\tau \to
\tau_\infty \;$ where $\tau_\infty \;$ is finite. Whether this
limit is reached at a finite or infinite physical time $t \;$
depends on the values of $u^I_i \;$, see below. $\Lambda(\tau)$
may be obtained in terms of $l^I(\tau)$ using equation
(\ref{c5}). Hence, in the limit $e^\Lambda \to \infty \;$, some
or all of the $e^{l^I}$s diverge. Consider the following ansatz
in the limit $\tau \to \tau_\infty \;$: 
\begin{equation}\label{asymptau}
e^{l^I} = \; e^{c^I} \; 
(\tau_\infty - \tau)^{- 2 \gamma^I} 
\; \; , \; \; \; \; 
e^{\lambda^i} = \; e^{c^i} \; 
(\tau_\infty - \tau)^{- 2 \gamma^i} \; \; , 
\end{equation}
where $c^I \;$ and $\gamma^I \;$ are constants, and some or all
of the $\gamma^I$s must be $> 0 \;$ so that some or all of the
$e^{l^I}$s diverge. Equation (\ref{c5}) now implies that
\begin{equation}\label{cigammai}
\gamma^i =  \sum_I u^i_I \; \gamma^I
\; \; \; , \; \; \; \; \; \;  
c^i = \sum_I u^i_I \; ( c^I - l^I_0 ) + L^i \; \tau_\infty 
\; \; . 
\end{equation}
Also, $e^\Lambda = \; e^c \; (\tau_\infty - \tau)^{- 2 \gamma}
\;$ where $c = \sum_i c^i \;$ and $\gamma = \sum_i \gamma^i
\;$. For the ansatz in equations (\ref{asymptau}) to be
consistent, it is necessary that $\gamma > 0 \;$ so that
$e^\Lambda \to \infty \;$ in the limit $\tau \to \tau_\infty
\;$. Now $t(\tau)$ follows from equation (\ref{tau}) and is
given by
\begin{equation}\label{ttauasymp}
t - t_s = \frac{1}{2 \gamma - 1} \; e^c \; 
(\tau_\infty - \tau)^{- (2 \gamma - 1)} 
\; \; \; , \; \; \; \; 
\gamma = \sum_{i, I} u^i_I \; \gamma^I
\end{equation}
where $t_s \;$ is a finite constant. If $2 \gamma < 1 \;$ then
$t \to t_s \;$ which means that $e^\Lambda \to \infty$ at a
finite physical time $t_s \;$. If $2 \gamma > 1 \;$ then $t \to
\infty \;$ in the limit $e^\Lambda \to \infty$. Which case is
realised, {\em i.e.} whether $2 \gamma < 1 \;$ or $> 1 \;$,
depends on the values of $u^I_i \;$.

Using equation (\ref{ttauasymp}), the asymptotic behaviour of
$e^{l^I} \;$ and $e^{\lambda^i} \;$ can be obtained in terms of
$t$. For example, let $2 \gamma > 1 \;$ and
\begin{equation}\label{asymp}
e^{l^I} = \; e^{b^I + 2 b} \; t^{\beta^I} 
\; \; , \; \; \;
e^{\lambda^i} = \; e^{b^i} \; t^{\beta^i} 
\; \; , \; \; \;
e^{\Lambda} = \; e^b \; t^{\beta} 
\end{equation}
in the limit $t \to \infty \;$. It then follows that 
\begin{equation}\label{betagamma}
\beta^I = \; \frac{2 \; \gamma^I}{2 \gamma - 1} 
\; \; , \; \; \;
\beta^i = \; \frac{2 \; \gamma^i}{2 \gamma - 1} 
\; \; , \; \; \;
\beta = \; \frac{2 \; \gamma}{2 \gamma - 1} \; \; . 
\end{equation}
Note that, in this case, we have $e^\Lambda \sim t^\beta \;$ in
the limit $t \to \infty \;$ with $\beta > 1 \;$. See the
discussion below equation (\ref{weak}) for the relevance of this
feature.

To obtain the values of $\gamma^I \;$, and thereby $\gamma^i
\;$, in equation (\ref{asymptau}), consider equation (\ref{c4})
from which it follows that
\begin{equation}\label{asymp2tau}
2 \; \sum_J {\cal G}_{I J} \; \gamma^J = \; e^{c^I} \;
(\tau_\infty - \tau)^{2 (1 - \gamma^I) } \; \; .  
\end{equation}
The left hand side in the above equation is a constant but the
right hand side depends on $\tau \;$. This is consistent if
$\gamma^I = 1 \;$ in which case the right hand side becomes a
positive constant, or if $\gamma^I < 1 \;$ in which case the
right hand side vanishes in the limit $\tau \to \tau_\infty
\;$. Thus, there are two possibilities:
\begin{eqnarray}
{\bf (i)} \; \; \;  \gamma^I = 1 
& \Longrightarrow & 
2 \; \sum_{J} {\cal G}_{I J} \; \gamma^J \; = \; e^{c^I} 
> 0 \label{eqtau} \\
{\bf (ii)} \; \; \; \gamma^I \ne 1 
& \Longrightarrow & 
\sum_{J} {\cal G}_{I J} \; \gamma^J = 0 
\; \; \; , \; \; \; \gamma^I < 1 \; \; .  \label{neqtau}
\end{eqnarray}

For a given ${\cal G}_{I J} \;$, the possible consistent
solutions for $(\gamma^I, \; e^{c^I})$ are to be obtained as
follows. Assume that some $I$'s are of type {\bf (i)} and the
remaining ones are of type {\bf (ii)}. Then solve equations
(\ref{eqtau}) and (\ref{neqtau}) for $e^{c^I}$ in type {\bf (i)}
and for $\gamma^I$ in type {\bf (ii)}. Such a solution is
consistent if the resulting $(e^{c^I}, \; \gamma^I)$ satisfy
$e^{c^I} > 0$ for $I$s in type {\bf (i)} and $\gamma^I < 1 \;$
for $I$s in type {\bf (ii)}. Also, some or all of the
$\gamma^I$s must be $> 0 \;$ as required in equation
(\ref{asymptau}). (It is further necessary that the resulting
$\gamma > 0 \;$ so that $e^\Lambda \to \infty \;$, but
calculating $\gamma$ requires $u^I_i \;$.)
 
Consider an example, which will be useful later, where ${\cal
G}^{I J} \;$ and ${\cal G}_{I J} \;$ are given by
\begin{equation}\label{ab}
{\cal G}^{I J} = a \; (b - \delta^{I J}) 
\; \; \; , \; \; \; \; \; \; 
{\cal G}_{I J} = \frac{1}{a} \; 
( \frac{b}{N b - 1} - \delta_{I J} ) 
\end{equation}
with $a > 0$ and $N b > 1 \:$. It is then easy to show that the
only possibility is the one given in {\bf (i)}. Also $\sum_J
{\cal G}_{I J} = \frac{1}{a (N b - 1)} > 0 \;$, and thus
$\gamma^I = 1 \;$ for all $I \;$ is a consistent solution as
required by equation (\ref{eqtau}). In the $N = 1 \;$ case, we
get ${\cal G}^{1 1} = {\cal G} = a \; (b - 1) > 0 \;$, and $e^l
\;$ in the limit $\tau \to \tau_\infty \;$ obtained as described
above agrees with that obtained from the explicit solution, see
below equation (\ref{1alpha}).

Thus $e^{c^I}$ and $\gamma^I$, and thereby $\gamma^i = \sum_I
u^i_I \; \gamma^I \;$ and $\gamma = \sum_{i, I} u^i_I \;
\gamma^I \;$, are all determined ultimately by $u^I_i \;$. The
constants $c^i $ are given by equation (\ref{cigammai}) and they
depend on $u^I_i \;$, on the initial values $l^I_0 \;$ and $L^i
\;$, and also on $\tau_\infty \;$. But determining $\tau_\infty
\;$, and hence determining $c^i $ when $L^i \;$ do not all
vanish, requires complete solution for $l^I (\tau) \;$.

\subsection{Deviations from $e^{l^I} \to \infty \;$
Asymptotics} \label{sec:deviations}
We consider the deviations of $l^I(\tau) \;$ from its asymptotic
behaviour given in equation (\ref{asymptau}), which will turn
out to be of interest. Let the deviations $s^I(\tau) \;$ for $I
= 1, 2, \cdots, N \;$ be defined, in the limit $\tau \to
\tau_\infty \;$, by
\begin{equation}\label{si}
e^{l^I} = \; e^{c^I} \; (\tau_\infty - \tau)^{- 2 \gamma^I} 
\; \; e^{s^I(\tau)}
\end{equation}
where $c^I$ and $\gamma^I$ are determined as described earlier.
For the purpose of illustration, and also for later use, we now
assume that all the $I$s are of type {\bf (i)}, namely that
$\gamma^I = 1 \;$ and $e^{c^I} = 2 \; \sum_{J} {\cal G}_{I J} >
0 \;$ for all $I \;$. It then follows straightforwardly from
equation (\ref{c3}) that
\begin{equation}\label{si2}
(\tau_\infty - \tau)^2 \; s^I_{\tau \tau} 
= 2 \; \sum_{K, L} {\cal G}^{I K} \; {\cal G}_{K L} \; 
(e^{s^K} - 1) \; \; .
\end{equation}

Consider the example of ${\cal G}^{I J} \;$ given in equation
(\ref{ab}). Then $\sum_J {\cal G}_{I J} = \frac{1}{a (N b - 1)}
\;$ and, for any $\sigma^K \;$, one has
\begin{equation}\label{sigma}
\sum_{K, L} {\cal G}^{I K} \; {\cal G}_{K L} \; \sigma^K
= - \; \frac{1}{N b - 1} \; (\sigma^I - b \; 
\sum_K \sigma^K ) \; \; .
\end{equation}
In equation (\ref{si2}), $\sigma^K = 2 \; (e^{s^K} - 1) \;$. It
now follows easily that, up to the leading order in $\{s^K\}
\;$, the difference $s^I - s^J \;$ obeys the equation
\begin{equation}\label{sisj}
(\tau_\infty - \tau)^2 \; (s^I - s^J)_{\tau \tau} 
+ \frac{2}{N b - 1} \; (s^I - s^J) = 0 \; \; . 
\end{equation}
The solutions to these equations are of the form
\begin{equation}\label{sisjsoln}
(s^I - s^J) \; \sim \; 
(\tau_\infty - \tau)^{ \frac{1}{2} \; (1 \pm \sqrt{\Delta})}
\; \; \; , \; \; \; \; 
\Delta = 1 - \frac{8}{N b - 1} \; \; .
\end{equation}
Note that $s^I - s^J = l^I - l^J \;$ since $\gamma^I$ and $c^I$
are same for all $I \;$, see equation (\ref{si}). Hence,
equations (\ref{sisj}) and (\ref{sisjsoln}) can be used to
understand in more detail the behaviour of $l^I$s as they all
diverge in the limit $\tau \to \tau_\infty \;$ as given in
equation (\ref{asymptau}). We will discuss these features in 
next two sections.

\section{Intersecting Branes} \label{sec:IB}
We now analyse the evolution of the universe dominated by
mutually BPS $N$ intersecting brane configurations of M
theory. The number of spacetime dimensions $D = 11 \;$.
We describe general case first, then we specialise on $22'55'$ 
configuration.
The
equations of state are assumed to be given by $p_{i I} = (1 -
u^I_i) \; \rho_I \;$ where, as a consequence of U duality
symmetries, $u^I_i \;$ are parametrised in terms of one constant
$u \;$. The indices $i, j, \cdots$ run from $1$ to $10$ and the
indices $I, J, \cdots$ from $1$ to $N \;$. For 2 branes, 5
branes, and waves, $N = 1 \;$ and the corresponding $u^I_i$ are
given in equations (\ref{25W}). For $22'55' \;$ configuration,
$N = 4 \;$ and the corresponding $u^I_i$ are given in equations
(\ref{I}).

\subsection{Evolution Equations} \label{sec:evoleq}
The evolution of $\lambda^i \;$ describing the scale factors is
given by the equations described earlier which, for ease of
reference, we summarise below:
\begin{eqnarray} 
\lambda^i_{\tau \tau} & = & \sum_J u^{i J} \; e^{l^J} 
\label{d2} \\
l^I_{\tau \tau} & = & \sum_J {\cal G}^{I J} \; e^{l^J} 
\label{d3} \\
\lambda^i & = & \sum_J u^i_J \; (l^J - l^J_0) + \; L^i \; \tau
\label{d5} 
\end{eqnarray}
where 
\begin{equation}\label{defn}
u^{i I} = \sum_j G^{i j} \; u^I_j 
\; \; \; , \; \; \; \; 
{\cal G}^{I J} = \sum_{i, j} G^{i j} \; u^I_i \; u^J_j 
\; \; \; , \; \; \; \; 
u^i_I = \sum_{j, J} {\cal G}_{I J} \; G^{i j} \; u^J_j 
\end{equation}
with $G^{i j} \;$ and ${\cal G}_{I J} \;$ as defined earlier,
and $L^i \;$ are arbitrary constants satisfying the constraints
$\sum_i u^I_i L^i = 0 \;$ for all $I \;$. Also, $l^I_\tau \;$
obey the constraint
\begin{equation}\label{d6}
\sum_{I, J} {\cal G}_{I J} \; l^I_\tau \; l^J_\tau = 
2 \; ( E + \sum_J e^{l^J} ) 
\end{equation}
where $2 E = - \; \sum_{i, j} G_{i j} L^i L^j \;$. Equations
(\ref{d3}) and (\ref{d6}) are to be solved for $l^I(\tau) \;$
with initial conditions $l^I(0) = l^I_0 = ln \; \rho_{I 0} \;$
and $l^I_\tau(0) = K^I \;$ where $\rho_{I 0} \;$ are initial
densities and
\begin{equation}\label{d6ic}
\sum_{I, J} {\cal G}_{I J} \; K^I \; K^J = 
2 \; ( E + \sum_J e^{l^J_0} ) \; \; . 
\end{equation}
Then $\lambda^i (\tau) \;$ follow from equation (\ref{d5}) and
the physical time $t(\tau) \;$ from $d t = e^\Lambda \; d \tau
\;$. Inverting $t(\tau) \;$ then gives $\tau(t) \;$, and thereby
$\lambda^i (t) \;$.

We can now calculate ${\cal G}^{I J} \;$ for the mutually BPS
intersecting brane configurations. As explained in footnote 
{\bf \ref{foot:bps}}, 
in the BPS configurations two stacks of 2 branes intersect
at a point; two stacks of 5 branes intersect along three common
spatial directions; a stack of 2 branes intersects a stack of 5
branes along one common spatial direction; and, waves, if
present, will be along a common intersection direction. With
these rules given, it is now straightforward to calculate ${\cal
G}^{I J} \;$ using equations (\ref{25W}) and (\ref{defn}). It
turns out because of the BPS intersection rules that the
resulting ${\cal G}^{I J} \;$ are given by
\begin{equation}\label{G^IJ}
{\cal G}^{I J} = 2 u^2 \; (1 - \delta^{I J}) \; \; . 
\end{equation}
The corresponding ${\cal G}_{I J} \;$ exists for $N > 1 \;$, and
is given by
\begin{equation}\label{G_IJ}
{\cal G}_{I J} = \frac{1}{2 u^2} \; 
( \frac{1}{N - 1} - \delta_{I J} ) \; \; . 
\end{equation}
Note that, for $N > 1 \;$, the above ${\cal G}^{I J} \;$ is a
special case of the example considered earlier in equation
(\ref{ab}), now with $a = 2 u^2 \;$ and $b = 1 \;$,

\subsection{$22'55'$ Case}
It is also straightforward to calculate $u^{i I} \;$ and $u^i_I
\;$ for the $22'55'$ configuration using the definitions in
equation (\ref{defn}) and the $u^I_i \;$ in equation
(\ref{I}). They are given by
\begin{eqnarray} 
2 & : & u^{i 1} \; \propto \; 
(-2, -2, \; \; 1, \; \; 1, \; \; 1, 
\; \; \; 1, \; \; \; 1, \; \; 1, \; \; 1, \; \; 1) 
\nonumber \\
2' & : & u^{i 2} \; \propto \;
(\; \; 1, \; \; 1, -2, -2, \; \; 1, 
\; \; \; 1, \; \; \; 1, \; \; 1, \; \; 1, \; \; 1) 
\nonumber \\
5 & : & u^{i 3} \; \propto \;
(-1, \; \; 2, -1, \; \; 2, -1, -
1, -1, \; \; 2, \; \; 2, \; \; 2) 
\nonumber \\
5' & : & u^{i 4} \; \propto \;
(\; \; 2, -1, \; \; 2, -1, -1, 
-1, -1, \; \; 2, \; \; 2, \; \; 2) 
\label{I2} 
\end{eqnarray} 
where the proportionality constant is $\frac{u}{3} \;$, and by
\begin{eqnarray} 
2 & : & u_1^i \; \propto \;  
(\; \; 2, \; \; 2, -1, -1, -1, 
-1, -1, \; \; 1, \; \; 1, \; \; 1) 
\nonumber \\
2' & : & u_2^i \; \propto \;   
(-1, -1, \; \; 2, \; \; 2, -1, 
-1, - 1, \; \; 1, \; \; 1, \; \; 1) 
\nonumber \\
5 & : & u_3^i \; \propto \;  
(\; \; 1, -2, \; \; 1, -2, \; \; \; 1, 
\; \; 1, \; \; 1, \; \; \; 0, \; \; 0, \; \; 0) 
\nonumber \\
5' & : & u_4^i \; \propto \;
(-2, \; \; 1, -2, \; \; 1, \; \; \; 1, 
\; \; 1, \; \; 1, \; \; \; 0, \; \; 0, \; \; 0) 
\label{I3}
\end{eqnarray}
where the proportionality constant is $\frac{1}{6 u} \;$. 

We are unable to solve equations (\ref{d3}), (\ref{d6}), and
(\ref{G^IJ}) for $N > 1 \;$.\footnote{ In the case of black
holes, the equations of motion for the corresponding harmonic
functions $H^I = 1 + \frac{Q^I}{r} \equiv e^{\tilde{h}_I} \;$
can also be written in a form similar to that of equation
(\ref{d3}). The main steps are indicated in Appendix {\bf \ref{sec:Ubh}}. 
The analogous ${\cal G}^{I J} \;$ in the black hole case
turns out to be $\propto \; \delta^{I J} \;$, and the equations
can then be solved.

Also, note that if $L^i = 0 \;$ for all $i \;$ then $\lambda^i
\;$ in equation (\ref{d5}) here may be written as in equation
(\ref{2255bh}) in Appendix {\bf \ref{sec:Ubh}}. The role of $\tilde{h}_I \;$
there is played by the functions $2 u h_I = 2 u \; \sum_J {\cal
G}_{I J} \; ( l^J - l^J_0 ) \;$ here. Such a similarity is
present for other intersecting brane configurations also.}
However, applying the general analysis described in section 
{\bf \ref{sec:GEvol}} 
and making further use of the explicit forms of $u^I_i \;$
and ${\cal G}^{I J} \;$ given in equations (\ref{I}) and
(\ref{G^IJ}), one can understand the qualitative features of the
evolution of the $22'55' \;$ configuration.

We first make several remarks which will lead to an immediate
understanding of the evolution of this configuration.

\vspace{2ex}

{\bf (1) $ \; \;$ }
Let $u_i = \sum_I u^I_i \;$. It can then be checked that
$\sum_{i, j} G^{i j} u_i u_j > 0 \;$.  Also, $\sum_i u_i L^i = 0
\;$ since $\sum_i u^I_i L^i = 0 \;$ for all $I \;$. Hence, as
shown in Appendix {\bf \ref{ap:Ege0}}, it follows that $E \;$ given in
equation (\ref{d6}) is $ \ge 0 \;$ and that it vanishes if and
only if $L^i \;$ all vanish.

\vspace{2ex}

{\bf (2) $ \; \;$ }
The constraints $\sum_i u^I_i L^i = 0 \;$ imply that
\begin{eqnarray}
& & L^1 - L^4 = L^2 - L^3 = L^5 + L^6 + L^7 = 0 \nonumber \\
& & L^8 + L^9 + L^{10} = - 3 \; (L^1 + L^2) \; \; . 
\label{LI4}
\end{eqnarray}
Thus, for example, we may take $(L^1, L^2, L^6, L^7, L^8, L^9)
\;$ to be independent. The remaining $L^i$s are then determined
by the above equations. Also, we have
\begin{equation}\label{LLc}
L \equiv \sum_i L^i = - (L^1 + L^2) \; \; .
\end{equation}

Using equations (\ref{LI4}), (\ref{LLc}), and the Schwarz
inequality (\ref{sch}) in Appendix {\bf \ref{ap:Ege0}}, we write $E \;$ as
\begin{eqnarray} 
2 E & = & \sum_i (L^i)^2 - (\sum_i L^i)^2 \nonumber \\
& = & 3 (L)^2 + \sum_{i = 5}^7 (L^i)^2 + 2 \sigma_2^2 
+ \sigma_3^2 \nonumber \\
& = & 3 (L^1)^2 + (L^1 + 2 L^2)^2 
+ \sum_{i = 5}^7 (L^i)^2 + \sigma_3^2  \label{E>}
\end{eqnarray}
where the first line is the definition of $E \;$, $\sigma_2 = 0
\;$ if and only if $L^1 = L^2 \;$, and $\sigma_3 = 0 \;$ if and
only if $L^8 = L^9 = L^{10} \;$. See the Schwarz inequality
given in equation (\ref{sch}). It is easy to show that the above
expressions for $E \;$ imply that $(L^i)^2 \;$ for all $i \;$
are bounded above by $E \;$ as follows: $E \ge c_i (L^i)^2 \ge 0
\;$ where $c_i$ are constants of ${\cal O}(1) \;$. In
particular, note the inequality $2 E \ge 3 (L)^2 \;$ which is
required in Appendix {\bf \ref{ap:Sign}}.

\vspace{2ex}

{\bf (3) $ \; \;$ }
It follows from equations (\ref{d5}), (\ref{I3}), and
(\ref{LLc}) that
\begin{equation}\label{d5ex}
\Lambda_\tau = \sum_i \lambda^i_\tau = \frac{1}{6 u} \;
(2 l^1_\tau + 2 l^2_\tau + l^3_\tau + l^4_\tau) + L \; \; .
\end{equation}
Using the explicit form of ${\cal G}_{I J} \;$ given in equation 
(\ref{G_IJ}) with $N = 4 \;$, equation (\ref{d6}) becomes
\begin{equation}\label{d6ex}
(\sum_I l^I_\tau)^2 - 3 \; \sum_I (l^I_\tau)^2 
= 12 u^2 \; (E + \sum_I e^{l^I}) \; \; > \; 0 
\end{equation} 
where the inequality follows since $E \ge 0 \;$ and $e^{l^I} > 0
\;$. We show in Appendix {\bf \ref{ap:Sign}} that this inequality implies
that none of $(\Lambda_\tau, \; l^I_\tau) \;$ may vanish, and
that they must all have same sign. Hence, for all $\tau \;$
throughout the evolution, $(\Lambda_\tau, \; l^I_\tau) \;$ must
all be non vanishing, and be all positive or all negative.

\section{Asymptotic Evolution} \label{sec:asevol}
With no loss of generality, let $\Lambda_t > 0$ initially at $t
= t_0 \;$. Then it follows from the above result that
$(\Lambda_\tau, \; l^I_\tau) \;$ must all be positive and non
vanishing for all $\tau \;$. Hence, $(\Lambda, \; l^I) \;$ are
all monotonically increasing functions for all $\tau \;$
throughout the evolution.

Equation (\ref{d3}) may be written, using equation (\ref{G^IJ}),
as
\begin{equation}\label{e3}
l^I_{\tau \tau} = 2 u^2 \; \sum_{J \ne I} \; e^{l^J} \; \; . 
\end{equation}
In the past, $\tau \;$ and all $l^I \;$ decrease continuously.
Hence, the right hand side in equation (\ref{e3}) becomes more
and more negligible. The asymptotic solution in the limit $\tau
\to - \infty \;$ is then given by $l^I = \tilde{c}^I \tau +
\tilde{d}^I \;$ where $\tilde{c}^I > 0 \;$. Thus $e^{l^I} \to 0
\;$ in this limit.

Similarly, in the future, $\tau \;$ and all $l^I \;$ increase
continuously. However, the right hand side in equation
(\ref{e3}) increases exponentially now. It is then obvious that
all $e^{l^I} \to \infty \;$ within a finite interval of $\tau
\;$, {\em i.e.} at a finite value $\tau_\infty \;$ of $\tau
\;$. In this context, see equations (\ref{1c1}) and (\ref{1c2}),
and the general analysis given in asymptotic evolution in 
section {\bf \ref{sec:ttoinfty}}.

We now analyse the corresponding asymptotic solutions.

\subsection{Asymptotic Evolution: $e^\Lambda \to 0 \;$}
It follows from the above discussion that $e^ \Lambda \to 0 \;$
in the limit $\tau \to - \infty \;$. Also, in this limit, we
have
\begin{equation}\label{x0l}
e^{l^I} \; = \; e^{\tilde{c}^I \tau} \; = \; 
t^{\tilde{b}^I} \; \; \; , \; \; \; \; 
e^{\lambda^i} \; = \; e^{\tilde{c}^i \tau} \; = \; 
t^{\tilde{b}^i} 
\end{equation}
up to multiplicative constants where $(\tilde{c}^I, \;
\tilde{c}^i, \; \tilde{b}^I, \; \tilde{b}^i) \;$ are constants.
The evolution is then of Kasner type and is similar to that
described in section {\bf \ref{sec:tto0}}. The constants $\tilde{c}^I$s
are determined by the initial values $l^I_0 \;$ and $K^I \;$,
but obtaining the exact dependence in the general case requires
complete solution for $l^I (\tau) \;$. However, if the initial
values $l^I_0 \;$ are large and negative then we have $e^{l^I}
\ll 1 \;$ for all $\tau < 0 \;$ and, hence, $\tilde{c}^I = K^I
\;$ to a good approximation.

\subsection{Asymptotic Evolution: $e^\Lambda \to \infty \;$}
It follows from the above discussion that $e^ \Lambda \to \infty
\;$ in the limit $\tau \to \tau_\infty \;$ where $\tau_\infty
\;$ is finite.  Also, $e^{l^I} \to \infty \;$ in this limit and
$\tau_\infty \;$ depends on the initial values $l^I_0 \;$ and
$K^I \;$.

Although solutions for $l^I(\tau) \;$ are not known, their
asymptotic forms in the limit $\tau \to \tau_\infty \;$, and
hence those of $\lambda^i (\tau) \;$, may be obtained following
the analysis given in section {\bf \ref{sec:ttoinfty}}. $\; {\cal G}^{I J}
\;$ in equation (\ref{G^IJ}) is a special case of the example
(\ref{ab}) where, now, $N = 4 \;$, $a = 2 u^2 \;$, and $b = 1
\;$. Hence, it can be shown to correspond to the possibility
{\bf (i)} given in equation (\ref{eqtau}). Therefore, we have
$\gamma^I = 1 \;$ and $e^{c^I} = 2 \sum_J {\cal G}_{I J} =
\frac{1}{3 u^2} \;$.

It then follows from equation (\ref{asymptau}) that $e^{l^I} \;$
and $e^{\lambda^i} \;$ are given in the limit $\tau \to
\tau_\infty \;$ by
\begin{eqnarray}
e^{l^I} & = &  \frac{1}{3 u^2} \; 
\frac{1}{(\tau_\infty - \tau)^2} \label{lItau} \\
e^{\lambda^i} & = & e^{v^i} \; \left( \frac{1}{3 u^2} \; \;
\frac{1}{(\tau_\infty - \tau)^2} \right)^{\sum_I u^i_I}
\label{lambdaitau}
\end{eqnarray}
where, since $\rho_{I 0} = e^{l^I_0} \;$, we have 
\begin{equation}\label{vi}
v^i = - \sum_J u^i_J \; l^J_0 + L^i \; \tau_\infty 
\; \; , \; \; \; \; \; 
e^{v^i} = e^{ L^i \; \tau_\infty } \; \; 
\prod_J (\rho_{J 0})^{- u^i_J} \; \; .
\end{equation}
Also, since $\gamma = \sum_{i, I} u^i_I = \frac{1}{u} \;$, we
have from equation (\ref{ttauasymp}) that the physical time $t
\;$ is given in this limit by
\begin{equation}\label{ttau}
t - t_s = A \; \left( \tau_\infty - \tau \right)^{- \; 
\frac{2 - u}{u}}
\end{equation}
where $t_s \;$ and $A$ are finite constants. Clearly, $t \to
\infty \;$ in the limit $\tau \to \tau_\infty \;$ since it is
assumed that $0 < u < 2 \;$. In this limit, the scale factors
$e^{\lambda^i} \;$ may be written in terms of $t \;$ as
\begin{equation}\label{lambdait}
e^{\lambda^i} \; = \; e^{v^i} \; \; (B \; t)^{\beta^i} 
\end{equation}
where $B$ is a constant and $\beta^i = \frac{2 u}{2 - u} \;
\sum_J u^i_J \;$. Using equation (\ref{I3}) for $u^i_I \;$, the
exponents $\beta^i \;$ are given by
\begin{equation}\label{betai}
\beta^i \; \propto \; 
(0, \; 0, \; 0, \; 0, \; 0, \; 0, \; 0, \; 1, \; 1, \; 1) 
\end{equation} 
where the proportionality constant is $\frac{2}{3 (2 - u)} \;$.
Note that $\beta = \sum_i \beta^i = \frac{2}{2 - u} > 1 \;$.
Hence, we have $e^\Lambda \sim t^\beta \;$ in the limit $t \to
\infty \;$ with $\beta > 1 \;$. See the discussion below
equation (\ref{weak}) for the relevance of this feature.

Thus, asymptotically in the limit $t \to \infty \;$, we obtain
that $e^{\lambda^i} \to t^{\frac{2}{3 (2 - u)}} \; $ for the
common transverse directions $i = 8, 9, 10 \;$. Hence, these
directions continue to expand, their expansion being precisely
that of a $(3 + 1)$ -- dimensional homogeneous, isotropic
universe containing a perfect fluid whose equation of state is
$p = (1 - u) \; \rho \;$. Also, $\; e^{\lambda^i} \to e^{v^i}
\;$ for the brane directions $i = 1, \cdots, 7 \;$. Hence, these
directions cease to expand or contract. Their sizes are thus
stabilised and are given by $e^{v^i} \;$. Note that this result
is in accord with the general result described in section {\bf \ref{sec:genr}} 
since, in the limit $\tau \to \tau_\infty \;$, the brane
densities $\rho_I \propto e^{l^I} \;$ all become equal and hence
the four types of branes all become identical; and, $t \to
\infty \;$ and $e^\Lambda \sim t^\beta \to \infty \;$ with
$\beta > 1 \;$.

\section{Mechanism of Stabilisation} \label{sec:Mec}
Using the asymptotic solutions, we can now give a physical
interpretation of the dynamical mechanism underlying the
stabilisation of the brane directions seen above for the $22'55'
\;$ configuration.

We first study the stabilisation process. Consider equation
(\ref{d2}) for $\lambda^1_{\tau \tau} \;$, for example. Using
the values of $u^{i I} \;$ given in equation (\ref{I2}), we have
\begin{equation}\label{1tautau}
\lambda^1_{\tau \tau} \; \propto \; 
(- 2 e^{l^1} + e^{l^2} - e^{l^3} + 2 e^{l^4}) \; \; . 
\end{equation}
In the $22'55' \;$ configuration, $x^1 \;$ direction is wrapped
by 2 branes and 5 branes and is transverse to $2'$ branes and
$5'$ branes. Thus, from the above equation for $\lambda^1 \;$
and from similar equations for $\lambda^2, \cdots, \lambda^7
\;$, we see that 2 brane and 5 brane directions `contract with a
force' proportional to $2 \rho_{(2)} \;$ and $\rho_{(5)} \;$
respectively, whereas the directions transverse to them `expand
with a force' proportional to $\rho_{(2)} \;$ and $2 \rho_{(5)}
\;$ respectively, where $\rho_{(*)} \propto e^{l^{(*)}} \;$ are
the time dependent densities of the corresponding branes.

When $\rho_{I} \propto e^{l^I} \;$ all become equal, the forces
of expansion cancel the forces of contraction resulting in
vanishing net force for the $x^1$ direction. Then, using 
equation (\ref{fttau}), one has
\begin{equation}\label{2tautau}
\lambda^1_{\tau \tau} = e^{2 \Lambda} \; 
(\lambda^1_{t t} + \Lambda_t \lambda^1_t) = 0 \; \; . 
\end{equation}
Now, as described earlier in the context of equations (\ref{K})
and (\ref{L}), the transient `velocity' $\lambda^1_t \;$ is
damped and $\lambda^1 \;$ reaches a constant value in the
expanding universe here since we have $e^\Lambda \sim t^\beta
\;$ in the limit $t \to \infty \;$ with $\beta > 1 \;$. The
result is the stabilisation of the $x^1$ direction.

The stabilised size $e^{v^1} \;$ of $x^1$ direction is given by
\begin{equation}\label{v1}
e^{v^1} = e^{ L^1 \; \tau_\infty } \; \; \left( 
\frac{\rho_{2 0} \; \rho_{4 0}^2} {\rho_{3 0} \; \rho_{1 0}^2}
\right)^{\frac{1}{6 u}} \; \; , 
\end{equation}
see equation (\ref{vi}). Note that $e^{v^1} \;$ can be
interpreted as arising from the imbalance among the initial
brane densities $\rho_{I 0} \;$, and from the parts $L^1 \;$ of
$\lambda^1_t (0) \;$ which indicate the transients. The above
analysis can be similarly applied to the stabilisation of other
brane directions $(x^2, \cdots, x^7) \;$ in the $22'55' \;$
configuration. 

Thus, three conditions need to be satisfied for stabilisation:
{\bf (1)} the time dependent brane densities $\rho_{I} \propto
e^{l^I} \;$ all become equal; {\bf (2)} the forces of expansion
and contraction for each of the brane directions be just right
so that the net force vanishes; {\bf (3)} the universe be
expanding as $e^\Lambda \sim t^\beta \;$ in the limit $t \to
\infty \;$ with $\beta > 1 \;$ so that the transient velocities
are damped and the corresponding scale factors reach constant
values.

For any mutually BPS $N > 1 \;$ intersecting brane
configurations with the equations of state as assumed here, it
is straightforward to show using the earlier analysis that the
evolution equations ensure that $e^{l^I} \;$ all become equal
asymptotically even if they were unequal initially, and that
$e^\Lambda \sim t^\beta \;$ in the limit $t \to \infty \;$ with
$\beta > 1 \;$. Thus conditions {\bf (1)} and {\bf (3)} are
satisfied. Condition {\bf (2)} requires the brane configuration
to be such that each of the brane directions is wrapped by, and
is transverse to, just the right number and kind of branes. This
condition is satisfied for the $N = 4 \;$ configurations $22'55'
\;$ and $55'5''W \;$, both of which result in the stabilisation
of seven brane directions and the expansion of the remaining
three spatial directions. To our knowledge, the only other
configurations which satisfy the condition {\bf (2)} are the $N
= 3 \;$ configurations $22'2'' \;$ and $25W \;$, both of which
result in the stabilisation of six brane directions and the
expansion of the remaining four spatial directions \cite{k0707}.
However it is the $N = 4 \;$ configurations that are
entropically favourable, see equation (\ref{sn}).

Note that, as described in section {\bf \ref{sec:genr}} and up to certain
technical assumptions regarding the equality of brane densities
and the asymptotic behaviour of $e^\Lambda \;$, the
stabilisation here follows essentially as a consequence of U
duality symmetries. In particular, it is independent of the
ansatz for energy momentum tensors, or of the assumptions about
equations of state, as long as the components of the energy
momentum tensors obey the U duality constraints of the type
given in equation (\ref{weak}). Obtaining the details of the
stabilisation, however, requires further assumptions {\em e.g.}
of the type made here.

Note also that the present mechanism of stabilisation of seven
brane directions, and the consequent emergence of three large
spatial directions, is very different from the ones proposed in
string theory or in brane gas models \cite{bv,gas,3+1,gm}.

\section{Stabilised Sizes of Brane Directions} \label{sec:size}
We thus see for the $22'55' \;$ configuration that,
asymptotically in the limit $e^\Lambda \to \infty \;$, the
initial $(10 + 1)$ -- dimensional universe effectively becomes
$(3 + 1)$ -- dimensional. Also, if $v^s = min \{ v^1, \cdots,
v^7 \} \;$ then a dimensional reduction of the $(10 + 1)$ --
dimensional M theory along the corresponding $x^s$ direction
gives type IIA string theory with its dilaton now stabilised.
Using the standard relations, one can obtain the string coupling
constant $g_s \;$, the string scale $M_s \;$, and the four
dimensional Planck scale $M_4 \;$ in terms of the M theory scale
$M_{11} \;$ and the stabilised values $e^{v^i} \;$. Defining $v^c
= \sum_{i = 1}^7 v^i \;$ and assuming, with no loss of
generality, that the coordinate sizes of all spatial directions
are of ${\cal O} (M_{11}^{- 1}) \;$, we obtain
\begin{equation}\label{411} 
g^2_s = e^{3 v^s} 
\; \; \; , \; \; \;  \; \; \;  
M_4^2 = e^{v^c - v^s} \; M^2_s = e^{v^c} \; M^2_{11} 
\end{equation} 
where the equalities are valid up to numerical factors of ${\cal
O} (1) \;$ only and
\begin{equation}\label{vc} 
e^{v^c} = e^{ L^c \; \tau_\infty } \; \; 
\left( \frac{\rho_{1 0} \; \rho_{2 0}} {\rho_{3 0} \; 
\rho_{4 0}} \right)^{\frac{1}{6 u}} 
\; \; \; , \; \; \;  \; \; \;  
L^c = \sum_{i = 1}^7 L^i 
\end{equation}
as follows from equations (\ref{I3}), (\ref{vi}), and $\rho_{I
0} = e^{l^I_0} \;$. Also, note that $g_s = (\frac {M_s}
{M_{11}})^3 \;$.

Since we have an asymptotically $3 + 1$ dimensional universe
evolving from a $10 + 1$ dimensional one, it is of interest to
study the resulting ratios $\frac{M_{11}}{M_4} \;$ and
$\frac{M_s}{M_4} \;$, and study their dependence on the initial
values $(l^I_0, K^I, L^i) \;$. In particular, one may like to
know the generic values of these ratios and to know whether
arbitrarily small values are possible. Setting $M_4 = 10^{19} \;
GeV \;$, one then knows the generic scales of $M_{11} \;$ and
$M_s \;$ and, for example, whether $M_{11} = 10^{- 15} \; M_4 =
10 \; TeV \;$ is possible.

In view of the relations between $(M_{11}, M_s, M_4) \;$ given
in equation (\ref{411}), this requires studying the stabilised
values $e^{v^c} \;$ and $e^{v^c - v^s} \;$, their dependence on
$(l^I_0, K^I, L^i) \;$, and knowing whether they can be
arbitrarily large. Note that if $L^i = 0 \;$ for all $i \;$ then
$v^i \;$ are all determined in terms of $l^I_0 \;$ only, see
equation (\ref{vi}). It is then obvious from equations
(\ref{vi}) and (\ref{vc}) that any values for $e^{v^c} \;$ and
$e^{v^c - v^s} \;$, no matter how large, may be obtained by fine
tuning $\rho_{I 0} \;$ correspondingly. \footnote{ It follows
from equation (\ref{d6ic}) and the definition of $E \;$ that the
generic ranges of the initial values may be taken to be given by
$\vert L^i \vert \simeq K^I \simeq \sqrt{E} \simeq \sqrt{\rho_{I
0}} \;$ within a couple of orders of magnitude. If the initial
values lie way beyond such a range then we consider it as fine
tuning.}

This statement remains true even when $L^i$s do not all
vanish. In this case, however, one may question the necessity of
fine tuning since, for example, the relation $e^{v^c} \propto
e^{L^c \tau_\infty} \;$ suggests that large values such as
$10^{30} \sim e^{70}$ may be obtained by tuning $L^i$s, or
$\tau_\infty \;$, or both to within a couple of orders of
magnitude only. It turns out, as we explain below, that fine
tuning is still necessary to obtain such large values.

Consider first the possibility of tuning $L^i \;$. Note that
equations (\ref{d3}) and (\ref{d6}) are invariant under the
scaling
\begin{equation}\label{scaling}
(E, \; e^{l^I}, \; \tau) 
\; \; \; \longrightarrow \; \; \;
(\sigma^2 E, \; \sigma^2 e^{l^I}, \; \frac{\tau}{\sigma})
\end{equation} 
where $\sigma \;$ is a positive constant. The initial values
scale correspondingly as
\begin{equation}\label{icscaling}
(e^{l^I_0}, \; K^I, \; L^i) 
\; \; \; \longrightarrow \; \; \;
(\sigma^2 e^{l^I_0}, \; \sigma K^I, \; \sigma L^i) 
\; \; . 
\end{equation} 
It then follows from equation (\ref{d5}) that $\lambda^i$, and
hence $e^{v^i} \;$, remain invariant.\footnote{ This invariance
is equivalent to that of equations (\ref{a1}) and (\ref{a2})
under the scaling $(\lambda^i, \rho, p_i, t) \to (\lambda^i,
\sigma^2 \rho, \; \sigma^2 p_i, \; \frac{t}{\sigma}) \;$.} This
scaling property merely reflects the choice of a scale for
time. For example, using this scaling, one may set $\sum_J
e^{l^J_0} = 1 \;$ or, when $E > 0 \;$ as is the case here, set
$E = 1 \;$. The corresponding $\sigma \;$ then provides a
natural time scale for evolution.  We set $E = 1 \;$ using the
above scaling.

With $E = 1 \;$, the value of $\tau_\infty \;$ now depends only
on $l^I_0 \;$ and $K^I \;$. Since $2 E = \sum_i (L^i)^2 -
(\sum_i L^i)^2 \;$, it is still plausible to have a range of non
zero measure where $L^i \;$ are large and $E = 1 \;$, and
thereby obtain large values for $e^{v^c} \;$ and $e^{v^c - v^s}
\;$. However, $L^i$s are further constrained by $\sum_i u^I_i
L^i = 0 \;$, $I = 1, \cdots, 4 \;$, and consequently their
magnitudes are all bounded from above. For example, with $E = 1
\;$, we obtain $(L^c)^2 \le \frac{8}{3} \;$. See remark {\bf
(2)} in section {\bf \ref{sec:evoleq}}. Thus, large values of $e^{v^i} \;$
cannot be obtained by tuning $L^i \;$ alone.

Consider now the possibility of tuning $\tau_\infty \;$.
Obtaining the dependence of $\tau_\infty \;$ on $(l^I_0, K^I)
\;$ requires explicit solutions which are not available. Hence,
we obtain $\tau_\infty \;$ numerically. We will present the
numerical results in the next section. Here we point out that an
approximate expression for $\tau_\infty \;$ can be given in the
limit when $e^{l^I_0} \ll E \;$ for all $I \;$. The reasoning
involved is analogous to that used in obtaining $\tau_a \;$ in
equation (\ref{tauapprox}). Using similar reasoning and setting
$E = 1 \;$ now, we have that if $e^{l^I_0} \ll 1 \;$ for all $I
\;$ then
\begin{equation}\label{tauaI}
\tau_\infty \simeq \tau_a = \; min \; \{\tau_I \} 
\; \; , \; \; \; \;
\tau_I = - \; \frac{l^I_0}{K^I} \; \; .
\end{equation}
Note that $\tau_a$ can be calculated easily and requires no
knowledge of explicit solutions. Our numerical results show that
$\tau_a \;$ given above indeed provides a good approximation to
$\tau_\infty \;$ when $e^{l^I_0} \ll 1 \;$ for all $I \;$.

Note also that $K^I \;$ must satisfy equation (\ref{d6ic}) with
$E = 1 \;$. It then follows from an analysis similar to that
given in Appendix {\bf \ref{ap:Sign}} that $K^I \;$ are all positive, cannot
be too small, and are of ${\cal O}(1) \;$ generically. Hence, in
the limit $e^{l^I_0} \ll 1 \;$ for all $I$, $\; \tau_a \;$ in
equation (\ref{tauaI}) are of ${\cal O}(min \{ - l^I_0 \}) \;$.
This indicates that large values of $\tau_\infty \;$, and hence
of $e^{v^i} \;$, cannot be obtained by tuning $K^I \;$ alone; a
tuning of $l^I_0 \;$, which translates to fine tuning of
$\rho_{I 0} = e^{l^I_0} \;$, is required. Our numerical analysis
also supports this conclusion.

We thus find that, even when $L^i$s do not all vanish, a fine
tuning of $\rho_{I 0} = e^{l^I_0} \;$ is necessary to obtain
large values for $e^{v^c - v^s} \;$ and $e^{v^c} \;$. We will see some example
in section {\bf \ref{sec:Num}}.

\section{Discussion with Other Intersecting Configurations} \label{sec:no2255}
In this section we will discuss some intersecting configuration which are not
$22'55'$ distributed in said way. We mentioned all our assumption in last 
section. There we say our $22'55'$ is actually a fine tuned initial condition,
and how this configuration comes out naturally form M theory is not understood.
To discuss this point we here give some example configuration where 
stabilisation of 7 directions is not achieved. 

Even $22'55'$ branes intersecting in some other way following BPS intersection
rule may not produces required stabilisation. As an example consider following 
configuration:
\begin{eqnarray}
2 : -, -, -, -, -, -, \times, \times, -, - \nonumber \\
2': \times, \times, -, -, -, -, -, -, -, - \nonumber \\
5 : \times, -, \times, \times, \times, -, -, \times, -, - \nonumber \\
5': -, \times, \times, \times, -, \times, -, \times, -, - \nonumber 
\end{eqnarray}
where $\times$ denotes brane direction and $-$ direction are perpendicular to 
brane. For this configuration general analysis of section {\bf \ref{sec:GEvol}}
gives 
\begin{equation}
 \beta^i =  \frac{1}{6-3u}\left\{0,0,0,0,1,1,1,-1,
 2,2 \right\} \;. 
\end{equation}
So one can see here only $x^1,x^2, x^3$ and $x^4$ get stabilised. 
This configuration is different from ours. Here stabilisation is achieved 
only in the directions where exactly two branes are presents. But in our
$22'55'$ case exactly 2 sets of branes are present in all 7 directions.   
Similarly we can also do the 
analysis for $N>4$ case. None of these cases stabilisation of 7 directions
occurs. 
As discussed in section {\bf \ref{sec:GEvol}}, using equations 
(\ref{betagamma}), (\ref{eqtau}), (\ref{neqtau}) and (\ref{ab})
one can find general formula for $\beta^i$. In our case all $\gamma^I = 1$.
So in this case if we define 
\begin{equation*}
 x = \sum_{i,I} u^i_I \;,
\end{equation*}
then $\beta^i$ will be given by
\begin{equation}
 \beta^i = \frac{2}{2 x -1} \sum_I u^i_I \;,
\end{equation}
where $u^i_I$ are defined in equation (\ref{c5}). $u_i^I$'s depend on
particular intersection. For our ansatz of: $T_{AB}$, mentioned in section
{\bf \ref{sec:TAB}}, $u_i^I$'s are given by equation (\ref{25W}).

All possible
configurations are given for example in \cite{bergshoeff}.
In Table {\bf \ref{Tab:N=3}}, {\bf \ref{Tab:N=4}}, {\bf \ref{Tab:N=5}},
{\bf \ref{Tab:N=6}} and {\bf \ref{Tab:N=7}} we list a few 
intersecting M brane configurations
and their corresponding late time evolution.
We use notation of \cite{bergshoeff}, 
which is following: in square bracket numbers of each type of branes are
indicated. For example $[2^n,5^m]$ means $n$ number of 2 branes and $m$ number 
of 5 branes. In curly bracket how many branes are there in each direction is 
indicated. Position of the number indicates number of branes and corresponding
number indicates number of directions. For example $\{p,q,r,s,\ldots\}$ means
there is only 1 brane in $p$ directions, there are 2 branes presents in $q$
directions and so on. For example take $([2^2,5^2], \{3,4,1,0\})$. This is the 
configuration we discuss in last page. This means 2 sets of 2 branes and 2 sets
of 5 branes. Then $\{3,4,1,0\})$ denotes following: 1)there are 1 brane in 
3 directions, 
in this case $x^5$, $x^6$ and $x^7$. 2) There are 2 branes in 4 directions, 
namely $x^1$, $x^2$, $x^3$ and $x^4$. 3) There are 3 branes in 1 direction,
namely $x^8$. 4) And there is no direction which is populated by 4 branes.
We also list the configurations explicitly for the shake
of reader's convenience.

We first give some example for $N\le 3$ 
case in Table {\bf \ref{Tab:N=2}} and {\bf \ref{Tab:N=3}}. 
Next we give some example for $N=4$ in
Table {\bf \ref{Tab:N=4}}. Finally we give example of $N>4$ in table
{\bf \ref{Tab:N=5}}, {\bf \ref{Tab:N=6}} and {\bf \ref{Tab:N=7}}.
\begin{table}[H]
\centering
 \begin{tabular}{||c|c||}
  \hline \hline
Type of intersection & $\beta^i$ \\
  \hline \hline
$[2^2], \{4, 0\}$ & $\frac{1}{8 - 3 u} \{-1,-1,-1,-1,2,2,2,2,2,2\}$ \\
\hline
$[2^1,5^1], \{5, 1\}$ &
$\frac{1}{3-u},\{-1,0,0,0,0,0,1,1,1,1\}$ \\
\hline
$[5^2], \{4, 3\}$ & $\frac{2}{10-3 u}\{-2,-2,-2,1,1,1,1,4,4,4\}$ \\
\hline \hline
 \end{tabular}
\caption{{\em Table for various type of intersection vs $\beta^i$,
for $N = 2$.}}
\label{Tab:N=2}
\end{table}
The explicit configuration mentioned in above table, {\bf \ref{Tab:N=2}} given
below: 
\begin{eqnarray*}
&[2^2], \{4, 0\}& \\
&2 : {\times, \times, -, -, -, -, -, -, -, -} &\\
&2': {-, -, \times, \times, -, -, -, -, -, -}&
\end{eqnarray*}
\begin{eqnarray*}
&[2^1,5^1], \{5, 1\}& \\
&2: {\times, \times, -, -, -, -, -, -, -, -} &\\
&5: {\times, -, \times, \times, \times, \times, -, -, -, -} &
\end{eqnarray*}
\begin{eqnarray*}
&[5^2], \{4, 3\}& \\
&5: {\times, \times, \times, \times, \times, -, -, -, -, -}& \\
&5':{\times, \times, \times, -, -, \times, \times, -, -, -} &
\end{eqnarray*}
\begin{table}[H]
\centering
 \begin{tabular}{||c|c||}
  \hline \hline
Type of intersection & $\beta^i$ \\
  \hline \hline
$[5^3], \{0, 6, 1\}$ &
$\frac{1}{5-2 u}\{-1,0,0,0,0,0,0,2,2,2\}$ \\
\hline
$[5^3], \{6, 0, 3\}$ & $\frac{1}{5-2 u}\{1,1,1,1,1,1,1,1,1,2\}$ \\
\hline
$[2^1, 5^2], \{5, 2, 1\}$ & $\frac{5}{14-6 u}\{-4,2,-2,-2,2,2,2,2,5,5\}$ \\
\hline
$[2^1, 5^2], \{2, 5, 0\}$ & $\frac{1}{14-6 u} \{-1,-1,-1,-1,-1,2,2,5,5,5\}$ \\
\hline
$[2^3], \{6, 0, 0\}$ &
$\frac{1}{4-2 u}\{0,0,0,0,0,0,1,1,1,1\}$ \\
\hline \hline
 \end{tabular}
\caption{{\em Table for various type of intersection vs $\beta^i$,
for $N = 3$.}}
\label{Tab:N=3}
\end{table}
The explicit configuration mention in above table, {\bf \ref{Tab:N=3}} given
below: 
\begin{eqnarray*}
&[5^3], \{0, 6, 1\}& \\
&5: {\times, \times, \times, \times, \times, -, -, -, -, -} & \\
&5': {\times, \times, \times, -, -, \times, \times, -, -, -} & \\
&5'':{\times, -, -, \times, \times, \times, \times, -, -, -} &
\end{eqnarray*}
\begin{eqnarray*}
&[5^3], \{6, 0, 3\}& \\
&5:  {\times, \times, \times, \times, \times, -, -, -, -, -}& \\
&5': {-, -, \times, \times, \times, \times, \times, -, -, -}& \\
&5'':{-, -, \times, \times, \times, -, -, \times, \times, -}&
\end{eqnarray*}
\begin{eqnarray*}
&[2^1,5^2], \{5, 2, 1\}& \\
&2:{\times, \times, -, -, -, -, -, -, -, -} & \\
&5': {\times, -, \times, \times, \times, \times, -, -, -, -} &\\
&5'': {\times, -, \times, \times, -, -, \times, \times, -, -} &
\end{eqnarray*}
\begin{eqnarray*}
&[2^1,5^2], \{2, 5, 0\}& \\
&2:{\times, \times, -, -, -, -, -, -, -, -} & \\
&5':{\times, -, \times, \times, \times, \times, -, -, -, -} &\\
&5'':{-, \times, \times, \times, \times, -, \times, -, -, -}&
\end{eqnarray*}
\begin{eqnarray*}
&[2^3], \{6, 0, 0\}& \\
&2: {\times, \times, -, -, -, -, -, -, -, -} &\\
&2': {-, -, \times, \times, -, -, -, -, -, -} &\\
&2'': {-, -, -, -, \times, \times, -, -, -, -} &
\end{eqnarray*}

\begin{table}[H]
\centering
 \begin{tabular}{||c|c||}
  \hline \hline
Type of intersection & $\beta^i$ \\
  \hline \hline
$[5^4],\{3, 3, 1, 2\}$ & $\frac{1}{20-9 u}\{-1,-4,-4,2,5,2,5,2,5,8\}$ \\
\hline
$[5^4],\{4, 0, 4, 1\}$ & $\frac{1}{20-9 u}\{-4,-1,-1,-1,5,-1,5,5,5,8\}$ \\
\hline
$[5^4],\{0, 6, 0, 2\}$ & $\frac{2}{20-9 u}\{-2,-2,1,1,1,1,1,1,4,4\}$ \\
\hline
$[2^1, 5^3],\{2, 3, 3, 0\}$ & $\frac{1}{19-9 u}\{-2,1,-2,-2,1,4,1,4,7,7\}$ \\
\hline
$[2^2, 5^2],\{0, 7, 0, 0\}$ &  
$\frac{2}{6-3 u}\{0,0,0,0,0,0,0,1,1,1\}$ \\
\hline
$[2^2, 5^2],\{3, 4, 1, 0\}$ & 
$\frac{1}{6-3 u}\{\frac{1}{6-3u}\{0,0,0,0,1,1,1,-1,2,2 \}$ \\
\hline
$[2^2, 5^2],\{6, 1, 2, 0\}$ & $\frac{1}{6-3 u}\{-1,1,-1,1,0,1,1,1,1,2\}$ \\
\hline \hline
 \end{tabular}
\caption{{\em Table for various type of intersection vs $\beta^i$,
for $N = 4$.}}
\label{Tab:N=4}
\end{table}
The explicit configuration mentioned in above table, {\bf \ref{Tab:N=4}} given
below: 
\begin{eqnarray*}
&[5^4], \{3,3,1,2\}& \\
&5:{\times, \times, \times, \times, \times, -, -, -, -, -} & \\
&5':{\times, \times, \times, -, -, \times, \times, -, -, -} & \\
&5'': {\times, \times, \times, -, -, -, -, \times, \times, -} &\\
&5''':{-, \times, \times, \times, -, \times, -, \times, -, -} &
\end{eqnarray*}
\begin{eqnarray*}
&[5^4], \{4,0,4,1\}& \\
&5:{\times, \times, \times, \times, \times, -, -, -, -, -} & \\
&5':{\times, \times, \times, -, -, \times, \times, -, -, -} & \\
&5'':{\times, -, \times, \times, -, \times, -, \times, -, -} &\\
&5''':{\times, \times, -, \times, -, \times, -, -, \times, -} &
\end{eqnarray*}
\begin{eqnarray*}
&[5^4], \{0,6,0,2\}& \\
&5:{\times, \times, \times, \times, \times, -, -, -, -, -} & \\
&5':{\times, \times, \times, -, -, \times, \times, -, -, -} & \\
&5'':{\times, \times, -, \times, -, \times, -, \times, -, -}&\\
&5''':{\times, \times, -, -, \times, -, \times, \times, -, -}&
\end{eqnarray*}
\begin{eqnarray*}
&[2^1,5^3], \{2,3,3,0\}& \\
&2:{\times, \times, -, -, -, -, -, -, -, -} & \\
&5:{\times, -, \times, \times, \times, \times, -, -, -, -} & \\
&5':{-, \times, \times, \times, \times, -, \times, -, -, -} &\\
&5': {\times, -, \times, \times, -, -, \times, \times, -, -} &
\end{eqnarray*}
\begin{eqnarray*}
&[2^2,5^2], \{0,7,0,0\}\;\footnotemark& \\
&2:{\times, \times, -, -, -, -, -, -, -, -} & \\
&2':{-, -, \times, \times, -, -, -, -, -, -} & \\
&5:{\times, -, \times, -, \times, \times, \times, -, -, -} & \\
&5':{-, \times, -, \times, \times, \times, \times, -, -, -} &
\end{eqnarray*}
\footnotetext{This configuration is our configuration.}

\begin{eqnarray*}
&[2^2,5^2], \{3,4,1,0\}& \\
&2 : -, -, -, -, -, -, \times, \times, -, - & \\
&2': \times, \times, -, -, -, -, -, -, -, - & \\
&5 : \times, -, \times, \times, \times, -, -, \times, -, - & \\
&5': -, \times, \times, \times, -, \times, -, \times, -, - & 
\end{eqnarray*}
\begin{eqnarray*}
&[2^2,5^2], \{6,1,2,0\}& \\
&2:{\times, \times, -, -, -, -, -, -, -, -}  & \\
&2':{-, -, \times, \times, -, -, -, -, -, -} & \\
&5:{\times, -, \times, -, \times, \times, \times, -, -, -} & \\
&5':{\times, -, \times, -, \times, -, -, \times, \times, -} &
\end{eqnarray*}

\begin{table}[H]
\centering
 \begin{tabular}{||c|c||}
  \hline \hline
Type of intersection & $\beta^i$ \\
  \hline \hline
$[5^5],\{0, 2, 4, 1, 1\}$ & $\frac{1}{25-12 u}\{-5,-2,1,4,1,1,4,1,10,10\}$ \\
\hline
$[2^1, 5^4],\{0, 4, 2, 2, 0\}$ & $\frac{3}{8-4 u}\{0,0,-1,-1,1,1,1,1,3,3\}$ \\
\hline
$[2^2, 5^3],\{1, 3, 4, 0, 0\}$ & 
$\frac{1}{23-12 u}\{-1,2,2,-1,-1,-1,2,5,8,8\}$ \\ 
\hline 
$[2^3, 5^2],\{4, 3, 2, 0, 0\}$ & 
$\frac{1}{22-12 u}\{-2,-4,-2,4,1,1,1,4,4,7\}$ \\ 
\hline \hline
 \end{tabular}
\caption{{\em Table for various type of intersection vs $\beta^i$, 
for $N = 5$.}}
\label{Tab:N=5}
\end{table}
The explicit configuration mentioned in above table, {\bf \ref{Tab:N=5}} given
below:
\begin{eqnarray*}
&[5^5], \{0,2,4,1,1\}& \\
&5:{\times, \times, \times, \times, \times, -, -, -, -, -} & \\
&5':{\times, \times, \times, -, -, \times, \times, -, -, -} & \\
&5'':{\times, \times, -, \times, -, \times, -, \times, -, -} &\\
&5''':{\times, \times, -, -, \times, -, \times, \times, -, -} & \\
&5'''':{\times, -, \times, -, \times, \times, -, \times, -, -} &
\end{eqnarray*}
\begin{eqnarray*}
&[2^1,5^4], \{0,4,2,2,0\}& \\
&2:{\times, \times, -, -, -, -, -, -, -, -} &\\
&5:{\times, -, \times, \times, \times, \times, -, -, -, -} & \\
&5':{-, \times, \times, \times, \times, -, \times, -, -, -} & \\
&5'':{\times, -, \times, \times, -, -, \times, \times, -, -} & \\
&5''':{-, \times, \times, \times, -, \times, -, \times, -, -} &
\end{eqnarray*}
\begin{eqnarray*}
&[2^2,5^3], \{1,3,4,0,0\}& \\
&2:{\times, \times, -, -, -, -, -, -, -, -} &\\
&2':{-, -, \times, \times, -, -, -, -, -, -} & \\
&5:{\times, -, \times, -, \times, \times, \times, -, -, -}& \\
&5':{-, \times, -, \times, \times, \times, \times, -, -, -}& \\
&5'':{\times, -, -, \times, \times, \times, -, \times, -, -} &
\end{eqnarray*}
\begin{eqnarray*}
&[2^3,5^2], \{4,3,2,0,0\}& \\
&2:{\times, \times, -, -, -, -, -, -, -, -} &\\
&2':{-, -, \times, \times, -, -, -, -, -, -} &\\
&2'': {-, -, -, -, \times, \times, -, -, -, -}&\\
&5:{\times, -, \times, -, \times, -, \times, \times, -, -} &\\
&5':{\times, -, \times, -, -, \times, \times, -, \times, -} &\\
\end{eqnarray*}

\begin{table}[H]
\centering
 \begin{tabular}{||c|c||}
  \hline \hline
Type of intersection & $\beta^i$ \\
  \hline \hline
$[5^6],\{0, 3, 4, 0, 0, 2\}$ & $\frac{1}{10-5 u}\{-2,-2,1,1,1,1,2,2,2,4\}$ \\
\hline
$[2^1, 5^5],\{1, 2, 4, 1, 0, 1\}$ & 
$\frac{1}{29-15 u}\{-7,-1,2,5,2,2,5,2,8,11\}$ \\
\hline
$[2^2, 5^4],\{0,2,4,2,0,0\}$ & 
$\frac{1}{28-15 u}\{1,1,1,1,-2,-2,4,4,10,10\}$ \\
\hline
$[2^3, 5^3],\{2, 3, 3, 1, 0, 0\}$ & $\frac{1}{9-5 u}\{-1,2,1,0,0,1,0,1,2,3\}$ \\
\hline \hline
 \end{tabular}
\caption{{\em Table for various type of intersection vs $\beta^i$, 
for $N = 6$.}}
\label{Tab:N=6}
\end{table}
The explicit configuration mention in above table, {\bf \ref{Tab:N=6}} given
below:
\begin{eqnarray*}
&[5^6], \{0, 0, 4, 3, 0, 1\}& \\
&5:{\times, \times, \times, \times, \times, -, -, -, -, -} & \\
&5':{\times, \times, \times, -, -, \times, \times, -, -, -} & \\
&5'':{\times, \times, -, \times, -, \times, -, \times, -, -} & \\
&5''':{\times, \times, -, -, \times, -, \times, \times, -, -} & \\
&5'''':{\times, -, \times, -, \times, \times, -, \times, -, -} & \\
&5''''':{\times, -, -, \times, \times, \times, \times, -, -, -} &
\end{eqnarray*}
\begin{eqnarray*}
&[2^1,5^5], \{1,2,4,1,0,1\}& \\
&2:{\times, -, -, -, -, -, -, -, \times, -} & \\ 
&5:{\times, \times, \times, \times, \times, -, -, -, -, -} & \\
&5':{\times, \times, \times, -, -, \times, \times, -, -, -} & \\ 
&5'':{\times, \times, -, \times, -, \times, -, \times, -, -} & \\ 
&5''':{\times, \times, -, -, \times, -, \times, \times, -, -} & \\ 
&5'''':{\times, -, \times, -, \times, \times, -, \times, -, -} & 
\end{eqnarray*}
\begin{eqnarray*}
&[2^2,5^4], \{0, 2, 4, 2, 0, 0\}& \\
&2:{\times, \times, -, -, -, -, -, -, -, -} & \\ 
&2':-, -, \times, \times, -, -, -, -, -, - & \\
&5:\times, -, \times, -, \times, \times, \times, -, -, - & \\ 
&5':-, \times, -, \times, \times, \times, \times, -, -, - & \\ 
&5'':\times, -, -, \times, \times, \times, -, \times, -, - & \\ 
&5''':-, \times, \times, -, \times, \times, -, \times, -, - & 
\end{eqnarray*}
\begin{eqnarray*}
&[2^3,5^3], \{2,3,3,1,0,0\}& \\
&2:{\times, \times, -, -, -, -, -, -, -, -} & \\
&2':{-, -, \times, \times, -, -, -, -, -, -} & \\
&2'':{-, -, -, -, \times, \times, -, -, -, -} & \\
&5:{\times, -, \times, -, \times, -, \times, \times, -, -} & \\
&5':{\times, -, -, \times, -, \times, \times, \times, -, -} & \\
&5'':{\times, -, -, \times, \times, -, \times, -, \times, -} & 
\end{eqnarray*}

\begin{table}[H]
\centering
 \begin{tabular}{||c|c||}
  \hline \hline
Type of intersection & $\beta^i$ \\
  \hline \hline
$[5^7],\{0, 0, 0, 7, 0, 0, 1\}$ & 
$\frac{1}{35-18 u}\{-7,2,2,2,2,2,2,2,14,14\}$ \\
\hline
$[2^3, 5^4],\{0, 0, 6, 2, 0, 0, 0\}$ &
$\frac{1}{32-18 u}\{2,2,2,2,-1,-1,2,2,2,11,11\}$ \\
\hline \hline
 \end{tabular}
\caption{{\em Table for various type of intersection vs $\beta^i$, 
for $N = 7$.}}
\label{Tab:N=7}
\end{table}
The explicit configuration mentioned in above table, {\bf \ref{Tab:N=7}} given
below:
\begin{eqnarray*}
&[5^7],\{0, 0, 0, 7, 0, 0, 1\}& \\
&5:{\times, \times, \times, \times, \times, -, -, -, -, -} & \\
&5':{\times, -, -, \times, \times, \times, \times, -, -, -} & \\
&5'':{\times, \times, \times, -, -, \times, \times, -, -, -} & \\
&5''':{\times, -, \times, \times, -, \times, -, \times, -, -} & \\
&5'''':{\times, \times, -, -, \times, \times, -, \times, -, -} & \\
&5''''':{\times, \times, -, \times, -, -, \times, \times, -, -} & \\
&5'''''':{\times, -, \times, -, \times, -, \times, \times, -, -} &
\end{eqnarray*}
\begin{eqnarray*}
&[2^3,5^4], \{0,0,6,2,0,0,0\}& \\
&2:{\times, \times, -, -, -, -, -, -, -, -} & \\
&2':{-, -, \times, \times, -, -, -, -, -, -} & \\
&2'':{-, -, -, -, -, -, \times, \times, -, -} & \\
&5:{\times, -, \times, -, \times, \times, \times, -, -, -} & \\
&5':{-, \times, -, \times, \times, \times, \times, -, -, -}& \\
&5'':{\times, -, -, \times, \times, \times, -, \times, -, -} & \\
&5''':{-, \times, \times, -, \times, \times, -, \times, -, -} &
\end{eqnarray*}

\section{Time Varying Newton's Constant} \label{sec:G(t)}
The evolution of the eleven dimensional early universe which is
dominated by the $22'55' \;$ configuration described here can
also be considered from the perspective of four dimensional
spacetime. Indeed, in general, let the eleven dimensional line
element $d s$ be given by
\begin{equation}\label{ds114}
d s^2 = g_{\mu \nu} \; d x^\mu d x^\nu 
+ \sum_{i = 1}^7 e^{2 \lambda^i} (d x^i)^2
\end{equation}
where $x^\mu = (x^0, x^8, x^9, x^{10})$, with $x^0 = t$, 
describes the four dimensional spacetime, and the fields $g_{\mu
\nu} \;$ and $\lambda^i \;$, $i = 1, \cdots, 7, \;$ depend on
$x^\mu \;$ only. Also, let $\Lambda^c = \sum_{i = 1}^7 \lambda^i
\;$. It is then straightforward to show that the gravitational
part of the eleven dimensional action $S_{11} \;$ given in
equation (\ref{s11}) becomes
\begin{equation}\label{s4} 
S_4 = \frac{V_7}{16 \pi G_{11}} \; \int d^4 x \; 
\sqrt{-g_{(4)}} \; e^{\Lambda^c} \; \{ \; R_{(4)} 
+ (\nabla_{(4)} \Lambda^c)^2 
- \sum_{i = 1}^7 (\nabla_{(4)} \lambda^i)^2 \; \}
\end{equation}
where $V_7 \;$ is the coordinate volume of the seven dimensional
space and the subscripts $(4)$ indicate that the corresponding
quantities are with respect to the four dimensional metric
$g_{\mu \nu} \;$. The action $S_4 \;$ describes four dimensional
spacetime in which the effective Newton's constant $G_4 \;$ is
spacetime dependent and is given by
\begin{equation}\label{tGN4}
G_4 (x^\mu) = e^{- \Lambda^c (x^\mu)} \; \; 
\frac{G_{11}}{V_7} \; \; .
\end{equation}
In the case of early universe, the fields $g_{\mu \nu} \;$ and
$\lambda^i \;$ depend on $t$ only.  Then $G_4 \;$ is time
dependent and we have, for $G_4 \;$ and its fractional time
derivative,
\begin{equation}\label{GN4}
G_4 (t) = e^{- \Lambda^c (t)} \; \; \frac{G_{11}}{V_7}
\; \; \; , \; \; \; \; 
\frac{(G_4)_t}{G_4} = - \; \Lambda^c_t \; \; . 
\end{equation}

For the four dimensional spacetime arising from the $22'55' \;$
configuration, the $g_{\mu \nu} \;$ fields are just the scale
factors $(e^{\lambda^8}, e^{\lambda^9}, e^{\lambda^{10}}) \;$
for $(x^8, x^9, x^{10}) \;$ directions, and all $\lambda^i
(\tau) \;$ are given in equation (\ref{d5}) in terms of $l^I
(\tau) \;$, $I = 1, \cdots, 4 \;$. Then, using equation
(\ref{I3}) and the definitions of $\Lambda^c \;$, $L^c \;$, and
$v^c \;$, we have
\begin{equation}\label{Lambdac}
\Lambda^c =  - \; \frac{1}{6 u} (l^1 + l^2 - l^3 - l^4)
- L^c (\tau_\infty - \tau) + v^c \; \; . 
\end{equation}
In the limit $t \to \infty \;$, we have from the results given
earlier that $\tau \to \tau_\infty \;$ and the fields $l^I \;$
all become equal. Then $\Lambda^c \to v^c \;$ where $e^{v^c} \;$
is given in equation (\ref{vc}), and the three dimensional scale
factors evolve as in the standard FRW case, namely
$e^{\lambda^8} = e^{\lambda^9} = e^{\lambda^{10}} \sim
t^{\frac{2}{3 (2 - u)}} \;$ as given in equations
(\ref{lambdait}) and (\ref{betai}).

It thus follows that the effective Newton's constant $G_4 \;$
varies with time in the early universe and, in the case of
$22'55' \;$ configuration, approaches a constant value $= e^{-
v^c} \; \frac{G_{11}}{V_7} \;$ as the four dimensional universe
expands to large size. The precise time dependence of $G_4 \;$
will follow from explicit solutions to equations (\ref{d3}) and
(\ref{d6}). The consequences of a such a time dependent $G_4 \;$
are clearly interesting, and are likely to be important too. But
their study is beyond the scope of the present thesis.

However, we like to point out here a characteristic feature of
the time dependence of $G_4 \;$ which arises in the case of
$22'55' \;$ configuration. Consider the behaviour of the
differences $l^I - l^J \;$ in the limit $\tau \to \tau_\infty
\;$ which, in our case, vanish to the leading order. These
quantities have been analysed in section {\bf \ref{sec:deviations}} and, for
the example of the ${\cal G}^{I J} \;$ given in equation
(\ref{ab}), they are given by equations (\ref{sisj}) and
(\ref{sisjsoln}) to the non trivial leading order. The case of
$22'55' \;$ configuration corresponds to $N = 4 \;$, $a = 2 u^2
\;$, and $b = 1 \;$. Noting that $s^I - s^J = l^I - l^J \;$ and
that $\Delta < 0 \;$ in our case, equation (\ref{sisjsoln}) now
gives
\begin{equation}\label{liljsoln}
(l^I - l^J) \; \sim \; (\tau_\infty - \tau)^{ 
\frac{1}{2} (1 \; \pm \; i \; \sqrt{\frac{5}{3}})}
\end{equation}
to the leading order. Clearly, $\Lambda^c (\tau) \;$ given in
equation (\ref{Lambdac}) will also have the same form as above
to the non trivial leading order. Thus, taking the real part and
writing in terms of $t$ using equation (\ref{ttau}), we have
\begin{equation}\label{lilj}
\Lambda^c = v^c + \frac{b}{t^\alpha} \; 
\; Sin (\omega \; ln \; t + \phi ) 
\end{equation}
to the non trivial leading order in the limit $t \to \infty \;$
where $b$ and $\phi$ are constants, $ \alpha = \frac{u}{2 (2 -
u)} \;$, and $\omega = \sqrt{ \frac{5}{3}} \; \frac{u}{2 (2 -
u)} \;$. Correspondingly, the time varying Newton's constant is
given by
\begin{equation}\label{G4}
G_4 \; \propto \; e^{- \Lambda^c} \; = \; e^{- v^c} \; 
(1 -  \frac{b}{t^\alpha} \; 
\; Sin (\omega \; ln \; t + \phi ) \; )
\end{equation}
to the leading order in the limit $t \to \infty \;$. Note that
the constants $b$ and $\phi$ depend on the details of
matching. The constants $\alpha$ and $\omega \;$ arise as real
and imaginary parts of an exponent on time variable, see
equation (\ref{liljsoln}). They do not depend on the initial
values $(l^I_0, K^I, L^i) \;$ and thus are independent of the
details of evolution, but depend only on the configuration
parameters $N \;$ and $u \;$.

The amplitude of time variation of $G_4 \;$ is dictated by
$\alpha \;$, and it vanishes in the limit $t \to \infty \;$.
Hence, the time variation of $G_4 \;$ in equation (\ref{G4}) is
unlikely to contradict any late time observations. The time
variation of $G_4 \;$ has log periodic oscillations also: $G_4
\;$ has an oscillatory behaviour where the $n^{th}$ and $(n +
1)^{th} \;$ nodes occur at times $t_n$ and $t_{n + 1}$ which are
related by $ln \; t_{n + 1} = \frac {\pi} {\omega} + ln \; t_n
\;$, {\em i.e.} by $t_{n + 1} = e^{\frac {\pi} {\omega}} \; t_n
\;$. The characteristic signatures and observational
consequences of such log periodic variations of $G_4 \;$ are not
clear to us.

Log periodic behaviour occurs in many physical systems with
`discrete self similarity' or `discrete scale symmetry': for
example, in quantum mechanical systems with strongly attractive
$\frac{1}{r^2} \;$ potentials near zero energy \cite{braaten};
in Choptuik scaling and brane -- black hole merger transitions
\cite{kol}; and in a variety of dynamical systems
\cite{sornette}. Algebraically, the log periodicity arises when
an exponent on an independent variable becomes complex for
certain values of system parameters. The relevant equations and
solutions can often be cast in a form given in equations
(\ref{sisj}) and (\ref{sisjsoln}). But we are not aware of a
physical reason which explains the ubiquity of the log
periodicity.

To our knowledge, this is the first time a log periodic
behaviour appears in a cosmological context. One expects such a
behaviour to leave some novel imprint in the universe. But it is
not clear to us which effects to look for, or which observables
are sensitive to the log periodic variations of $G_4 \;$.

\section{Numerical Results} \label{sec:Num}
We are unable to solve explicitly the equations (\ref{d3}) --
(\ref{d6}) describing the early universe evolution. Hence, we
have analysed these equations numerically. In this section, we
briefly describe our procedure and present a few illustrative
results. We have analysed both the $u = \frac{2}{3} \;$ and $u =
1 \;$ cases which would correspond to four dimensional universe
dominated by radiation and pressureless dust respectively. The
results are qualitatively the same and, hence, we take $u =
\frac{2}{3} \;$ in the following. Note that $\omega \;$ in
equation (\ref{G4}) is then determined and, for $u = \frac{2}{3}
\;$, the $n^{th}$ and $(n + 1)^{th} \;$ nodes in the log
periodic oscillations occur at times $t_n$ and $t_{n + 1}$
related by $ln (\frac{t_{n + 1}}{t_n}) = 4 \pi \sqrt{ \frac {3}
{5} } \simeq 9.734 \;$.

We proceed as follows. We start at an initial time $\tau = 0 \;$
and choose a set of initial values $l^I_0 = ln \; \rho_{I 0}
\;$. For each set of $l^I_0\;$, we further choose numerous
arbitrary sets of $(K^I, L^i) \;$ such that $K^I > 0 \;$, $E = 1
\;$, and equations (\ref{d6ic}) and (\ref{LI4}) are satisfied.
\footnote{ There are two special choices for the set of $K^I
\;$. One is where $K^1 = \cdots = K^4 \;$ and another is the one
which maximises the approximation $\tau_a \;$ given in equation
(\ref{tauaI}). The later set may be determined by the algorithm
given in Appendix {\bf \ref{ap:KI}}.}  For each set of initial values
$(l^I_0, K^I, L^i) \;$, we then numerically analyse the
evolution for $\tau > 0 \;$ and obtain the value of $\tau_\infty
\;$; the evolution of $l^I$, $(\lambda^1, \cdots, \lambda^{10})
$, and $t \;$; the stabilised values $(v^1, \cdots, v^7) \;$;
and the resulting values for $(g_s, \frac {M_{11}} {M_4}, \frac
{M_s} {M_4}) \;$. For a few sets of initial values, we have
analysed the evolution for $\tau < 0 \;$ also.

We find that the numerical results we have obtained confirm the
asymptotic features described in this thesis:

{\bf (1)} $e^{\lambda^i} \;$ and $l^I \;$ all vanish in the
limit $\tau \to - \infty \;$. In this limit, the evolution of
the scale factors $e^{\lambda^i} \;$ is of Kasner type.

{\bf (2)} $l^I \;$ and the physical time $t \;$ all diverge in
the limit $\tau \to \tau_\infty \;$ where $\tau_\infty \;$ is
finite. In this limit, the scale factors $(e^{\lambda^8},
e^{\lambda^9}, e^{\lambda^{10}}) \;$ evolve as in the standard
FRW case and $(e^{\lambda^1}, \cdots, e^{\lambda^7}) \;$ reach
constant values.

{\bf (3)} $\tau_a \;$ given in equation (\ref{tauaI}) provides a
good approximation to $\tau_\infty \;$ when $e^{l^I_0} \ll 1 \;$
for all $I \;$.

{\bf (4)} Any values for the ratios $\frac{M_{11}}{M_4} \;$ and
$\frac{M_s}{M_4} \;$ can be obtained, but a corresponding fine
tuning of $\rho_{I 0} = e^{l^I_0} \;$ is necessary.

{\bf (5)} The log periodic oscillations of $l^I - l^J \;$,
equivalently of $(\lambda^1, \cdots, \lambda^7) \;$, can also be
seen in the limit $\tau \to \tau_\infty \;$. They can be matched
to solutions of the type given in equation (\ref{liljsoln}).

To illustrate the values of $\tau_\infty \;$ and the ratios
$(\frac {M_{11}} {M_4}, \; \frac{M_s}{M_4}) \;$ one obtains, and
to give an idea of their dependence on the initial values $l^I_0
\;$, we tabulate these quantities in Table {\bf \ref{I}} for a few
sets of initial values $(l^I_0, K^I, L^i) \;$. We have also
tabulated the values of $\tau_a \;$ as given by equation
(\ref{tauaI}). The value of $g_s \;$ follows from $g_s = (\frac
{M_s} {M_{11}})^3 \;$ and, hence, is not tabulated.


\begin{table}[H]
\centering
\begin{tabular}{||c||c||c|c||c|c||} 
\hline \hline 
& & & & & \\ 

& $\; - \; (l^1_0, l^2_0, l^3_0, l^4_0)$
& $\tau_a \;$
& $\tau_\infty \;$
& $\frac{M_{11}}{M_4} \;$
& $\frac{M_s}{M_4} \;$
\\ 
& & & & & \\ 
\hline  \hline 

& & & & & \\ 
(1) 
& $(2, 5, 8, 8) \;$
& $ 1.88 \;$
& $ 3.21 \;$
& $ 5.77 * 10^{- 2} \;$
& $ 2.86 * 10^{- 2} \;$
\\
& & & & & \\ 
\hline 

& & & & & \\ 
(2) 
& $(5, 4, 6, 9) \;$
& $ 2.96 \;$
& $ 4.16 \;$
& $ 4.56 * 10^{- 2} \;$
& $ 1.93 * 10^{- 2} \;$
\\
& & & & & \\ 
\hline 

& & & & & \\ 
(3) 
& $(15, 12, 10, 16) \;$
& $4.88 \;$
& $6.61 \;$
& $ 5.95 * 10^{- 2} \;$
& $ 1.96 * 10^{- 2} \;$
\\
& & & & & \\ 
\hline  

& & & & & \\ 
(4) 
& $(25, 26, 27, 28) \;$
& $ 22.00 \;$
& $ 22.59 \;$
& $ 1.99 * 10^{- 7} \;$
& $ 7.30 * 10^{- 10} \;$
\\
& & & & & \\ 
\hline 

& & & & & \\ 
(5) 
& $(41, 30, 50, 43) \;$
& $ 25.80 \;$
& $ 28.30 \;$
& $ 1.87 * 10^{- 10} \;$
& $ 2.92 * 10^{- 11} \;$
\\
& & & & & \\ 
\hline 

& & & & & \\ 
(6) 
& $(44.5, 34, 49, 49.5) \;$
& $ 34.82 \;$
& $ 36.20 \;$
& $ 2.59 * 10^{- 14} \;$
& $ 3.80 * 10^{- 15} \;$
\\
& & & & & \\ 
\hline 

\hline \hline
\end{tabular}
\caption{ {\em 
The initial values $\; - \; (l^1_0, l^2_0, l^3_0, l^4_0)$ and
the resulting values of $\tau_a \;$, $\tau_\infty \;$, $\frac
{M_{11}} {M_4} \;$, and $\frac{M_s}{M_4} \;$. The values in the
last four columns have been rounded off to two decimal places.
}}
\label{I}
\end{table}



In Table {\bf \ref{II}}, the corresponding initial values $(K^I,
L^i)$, $i = 1, 2, 6, 7, 8, 9$, are tabulated up to overall
positive constants. The remaining $L^i$s are given by equations
(\ref{LI4}) and the overall positive constants are determined by
$E = 1 \;$ and equation (\ref{d6ic}). All the sets of initial
values $(l^I_0, K^I, L^i) \;$ are chosen arbitrarily with no
particular pattern and are presented here to give an idea of the
typical results.

\begin{table}[H] 
\centering
\begin{tabular}{||c||c||c||} 
\hline \hline 
& & \\ 

& $\; (K^1, K^2, K^3, K^4) \; \propto \; $
& $\; (L^1, L^2, L^6, L^7, L^8, L^9) \; \propto \; $
\\ 
& & \\ 
\hline  \hline 


& & \\ 
(1) 
& $(4.65, \; 9.14, \; 4.57, \; 6.87) \;$
& $(0.60, \; 0.62, \; 0.76, \; 0.72, - 0.94, - 0.26) \;$
\\
& & \\ 
\hline 

& & \\ 
(2) 
& $(8.86, \; 8.26, \; 6.01, \; 6.62) \;$
& $- (0.08, -0.93, \; 0.08, - 0.72, \; 0.54, \; 0.63) \;$
\\
& & \\ 
\hline 

& & \\ 
(3) 
& $(1.61, \; 2.65, \; 0.69, \; 2.1) \;$
& $- (0.2, - 0.68, - 0.14, \; 0.3, \; 0.08, \; 0.19) \;$
\\
& & \\ 
\hline 

& & \\ 
(4) 
& $(1.03, \; 1.18, \; 1.17, \; 1.27) \;$
& $(0.08, \; 0.58, \; 0.27, \; 0.27, - 0.66, - 0.66) \;$
\\
& & \\ 
\hline 

& & \\ 
(5) 
& $(5.24, \; 4.83, \; 4.30, \; 4.96) \;$
& $(0.74, \; 0.02, \; 0.24, - 0.22, - 0.61, - 0.75) \;$
\\
& & \\ 
\hline 

& & \\ 
(6) 
& $(33.79, \; 24.23, \; 35.4, \; 32.29) \;$
& $(11.72, \; 9.31, \; 4.59, - 6.46, - 21.02, - 21.02) \;$
\\
& & \\ 
\hline 

\hline \hline
\end{tabular}
\caption{{\em 
The initial values of $(K^I, L^i) \;$ for the data shown in
Table {\bf \ref{I}}, tabulated here up to overall positive constants.
These constants and the remaining $L^i$s are to be fixed as
explained in the text.
}}
\label{II}
\end{table}
%


We find, by analysing numerous sets of initial values, that
changing the values of $(K^I, L^i) \;$ for a given set of $l^I_0
\;$ changes the values of $\frac {M_{11}} {M_4} \;$ and $\frac
{M_s} {M_4} \;$ only up to about four orders of magnitude. Any
bigger change requires changing $e^{l^I_0} \;$ to a similar
order, confirming that any values for $\frac {M_{11}} {M_4} \;$
and $\frac {M_s} {M_4} \;$ can be obtained but only by fine
tuning $\rho_{I 0} = e^{l^I_0} \;$.

\begin{figure}[t]
 \centering

\includegraphics[width=0.8\textwidth]{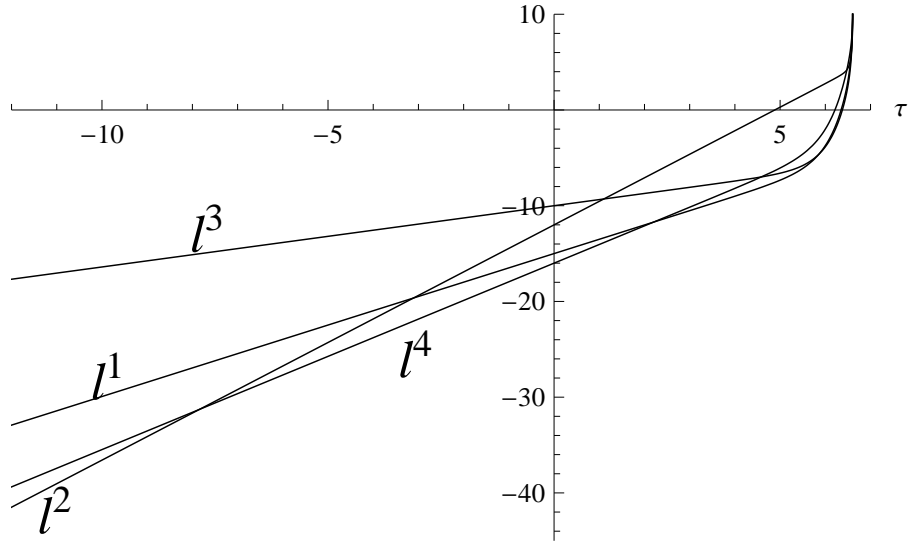}
 
 \caption{ 
{\em 
The plots of $l^I \;$ with respect to $\tau \;$. The lines
continue with no further crossings for negative values of $\tau
\;$ not shown in the figure. All $l^I \;$ diverge at
$\tau_\infty \simeq 6.612 \;$. All figures in this thesis are for
the data set (3) given in Tables {\bf \ref{I}} and {\bf \ref{II}}.
}
}
\label{fig:ltau1}
\end{figure}

\begin{figure}[h]
 \centering

 \subfloat[]{
 \includegraphics[width=0.5\textwidth]{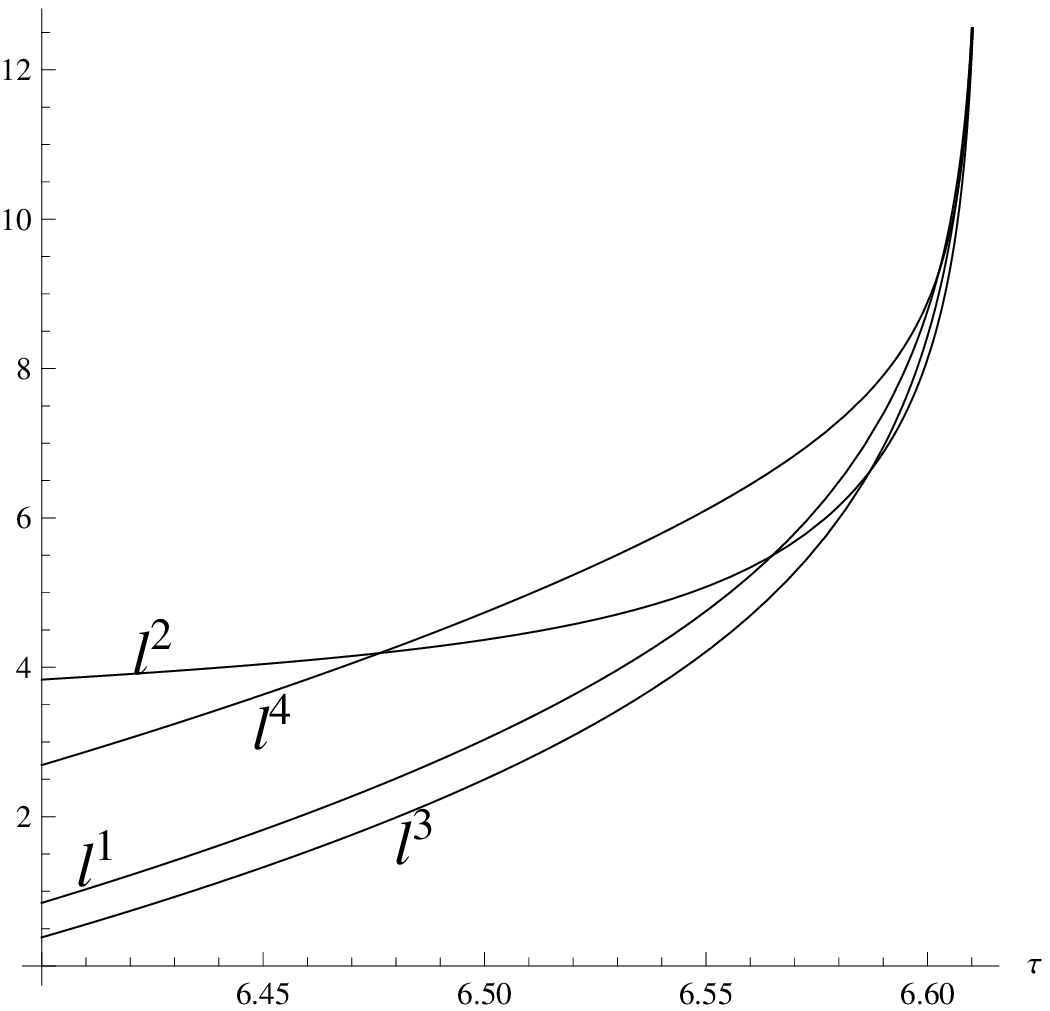}}
 ~~~~ ~~ 
 \subfloat[]{
 \includegraphics[width=0.5\textwidth]{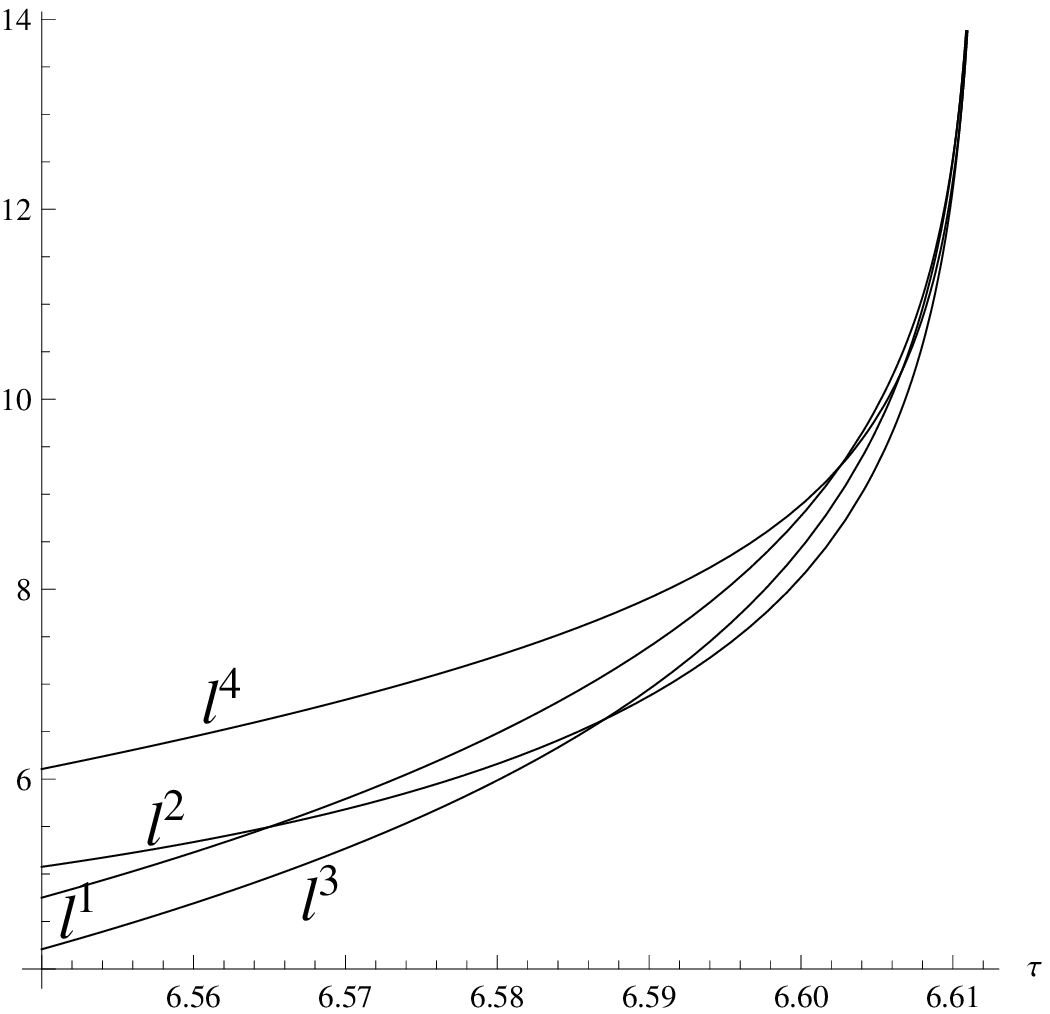}}
 \caption{ 
{\em 
{\bf (a), (b)} The magnified plots of $l^I \;$ with respect to
$\tau \;$ for $\; \tau > 6.40 \;$ and for $\; \tau > 6.55 \;$
showing the continually criss-crossing evolution of $\; l^I
\;$. Near $\tau_\infty \simeq 6.612 \;$, these crossings are
well described by equation (\ref{liljsoln}).
}
}
\label{fig:ltau2}
\end{figure}

We illustrate the evolution of the universe for the data set (3)
given in Tables {\bf \ref{I}} and {\bf \ref{II}} where many features can be
seen clearly. The evolution with respect to $\tau \;$ of $l^I
\;$ is shown in Figures {\bf \ref{fig:ltau1}}, {\bf \ref{fig:ltau2} a}, and 
{\bf \ref{fig:ltau2} b}. For
negative values of $\tau \;$ not shown in Figure {\bf \ref{fig:ltau1}}, all
$l^I \;$ evolve along straight lines with no further crossings
and their evolution is of Kasner type. Also, all $l^I \;$
diverge at a finite value $\tau_\infty \simeq 6.612 \;$ of $\tau
\;$. The magnified plots in Figures {\bf \ref{fig:ltau2} a} and 
{\bf \ref{fig:ltau2} b} 
for $\tau > 6.40 \;$ and for $\tau > 6.55 \;$ respectively show the
continually criss-crossing evolution of $l^I \;$ which, near
$\tau_\infty \;$, represent the log periodic oscillations and
are well described by equation (\ref{liljsoln}).


\begin{figure}[H]
 \centering

\includegraphics[width=0.8\textwidth]{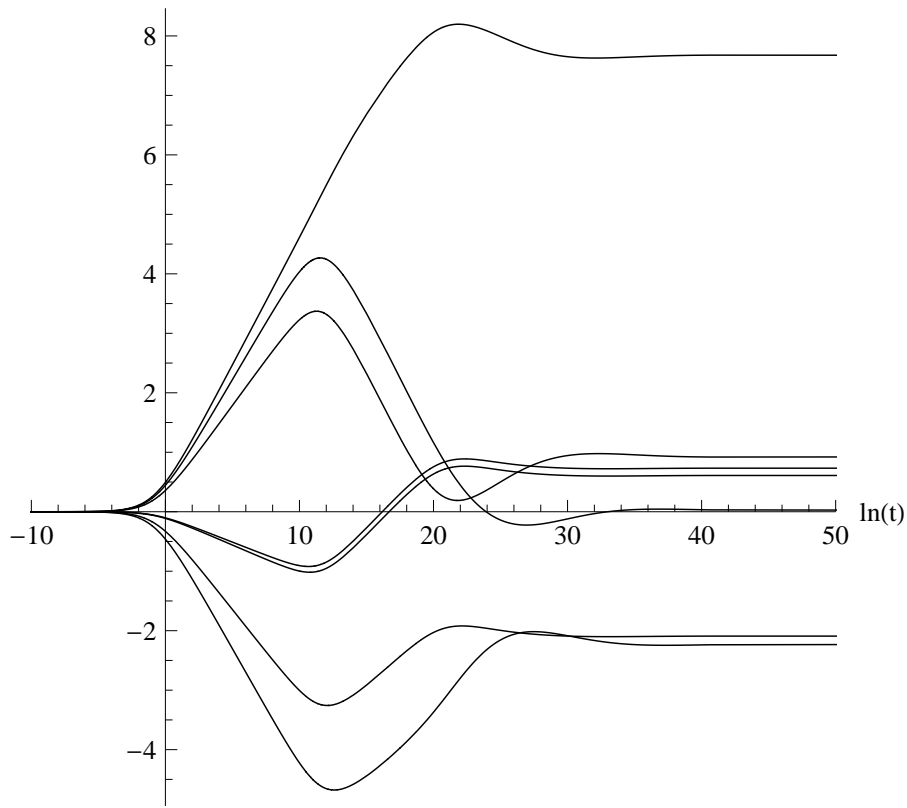}

 \caption{
{\em 
The plots of $(\lambda^1, \cdots, \lambda^7) \;$ with respect to
$ln \; t \;$. The lines, from top to bottom at the right most
end, correspond to $(\lambda^2, \lambda^3, \lambda^5, \lambda^6,
\lambda^4, \lambda^7, \lambda^1) \;$. $ \; (\lambda^1, \cdots,
\lambda^7) \;$ all stabilise to constant values as $t \to \infty
\;$.
}
}
\label{fig:labdac}
\end{figure}

The evolution with respect to $ln \; t \;$ of $(\lambda^1,
\cdots, \lambda^7) \;$ is shown in Figure {\bf \ref{fig:labdac}}. It can be
seen that $(\lambda^1, \cdots, \lambda^7) \;$, and hence the
scale factors $(e^{\lambda^1}, \cdots, e^{\lambda^7})$ of the
brane directions, all stabilise to constant values as $t \to
\infty \;$.


\begin{figure}[H]
 \centering

\includegraphics[width=0.8\textwidth]{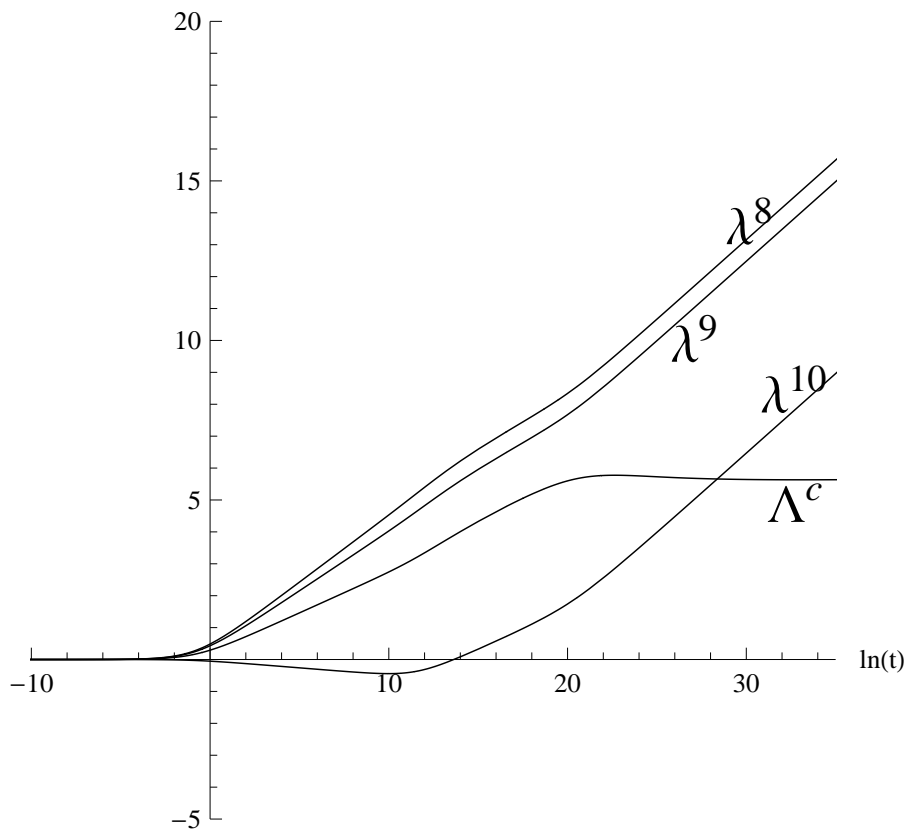}

 \caption{
{\em 
The plots of $(\lambda^8, \lambda^9, \lambda^{10}, \Lambda^c)
\;$ with respect to $ln \; t \;$. The seven dimensional volume
of the brane directions $\propto e^{\Lambda^c} \;$. The
evolution of $(\lambda^8, \lambda^9, \lambda^{10}) \;$ as $t \to
\infty \;$ is same as that of the corresponding ones in a four
dimensional radiation dominated FRW universe.
}
}
\label{fig:lambdanc}
\end{figure}

The evolution with respect to $ln \; t \;$ of $(\lambda^8,
\lambda^9, \lambda^{10})$ and $\Lambda^c = \sum_{i = 1}^7
\lambda^i \;$ is shown in Figure {\bf \ref{fig:lambdanc}}. Note that the seven
dimensional volume of the brane directions $\propto
e^{\Lambda^c} \;$ and that it stabilises to a constant value
$e^{v^c} \;$ as $t \to \infty \;$. We have also verified that
the evolution of $(\lambda^8, \lambda^9, \lambda^{10}) \;$ as $t
\to \infty \;$ is same as that of the corresponding ones in a
four dimensional radiation dominated FRW universe.


\begin{figure}[H]
 \centering

 \subfloat[]{
 \includegraphics[width=0.5\textwidth]{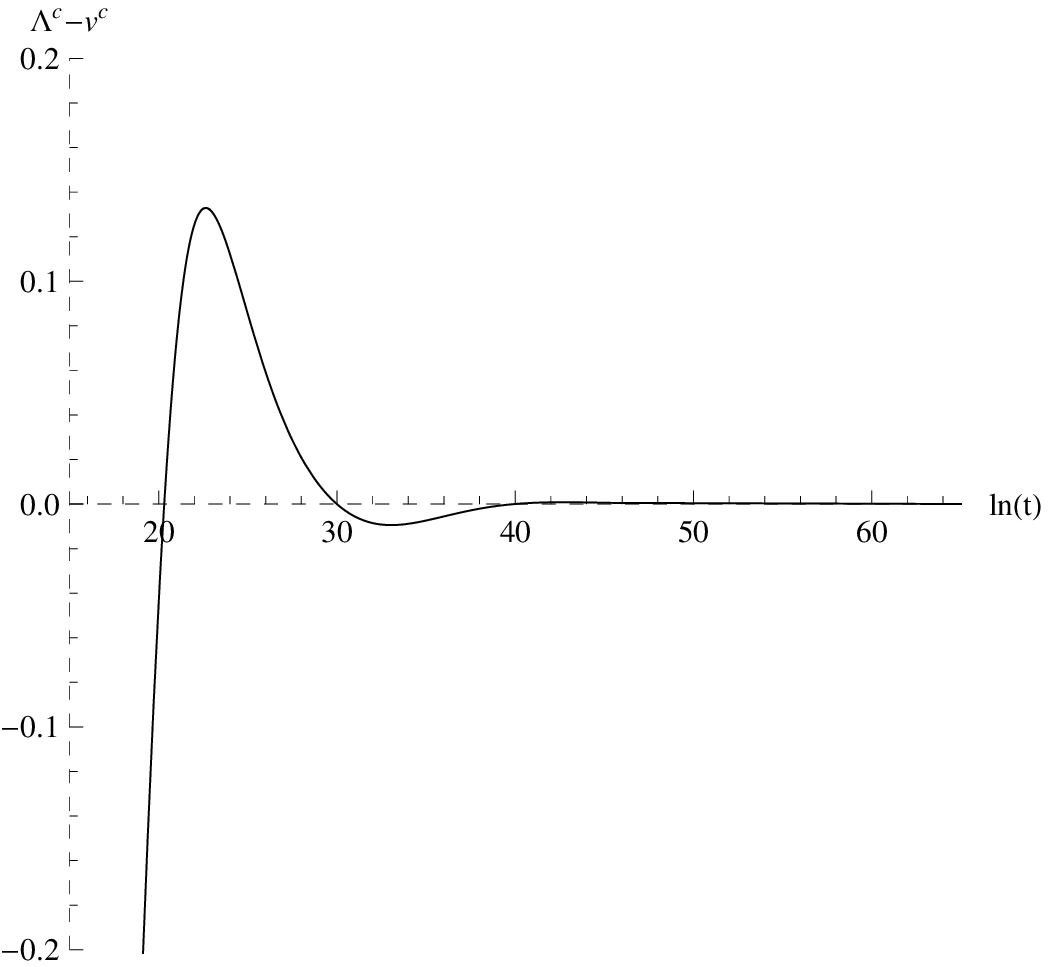}} ~~~~ ~~ 
 \subfloat[]{
 \includegraphics[width=0.5\textwidth]{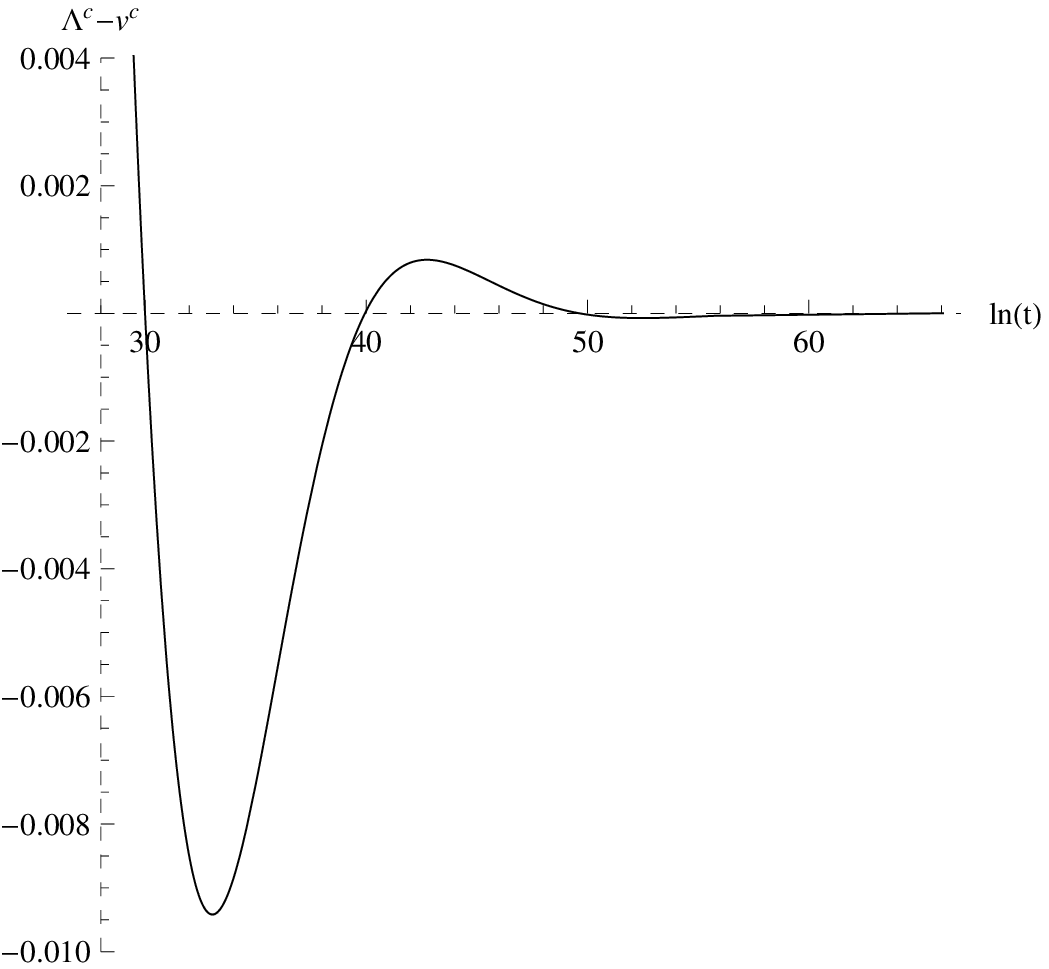}}

 \caption{
{\em
{\bf (a), (b)}
The magnified plots of $(\Lambda^c - v^c) \;$ with respect to
$ln \; t \;$ for $ln \; t > 20 \;$ and for $ln \; t > 30 \;$
showing log periodic oscillations and the internode separation
which is $\simeq 9.734 $.
}
}
\label{fig:osc}
\end{figure}

The log periodic oscillations of $\Lambda^c \;$ are illustrated
in Figures {\bf \ref{fig:osc} a} and 
{\bf \ref{fig:osc} b} by magnifying the plots of
$(\Lambda^c - v^c) \;$ with respect to $ln \; t \;$ for $ln \; t
> 20 \;$ and for $ln \; t > 30 \;$. The internode separations
can be seen in these figures, and they match the value $\simeq
9.734 \;$ obtained in equation (\ref{lilj}) from the asymptotic
analysis.


In all the cases we have analysed, the evolutions of $(l^I,
\lambda^i) \;$ are qualitatively similar to the ones shown in
the figures above. The details, such as the rise and fall of
$\lambda^i \;$ in the initial times or the value of $\tau_\infty
\;$ or the stabilised values of the brane directions, depend on
the initial values but the asymptotic features described in the
beginning of this section are all the same. Hence, we have
presented the plots for one illustrative set of initial values
only.

\section{Summary and Conclusions} \label{sec:concl}

We summarise the main results of this thesis. We assume that the
early universe in M theory is homogeneous, anisotropic, and is
dominated by $N = 4$ mutually BPS $22'55'$ intersecting brane
configurations which are assumed to be the most entropic ones.
Also, the ten dimensional space is assumed to be toroidal. We
further assumed that the brane antibrane annihilation effects
are negligible during the evolution of the universe at least
until the brane directions are stabilised resulting in an
effective $3 + 1$ dimensional universe.

We then present a thorough analysis of the evolution of such an
universe. We obtain general relations among the components of
the energy momentum tensor $T_{A B} \;$ using U duality
symmetries of M theory and show that these relations alone
imply, under a technical assumption, that the $N = 4$ mutually
BPS $22'55'$ intersecting brane configurations with identical
numbers of branes and antibranes will asymptotically lead to an
effective $(3 + 1)$ dimensional expanding universe.

To obtain further details of the evolution, we make further
assumptions about $T_{A B} \;$. We then analyse the evolution
equations in D dimensions in general, and then specialise to the
eleven dimensional case of interest here. Since explicit
solutions are not available, we apply the general analysis and
describe the qualitative features of the evolution of the $N =
4$ brane configuration : In the asymptotic limit, three spatial
directions expand as in the standard FRW universe and the
remaining seven spatial directions reach constant, stabilised
values. These values depend on the initial conditions and can be
obtained numerically. Also, any stabilised values may be
obtained but it requires a fine tuning of the initial brane
densities.

We also present a physical description of the mechanism of
stabilisation of the seven brane directions. The stabilisation
is due, in essence, to the relations among the components of
$T_{A B} \;$ which follow from U duality symmetries, and to each
of the brane directions in the $N = 4 \;$ configuration being
wrapped by, and being transverse to, just the right number and
kind of branes. This mechanism is very different from the ones
proposed in string theory or in brane gas models.

In the asymptotic limit, from the perspective of four
dimensional spacetime, we obtain an effective four dimensional
Newton's constant $G_4 \;$ which is now time varying. Its
precise time dependence will follow from explicit solutions of
the eleven dimensional evolution equations. We find that, in the
case of $N = 4$ brane configuration, $G_4 \;$ has characteristic
log periodic oscillations. The oscillation `period' depends only
on the configuration parameters.

Using numerical analysis, we have confirmed the qualitative
features mentioned above.

We now make a few comments on the assumptions made in this
thesis. Note that the assumptions mentioned above in the first
paragraph of this section pull a rug over many important
dynamical questions that must be answered in a final analysis.
Some of these questions, \footnote{Many of the questions listed
below have been raised by the referee also.} in the context of M
theory, are:

\vspace{1ex}

* Starting from the highly energetic and highly interacting M
theory excitations, which are expected to describe the high
temperature state of the universe, how does a eleven dimensional
spacetime emerge?

\vspace{1ex}

* What determines the topology of the ten dimensional space?
Here, we assumed it to be toroidal. How does the universe evolve
if its spatial topology is not toroidal?

\vspace{1ex}

* From what stage onwards, is the eleven dimensional `low
energy' effective action a good description of further
evolution?

\vspace{1ex}

* What are the relevant `low energy' configurations of M theory?
Here, based on the black hole studies, we have assumed that the
$N = 4$ mutually BPS $22'55'$ intersecting brane configurations
are the most entropic ones and, hence, that they are the
dominant configurations in the early universe studied here.

This raises further questions: Are the $22'55' \;$, and not some
other mutually BPS $N \ge 4 \;$ or some other non BPS,
configurations really the most entropic and the dominant ones?
Even assuming that mutually BPS $N = 4 \;$ is the answer, are
there other $N = 4$ configurations beside the $22'55'$ ones and,
if so, how do they affect the evolution described here? What are
the effects of the sub-dominant configurations? In particular,
will the effects of other brane configurations mentioned above
undo the stabilisation of seven directions presented here?

Note that unless these questions are answered and, furthermore,
it is shown that other brane configurations mentioned above do
not undo the stabilisation presented here, our assumption that
the evolution of the universe is dictated by the $22'55'$
configuration amounts to a fine tuning: The $22'55'$
configuration assumed here, where the sets of 2 branes and 5
branes wrap the directions $(x^1, \cdots, x^7) \;$ homogeneously
everywhere in the mutually transverse three dimensional space,
may not arise generically. Also, the implicitly required absence
of other brane configurations is not natural in the context of
early universe. Then the problem of the emergence of an
effective $3 + 1$ dimensional universe, a solution for which is
presented here, gets shifted to answering how the required,
finely tuned, initial conditions may arise naturally from M
theory.

\vspace{1ex}

* What is the time scale of brane antibrane annihilations in the
$22'55'$ configuration studied here? Is it long enough for the
brane directions to be stabilised as described in this thesis?
Here, based on the black hole studies, we have assumed it to be
long enough.

\vspace{1ex}

* A related question, but applicable after stabilisation of
brane directions, is the following: If all the branes and
antibranes will eventually decay, as seems natural, then what
are the decay products? How can one obtain the known
constituents of our present universe?

Although one of us have presented a principle in \cite{k0610}
that may be of help, the fact is that we do not know even where
to begin in answering these questions quantitatively, much less
know the answers. Nevertheless we present the above list of
questions, unlikely to be complete, in order to emphasise the
further work required to understand how our known $3 + 1$
dimensional universe may emerge from M theory.

In the present work, with many attendant assumptions, we
considered the $22'55'$ configurations and explained a mechanism
by which seven directions stabilise and an effective $3 + 1$
dimensional universe results. Clearly, it is important to answer
the questions listed above and thereby determine the relevance
of this mechanism.

Within the present framework, there are many other issues that
may be studied further. We conclude by mentioning a sample of
them. We have shown here that a large stabilised seven
dimensional volume can be obtained but it requires a
corresponding fine tuning of initial brane densities. This is
within the context of our ansatzes for $T_{A B} \;$ and the
equations of state. It will be of interest to prove or disprove
the necessity of such a fine tuning in more general contexts.

The $N = 4 \;$ intersecting brane configuration studied here is
the entropically favourable one and, as proposed in
\cite{k0610}, may be thought of as emerging from the high
temperature phase of M theory in the early universe. Such an
emergence suggests that there may be novel solutions to the
horizon problem and to the primordial density fluctuations,
perhaps similar to those explored recently in the Hagedorn phase
of string theory by Nayeri et al \cite{nayeri1} (see also \cite{brand1}). 
Note that this
involves answering many of the questions listed above.

It may be of interest to study further the consequences of time
varying Newton's constant which appears here, in particular
possible imprints of its asymptotic log periodic oscillations.

In the case of a class of black holes, the brane configurations
describe well their entropy and Hawking radiation. In the
present description of a four dimensional early universe in
terms of $N = 4 \;$ intersecting branes, it is not clear which
quantities to calculate which, analogously to entropy or Hawking
radiation in the black hole case, may provide further
validation. It is important to study this further.

\appendix

\chapter{T and S duality rules} \label{ap:TS}

\section{T duality rules for background fields}
Lets take a solution of type II supergravity in string frame
\begin{equation}
 \label{A1}
 ds^2 = g_{\mu\nu} \;dx^\mu dx^\nu
\end{equation}
where $\mu$, $\nu$ vary from 0 to 9. For simplicity assume all special 
dimensions are compact. NS sector fields $B_{\mu\nu}$ and dilaton $\phi$
are switched on. Consider a metric such that it does not depend on $x^9$. we
denote $x^9$ by $z$. Now $\mu$, $\nu$ run over $(0, \cdots, 8)$.
Under T duality background fields transform to 
a new set of fields. Rule for this transformation is known as 
Buscher's rules. See for more details \cite{duality,bOrtin,pol} for example.
According to these set of rules fields goes like following:
\begin{eqnarray}
 \label{TD1}
 g_{zz} & \to & g'_{zz} = \frac{1}{g_{zz}} \\
 \label{TD2}
 g_{\mu z} & \to & g'_{\mu z} = \frac{B_{\mu z}}{g_{zz}} \\
 \label{TD3}
 g_{\mu\nu} & \to & g'_{\mu\nu} = g_{\mu\nu} 
 - \frac{g_{\mu z} g_{\nu z} - B_{\mu z} B_{\nu z}}{g_{zz}} \\
 \label{TD4}
 B_{\mu z} & \to & B'_{\mu z} = \frac{g_{\mu z}}{g_{zz}} \\
 \label{TD5}
 B_{\mu\nu} & \to & B'_{\mu\nu} = B_{\mu\nu} 
 + \frac{2g_{[\mu |z|} B_{\nu] z}}{g_{zz}} \\
 \label{TD6}
 \phi & \to & \phi' = \phi -\frac{1}{2}\, \ln (g_{zz})
\end{eqnarray}

When R-R fields are switched on they also transform under T-duality. As T 
duality transform IIA theory to IIB theory and vice versa, it transform
odd form fields to even form and even to odd.
If we start with a type IIA theory, transformation rules are given by \cite{pol}
\begin{eqnarray}
 C_z & \to & C \\ 
 C_\mu\, , \, C_{\mu\nu z} & \to & C_{\mu z} \,,\, C_{\mu\nu} \\
 C_{\mu\nu\lambda} & \to & C_{\mu\nu\lambda z} 
\end{eqnarray}
So we see the theory goes to type IIB theory.

\section{S duality rules for background fields}
S duality is a duality of type IIB theory. This duality relates strongly
coupled theory to weakly coupled theory. Assuming solution of the form 
equation (\ref{A1}) 
in string frame rules for transformation of 
background fields under S duality are following:
\begin{eqnarray}
 \label{SD1}
 g_{\mu\nu} & \to & g'_{\mu\nu} = e^{-\phi} \, g_{\mu\nu} \\
 \label{SD2}
 B_{\mu\nu} & \to & B'_{\mu\nu} = B_{\mu\nu}\;,\;\;
 (B^{NS}_{\mu\nu} \leftrightarrow B^{R}_{\mu\nu})\;\footnotemark \\
 \label{SD3}
 D_{\mu\nu\lambda\sigma} & \to & D'_{\mu\nu\lambda\sigma} = 
 D_{\mu\nu\lambda\sigma} \\
 \label{SD4}
 \phi & \to & \phi' = -\phi
\end{eqnarray}
\footnotetext{This notation means S duality transform NS-NS 2-form
potential to R-R 2-form, and back.}
So we see it transform $D1$ brane to $F1$ string and $D5$ brane to
$NS5$ brane. $D3$ brane remains unchanged and coupling constant 
goes to $(\mbox{coupling constant})^{-1}$.

\chapter{BPS Intersection Rules} \label{ap:BPS}

In general a single $p$-brane solutions break half of the supersymmetries 
present in the theory. If we add another set of $q$-branes intersecting the 
first set, the new solution breaks another half of the supersymmetries. But it 
is possible to add second set in such a way that, it breaks the same 
supersymmetries as the first set does. Then we have a solution which preserves
half of the total supersymmetries. We call, this particular way of adding more 
sets of branes, BPS intersection rules. A compact form of these rules can be 
found in \cite{bOrtin}, [See also reference given in \cite{bOrtin}.]

We list a few of these rules hare. We give for both 10-dimensional 
and 11-dimensional theory. 
\begin{table}[H]
 \centering
 \begin{tabular}{|c|}
  \hline \hline
  $F1 \parallel NS5$, $F1 \perp Dp \,(0)$, \\
  $NS5 \perp NS5 \,(1)$, $NS5 \perp NS5 \,(3)$, $NS5 \perp Dp \,(p-1)\; 
(p>1)$, \\
  $Dp \perp Dq \,(m) \;:\;\; p+q = 4 +2m$, \\
  $W \parallel F1$, $W \parallel NS5$, $W \parallel Dp$, \\
  $KK6 \perp Dp \, (p-2)$ \\
  \hline \hline
 \end{tabular}
\caption{{\em BPS intersections in 10 dimensions.}}
\label{tab:bps10}
\end{table}

\begin{table}[H]
 \centering
 \begin{tabular}{|c|}
  \hline \hline
  $M2 \perp M2 \, (0)$, $M2 \perp M5 \, (1)$, $M5 \perp M5 \, (1)$,
$M2 \perp M2 \, (3)$, \\
  $W  \parallel M2$, $W  \parallel M5$, \\
  $KK7M \parallel M2$, $KK7M \perp M2 \, (0)$, $KK7M \parallel M5$, 
$KK7M \perp M5 \, (1)$, $KK7M \perp M5 \, (3)$\\
  $W \parallel KK$, $W \perp KK7M \, (2)$, $W \perp KK7M \, (4)$ \\ 
  \hline \hline
 \end{tabular}
\caption{{\em BPS intersections in 11 dimensions.}}
\label{tab:bps11}
\end{table}
The notation of the above tables is follows: 

$Xp \parallel Yq$ means $X$ and $Y$ type branes are considered. $X$ brane is 
$p$ dimensional and $Y$ brane is $q$ dimensional. $\parallel$ indicates all
of the worldvolume directions of $Xp$ and $Yq$ are parallel. 
$Xp \perp Yq \, (m)$ means $Xp$ and $Yq$ branes worldvolume intersects 
only in $m$ directions.
\chapter{To Show $E \ge 0$} \label{ap:Ege0}

Let $\vec{1} = (1, \; 1, \; \cdots, 1) \;$ and $\vec{v} = (v_1,
\; v_2, \cdots, \; v_n)$ be the standard $n$ -- component
vectors with the standard vector product. Let $\theta_n \;$ be
the angle between them. Then $\vec{1} \cdot \vec{1} = n \;$,
$\vec{v} \cdot \vec{v} = \sum_a v^2_a \;$, $(\vec{1} \cdot
\vec{v})^2 = (\sum_a v_a)^2 = n \; cos^2 \theta_n \; \sum_a
v_a^2 \;$, and we have the Schwarz inequality in the form
\begin{equation}\label{sch}
n \; \sum_{a = 1}^n v_a^2 - (\sum_{a = 1}^n v_a)^2 = 
n \; \sigma_n^2 \; \ge \; 0
\end{equation}
where $\sigma_n^2 = sin^2 \theta_n \; \sum_{a = 1}^n v_a^2 \;$.
The equality is valid, {\em i.e.}  $\sigma_n = 0 \;$, if and
only if $sin \; \theta_n = 0 \;$, equivalently $v_1 = \cdots =
v_n \;$.

\vspace{6ex}

We now show the following: 

\vspace{2ex}

\noindent 

{\em Let $G^{i j}$ and $G_{i j}$ be given by equation
(\ref{Gij}). If $u_i$ and $L^i$ satisfy the relations $\sum_i
u_i L^i = 0 \;$ and $\sum_{i j} G^{i j} u_i u_j > 0 \;$ then $2
E = - \sum_{i j} G_{i j} L^i L^j \ge 0 \;$. $E$ vanishes if and
only if $L^i \;$ all vanish.}

\vspace{2ex}

{\bf Proof :} It is clear that $E$ vanishes if $L^i \;$ all
vanish. Now, let $\vec{1} = (1, \; 1, \; \cdots, 1) \;$,
$\vec{u} = (u_1, \; \cdots, u_{D -1}) \;$, and $\theta \;$ be
the angle between them. Then $(\sum_i u_i)^2 = (D - 1) \; cos^2
\theta \; \sum_i u_i^2 \;$. Hence, $\sum_{i j} G^{i j} u_i u_j =
\frac{1} {D - 2} \; (\sum_i u_i)^2 - \sum_i u^2_i \; \; > 0 \;$
implies that
\begin{equation}\label{theta}
1 - (D - 1) \; sin^2 \theta > 0 \; \; .
\end{equation}
The vector $\vec{L} = (L^1, \; \cdots , \; L^{D - 1})$ is
perpendicular to $\vec{u} \;$ since $\sum_i u_i L^i = 0 \;$. Let
$\vec{L} = \vec{L}_\perp + \vec{L}_\parallel \;$ where
$\vec{L}_\perp$ is perpendicular to the plane defined by
$\vec{1}$ and $\vec{u} \;$, and $\vec{L}_\parallel$ lies in
it. Then $\sum_i (L^i)^2 = L^2_\perp + L^2_\parallel \;$ where
$L^2_\perp = \vec{L}_\perp \cdot \vec{L}_\perp \;$ and $
L^2_\parallel = \vec{L}_\parallel \cdot \vec{L}_\parallel
\;$. Since $\vec{L}$ and $\vec{u}$ are perpendicular and
$\vec{L}_\parallel$ lies in the plane defined by $\vec{1}$ and
$\vec{u} \;$, it follows that $\vec{L}_\parallel$ is
perpendicular to $\vec{u}$, and that the angle between the
vectors $\vec{1}$ and $\vec{L}_\parallel$ is $ \frac{\pi}{2} \pm
\theta \;$. We then have
\begin{eqnarray*}
2 E & = & - \sum_{i j} G_{ij} L^i L^j 
= \sum_i (L^i)^2  - (\sum_i L^i)^2 \\
& = & L^2_\perp + L^2_\parallel 
- (D - 1) \; L^2_\parallel \; sin^2 \theta \; \; \ge \; 0 
\end{eqnarray*}
where the inequality follows from equation (\ref{theta}). The
equality holds, and hence $E$ vanishes, only when $L^2_\perp =
L^2_\parallel = 0 \;$, {\em i.e.} only when $L^i \;$ all
vanish. This completes the proof.

\chapter{Signs and non vanishing of 
$(\Lambda_\tau, \; l^I_\tau) \;$} \label{ap:Sign}

Here, we show that the inequality in equation (\ref{d6ex})
implies that none of $(\Lambda_\tau, \; l^I_\tau) \;$ may
vanish, and that they must all have same sign.

Setting $x_I = l^I_\tau \;$, equation (\ref{d6ex}) becomes $X =
12 u^2 \; ( E + \sum_I e^{l^I} ) \; > 0 \;$ where the polynomial
$X = (x_1 + x_2 + x_3 + x_4)^2 - 3 (x^2_1 + x^2_2 + x^3_3 +
x^2_4) \;$. Now, if any of the $x_I$ vanishes then $X \le 0 \;$,
see the Schwarz inequality given in equation (\ref{sch}). Hence,
none of the $x_I$ may vanish. Rewrite $X$ as
\begin{eqnarray*}
X & = & 
\left\{ (x_1 + x_2 + x_3)^2 - 3 (x^2_1 + x^2_2 + x^3_3) 
\right\} - 2 x^2_4 + 2 x_4 \; (x_1 + x_2 + x_3) \\
& = & 
\left\{ (x_1 + x_2)^2  - 2 (x^2_1 + x^2_2) \right\} 
+ \left\{ (x_3 + x_4)^2 - 2 (x^2_3 + x^2_4) \right\} \\
& & 
- (x^2_1 + x^2_2 + x^3_3 + x^2_4) 
+ 2 (x_1 + x_2) \; (x_3 + x_4) 
\end{eqnarray*} 
and note that $ \left\{ \cdots \right\} \le 0 \;$ for each curly
bracket, see equation (\ref{sch}). Hence, the necessary
conditions for $X > 0$ are
\[
x_4 \; (x_1 + x_2 + x_3) > 0 
\; \; \; \; \; \; , \; \; \; \; \; \; 
(x_1 + x_2) \; (x_3 + x_4) > 0 \; \; . 
\]
Let one of the $x_I \;$, {\em e.g.} $x_4 \;$, be negative and
the other three positive. This violates the first inequality
above and, hence, is not possible. Let two of the $x_I \;$, {\em
e.g.} $x_3 \;$ and $x_4 \;$, be negative and the other two
positive. This violates the second inequality above and, hence,
is not possible. Similarly, three of the $x_I \;$ being negative
and one positive is also not possible. Thus, the only
possibility is that all $x_I$ have same sign. Thus we have that
none of the $l^I_\tau \;$ may vanish, and that they must all
have same sign.

With $l^I_\tau \;$ denoted as $x_I \;$, equation (\ref{d5ex})
for $\Lambda_\tau \;$ becomes
\[
6 u \; \Lambda_\tau = 2 x_1 + 2 x_2 + x_3 + x_4 + 6 u \; L 
\; \; .
\]
Note that $u > 0 \;$. If $L = 0 \;$ then it follows that
$\Lambda_\tau $ does not vanish and has the same sign as $x_I
\;$. Consider now the case where $L \ne 0 \;$. Using equation
(\ref{sch}) to eliminate $\sum_I x_I^2 \;$ in the polynomial
$X$, we obtain
\[
X = \frac{1}{4} \; (x_1 + x_2 + x_3 + x_4)^2 - 3 \sigma_4^2 = 
12 u^2 \; ( E + \sum_I e^{l^I} ) \; \; .
\]
Using the inequality $2 E > 3 (L)^2 \;$, see equation
(\ref{E>}), it follows that $(x_1 + x_2 + x_3 + x_4)^2 > 72 u^2
(L)^2 \;$. Combined with the earlier result on $l^I_\tau \;$,
this inequality implies that $(x_1 + x_2 + x_3 + x_4 + 6 u L)
\;$, and hence $\Lambda_\tau \;$ given above, may not vanish and
must have the same sign as $x_I = l^I_\tau \;$, irrespective of
whether $L$ is positive or negative. This completes the proof.

\chapter{Set of $K^I \;$ which maximises $\tau_a \;$} \label{ap:KI}

With no loss of generality, let $0 < - l^1_0 \le \cdots \le -
l^4_0 \;$. The corresponding set of $K^I \;$ which satisfies
equation (\ref{d6ic}), with $E = 1 \;$, and which maximises
$\tau_a = min \{ \tau_I \} \;$, where $\tau_I = - \frac {l^I_0}
{K^I} \;$, may be obtained by the following algorithm. The
required analysis is straightforward but a little tedious and,
hence, is omitted.

\begin{itemize}

\item Let $K^1 = - l^1_0 \; K \;$. It will turn out that $\tau_a
= \tau_1 = \frac{1}{K} \;$.

\item Choose $K^2 = - l^2_0 \; K \;$. Then $\tau_2 = \tau_1 \;$. 

\item If $- l^1_0 - l^2_0 \le - l^3_0 \;$ then choose $K^3 = K^4
= - (l^1_0 + l^2_0) \; K \;$. Then $\tau_4 \ge \tau_3 \ge
\tau_2 = \tau_1 \;$.

\item If $- l^1_0 - l^2_0 > - l^3_0 \;$ then choose $K^3 = -
l^3_0 \; K \;$. Then $\tau_3 = \tau_2 = \tau_1 \;$.

\item If $- l^1_0 - l^2_0 > - l^3_0 \;$ and if $- l^1_0 - l^2_0
- l^3_0 \le - 2 l^4_0 \;$ then choose $K^4 = - \frac{1}{2} \;
(l^1_0 + l^2_0 + l^3_0) \; K \;$. Then $\tau_4 \ge \tau_3 =
\tau_2 = \tau_1 \;$.

\item If $- l^1_0 - l^2_0 > - l^3_0 \;$ and if $- l^1_0 - l^2_0
- l^3_0 > - 2 l^4_0 \;$ then choose $K^4 = - l^4_0 \; K \;$.
Then $\tau_4 = \cdots = \tau_1 \;$.

\item $K^I \;$ are all thus determined in terms of $K \;$.
Equation (\ref{d6ic}), with $E = 1 \;$, will now determine $K
\;$.

\end{itemize}

\bibliographystyle{hbni}
\bibliography{thesis}
\end{document}